\UseRawInputEncoding
\documentclass[12pt]{article}
\pdfoutput=1 
\usepackage{amssymb, amsmath,amsfonts}
\usepackage{graphicx}
\usepackage[pdftex, bookmarks=true,colorlinks,linkcolor=red,urlcolor=blue,citecolor=blue]{hyperref}
\usepackage[normalem]{ulem}
\usepackage{cite}

\def\arXiv#1{\href{http://arxiv.org/abs/#1}{arXiv:#1}}
\def\arXiv#1#2{\href{http://arxiv.org/abs/#1}{arXiv:#1}}

\def\doi#1#2{\href{http://doi.org/#1}{#2}}

\makeatletter\@addtoreset{equation}{section}\makeatother

\newcommand{\preprint}[1]{\begin{table}[t]  
             \begin{flushright}               
             {#1}                             
             \end{flushright}                 
             \end{table}}                     
\renewcommand{\title}[1]{\vbox{\center\LARGE{#1}}\vspace{3mm}}
\renewcommand{\author}[1]{\vbox{\center#1}\vspace{3mm}}
\newcommand{\address}[1]{\vbox{\center\em#1}}

\usepackage{bm}
\def\be{\begin{eqnarray}}
\def\ee{\end{eqnarray}}
\def\bea{\begin{eqnarray}}
\def\eea{\end{eqnarray}}
\newcommand{\nn}{\nonumber}

\def\Dslash{\,\,{\raise.15ex\hbox{/}\mkern-12mu D}}
\def\Dbarslash{\,\,{\raise.15ex\hbox{/}\mkern-12mu {\bar D}}}
\def\delslash{\,\,{\raise.15ex\hbox{/}\mkern-9mu \partial}}
\def\delbarslash{\,\,{\raise.15ex\hbox{/}\mkern-9mu {\bar\partial}}}
\def\pslash{\,\,{\raise.15ex\hbox{/}\mkern-9mu p}}
\def\calDslash{\,\,{\raise.15ex\hbox{/}\mkern-12mu {\cal D}}}

\newcommand{\bra}{\langle}
\newcommand{\ket}{\rangle}

\def\lae{\mathrel{\mathop{\smash{\lower .5 ex \hbox{$\stackrel<\sim$}}}}}
\def\lae{\mathrel{\mathop{\smash{\lower .5 ex \hbox{$\stackrel>\sim$}}}}}

\begin{document}

\unitlength = .8mm

\begin{titlepage}
\vspace{.5cm}
\preprint{}

\begin{center}
\hfill \\
\hfill \\
\vskip 0.5cm

\title{ Topological hydrodynamic modes and holography}
\vskip 0.5cm


{Yan Liu$^{a}$}\footnote{Email: {\tt yanliu@buaa.edu.cn}} and
 {Ya-Wen Sun$^{b,c}$}\footnote{Email: {\tt yawen.sun@ucas.ac.cn}}
\address{${}^a$Center for Gravitational Physics, Department of Space Science, Beihang University, Beijing 100191, China\\
}
 \address{${}^b$School of physics $\&$ CAS Center for Excellence in Topological Quantum Computation, University of Chinese Academy of Sciences, Beijing 100049, China\\
 \vspace{.2cm}
 ${}^c$Kavli Institute for Theoretical Sciences,  University of Chinese Academy of Sciences, Beijing 100049, China}

\end{center}

\abstract{ We study topological modes in relativistic hydrodynamics by weakly breaking the conservation of energy momentum tensor. Several systems have been found to have topologically nontrivial crossing nodes in the spectrum of hydrodynamic modes and some of them are only topologically nontrivial with the protection of reflection symmetries in two directions. The nontrivial topology for all these systems is further confirmed from a calculation of the topological invariant. Associated transport properties and second order effects have also been studied for these systems. The non-conservation terms of the energy momentum tensor could come from an external effective symmetric tensor matter field or a gravitational field which could be generated by a specific non-inertial reference frame transformation from the original inertial reference frame. Finally we introduce a possible holographic realization of one of these systems. We propose a new method to calculate the holographic Ward identities for the energy momentum tensor without calculating out all components of the Green functions and match the Ward identities of both sides.}

\vfill

\end{titlepage}

\begingroup
\hypersetup{linkcolor=black}
\tableofcontents
\endgroup

\section{Introduction} 

Topological states of quantum matter have been studied extensively in condensed matter physics during the last decade \cite{Wen:2016ddy,Witten:2015aoa}. Well known topological states include anomalous Hall effect, topological insulators, topological superconductors, which are gapped, and various kinds of topological semimetals, e.g. Weyl/nodal line/Dirac semimetals, which are gapless topological states of matter. 

The earliest topological states of matter were all found in quantum electronic systems, however, soon it was realized that the topological property is not fundamentally quantum but is associated with the wave property of the system. In 2008, topological states in classical/bosonic systems have been predicted in several systems in \cite{Haldane20, Haldane21, Wang22}. Moreover, classical topological states of matter were realized experimentally in gyromagnetic photonic crystals at microwave frequencies \cite{Wang22}. Ever since then, the study in classical topological states of matter has been grown rapidly, see e.g. \cite{review23} and references therein. It has been found that many classical systems have nontrivial topological states too, e.g. topological optical/sound systems (see e.g. \cite{topphoton, topphoton2, natphy} and references therein), which have also been confirmed experimentally. 

In a recent paper \cite{Liu:2020ksx}, we have shown that semimetal-like topologically nontrivial modes could be found in relativistic hydrodynamics by weakly breaking the conservation of energy momentum. Hydrodynamics describes the classical behavior of a system of continuum away from its local equilibrium in the long wavelength and low frequency limit. The dynamics of a system in the hydrodynamic limit is governed by conservation equations of conserved currents, including the energy momentum tensor and the conserved charges of internal symmetries. In the simplest case with the only conserved quantity being the energy momentum tensor, we found that a special form of non-conservation terms of the energy momentum tensor in the conservation equation could deform the spectrum of hydrodynamics to a shape similar to that of topological semimetals. 

We provided an explicit example with this kind of behavior in a 4D hydrodynamic system in \cite{Liu:2020ksx} and in that case, the nontrivial topology of the system requires the protection of a special spacetime symmetry. In this paper, we generalize the findings to more systems with similar behavior and provide more detailed calculations on various aspects. Besides the single 4D system found in \cite{Liu:2020ksx}, we find several interacting hydrodynamic systems with two energy momentum sectors exchanging energy and momentum with each other as well as with the environment. Besides the fact that the crossing nodes in these systems will not become gapped under small perturbations, we will also show evidence of the nontrivial topology from the calculation of topological invariants. We also calculate the second order effects in detail in this paper for the single 4D system and show explicitly that the dissipative second order part contributes to an imaginary part in the dispersion relation of the hydrodynamic modes. This imaginary part shows a jump at the topologically nontrivial crossing node, seeming to provide another piece of evidence that the states near the nodes are not adiabatically connected to each other and the crossing nodes are topologically nontrivial.

The non-conservation terms of the energy momentum tensor could come from an appropriately chosen external matter field or a gravitational field. In the latter case, we could start from a flat spacetime and perform a specific reference frame transformation. The new spacetime could be viewed as a non-inertial reference frame. This suggests that topologically trivial hydrodynamic modes could become topologically nontrivial viewed by an observer accelerating in a specific way.

As hydrodynamics has been studied extensively for strongly coupled systems in holography with many important results \cite{Son:2007vk, Rangamani:2009xk}, e.g. the prediction of a KSS bound as the lower bound for the shear viscosity over entropy ratio \cite{Kovtun:2004de}, which has been confirmed in experiments, we also propose a holographic realization for the single 4D system that have topologically nontrivial hydrodynamic modes. We start from ordinary AdS/CFT correspondence and perform a specific reference frame transformation to get a non-inertial reference frame version of AdS/CFT correspondence. We develop a new method to calculate holographic Ward identities for the energy momentum tensor without calculating out all components of the Green functions. Using this method we show that the holographic Ward identities from our new holographic set-up match to the Ward identities of the hydrodynamic systems with specific non-conservation terms of the energy momentum tensor. This provides the evidence that we have found a strongly coupled holographic system which has topologically nontrivial hydrodynamic modes.



In the rest of the paper, we will first review some basics of relativistic hydrodynamics in section \ref{sec2}. In section \ref{sec:tophm}, we introduce the notion of effective Hamiltonians and show several systems which have topologically nontrivial hydrodynamic modes by adding non-conservation terms of the energy momentum tensor. In section \ref{sec:origin} we show the possible origins for the non-conservation terms of the energy momentum tensor, emphasizing the role of non-inertial reference frames. The symmetry of the single 4D system is derived and the symmetry needed for the protection of the topological nodes is also obtained in this section with the help of the external metric field.  We calculate the nontrivial topological invariants in section \ref{sec:ti} and the effects of these extra terms for transport properties in section \ref{sec:tp}. In section \ref{sec:2nd}, we consider second order effects to these systems. In section \ref{sec:wihr} we introduce the holographic realization for the single 4D system and match the Ward identities from holography to the hydrodynamic ones. Section \ref{sec:cd} devotes to conclusions and open questions. 

\section{A short review of relativistic hydrodynamics}
\label{sec2}

Hydrodynamics is the universal low energy theory for systems at long distance and late time. It could describe a variety of physical systems ranging from matter at large scales in the universe, the quark-gluon plasma \cite{CasalderreySolana:2011us}, to Weyl semimetals \cite{Landsteiner:2014vua,Lucas:2016omy} and graphenes \cite{Lucas:2017idv} in the laboratory. At small momentum and frequency, perturbations of a hydrodynamic system away from the equilibrium would produce sound and transverse modes \cite{Kovtun:2012rj}. These modes are gapless whose poles are at $\omega={\bf k}=0$, which reflects the fact that energy momentum is conserved. In this paper we focus on the simplest hydrodynamic systems with no internal charges whose only conserved quantity is the energy momentum tensor. 

The conservation equation for the energy momentum tensor in $d+1$ dimensions is 
\be\label{conserve}
\partial_\mu T^{\mu\nu}&=&0\,.
\ee
Up to the first order in derivative, the constitutive equation for the energy momentum tensor in the Landau frame is
\bea
T^{\mu\nu}&=&\epsilon u^\mu u^\nu+P \Delta^{\mu\nu}-\eta \Delta^{\mu\alpha}\Delta^{\nu\beta}\big(\partial_\alpha u_\beta +\partial_\beta u_\alpha -\frac{2}{d}\eta_{\alpha\beta}\partial_\sigma u^\sigma\big)-\zeta \Delta^{\mu\nu}\partial_\alpha u^\alpha+\mathcal{O}(\partial^2)\,,\nn 
\eea  
where $\Delta_{\mu\nu}=\eta_{\mu\nu}+u_\mu u_\nu$, $\epsilon$, $P$ are the energy densities and pressure and $\eta$, $\zeta$ are the shear and bulk viscosities. 
For conformally invariant systems, the bulk viscosity $\zeta=0.$ For $d=1$ conformal fluid, the first order transport coefficients $\eta$ and $\zeta$ are both zero. 

At equilibrium the system has an energy density $T^{00}=\epsilon$ and pressure $T^{ii}=P$. With small perturbations slightly away from equilibrium, the system would respond to the perturbations and develop hydrodynamic modes. The equations for perturbations of the energy momentum tensor are
\be
\partial_{\mu}\delta T^{\mu \nu}=0\,,
\ee where in four dimensions 
\bea\label{eq:constitutive}
\begin{split}
\delta T^{00} &=\delta \epsilon\,,\\
\delta T^{0i} &=\delta T^{i0}=(\epsilon+p) \delta u^i=\delta\pi^i \,,\\
\delta T^{ij}  &=
g^{ij} v_s^2\delta\epsilon-\frac{\eta}{\epsilon+P} (\partial^i \delta \pi^j+\partial^j \delta \pi^i)+\frac{\frac{2}{3}\eta-\zeta}{\epsilon+P}g^{ij}\partial_\alpha  \delta\pi^\alpha \,
\end{split}
\eea with $v_s=\sqrt{\frac{\partial P}{\partial \epsilon}}$.
We keep the momenta in all directions nonzero for later convenience, i.e. ${\bf k}=(k_x,k_y,k_z)$ in four dimensions. After solving the conservation equation and diagonalizing the solutions, there will be four eigenmodes of the system. Two of them are the sound modes propagating in the direction of ${\bf k}=(k_x,k_y,k_z)$ with the dispersion relation $\omega=\pm v_s k-i\frac{\Gamma_s}{2} k^2$
, where $\Gamma_s=(\frac{4}{3}\eta+\zeta)/(\epsilon+P)$. 
The other two are transverse modes with $\omega=-i\frac{\eta}{\epsilon+P} k^2$. These modes also lead to poles in the Green functions of various components of the energy momentum tensor at low frequency. 

\begin{figure}[h!]
  \centering
  \includegraphics[width=0.270\textwidth]{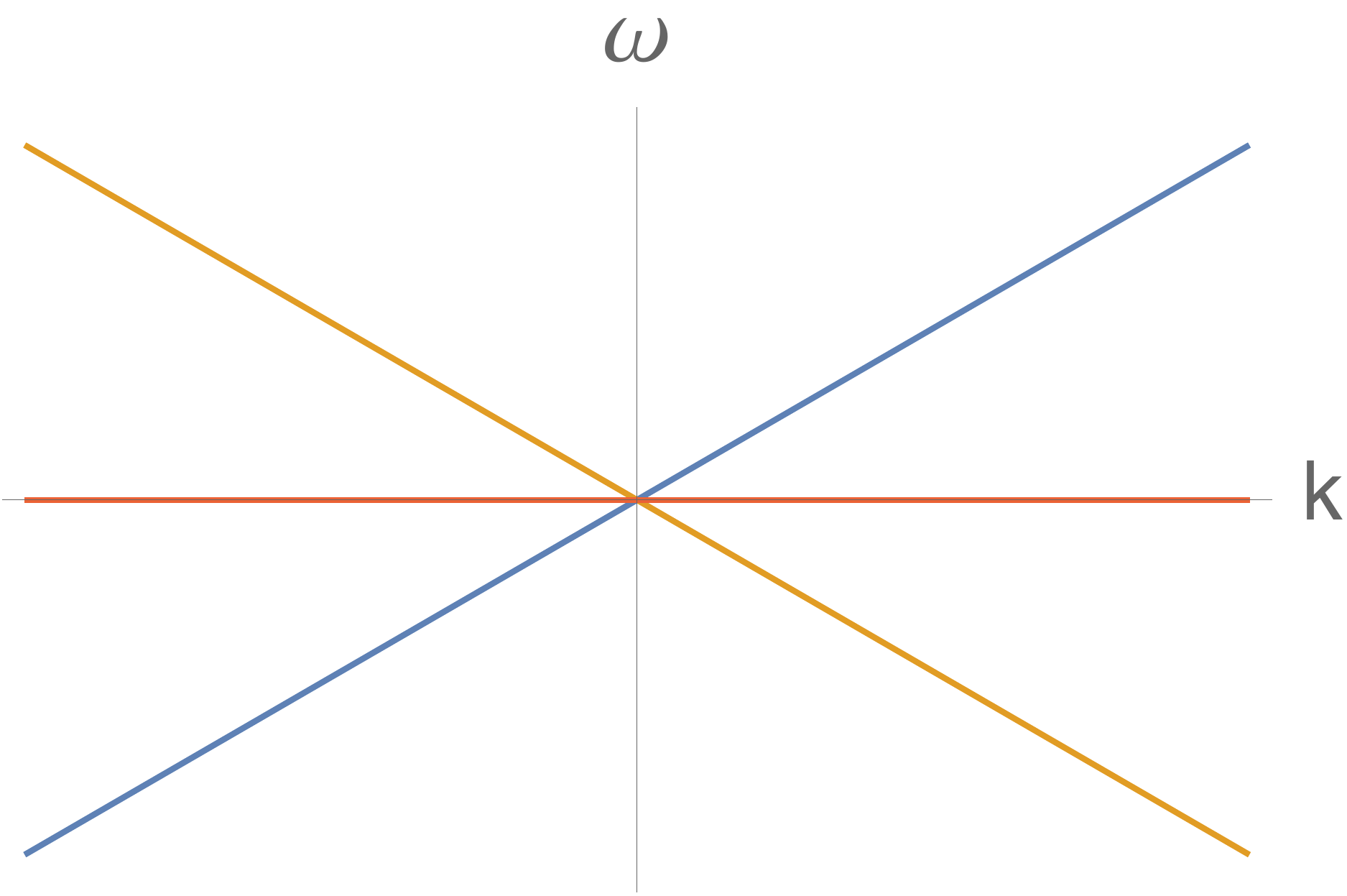}    ~~~
    \includegraphics[width=0.270\textwidth]{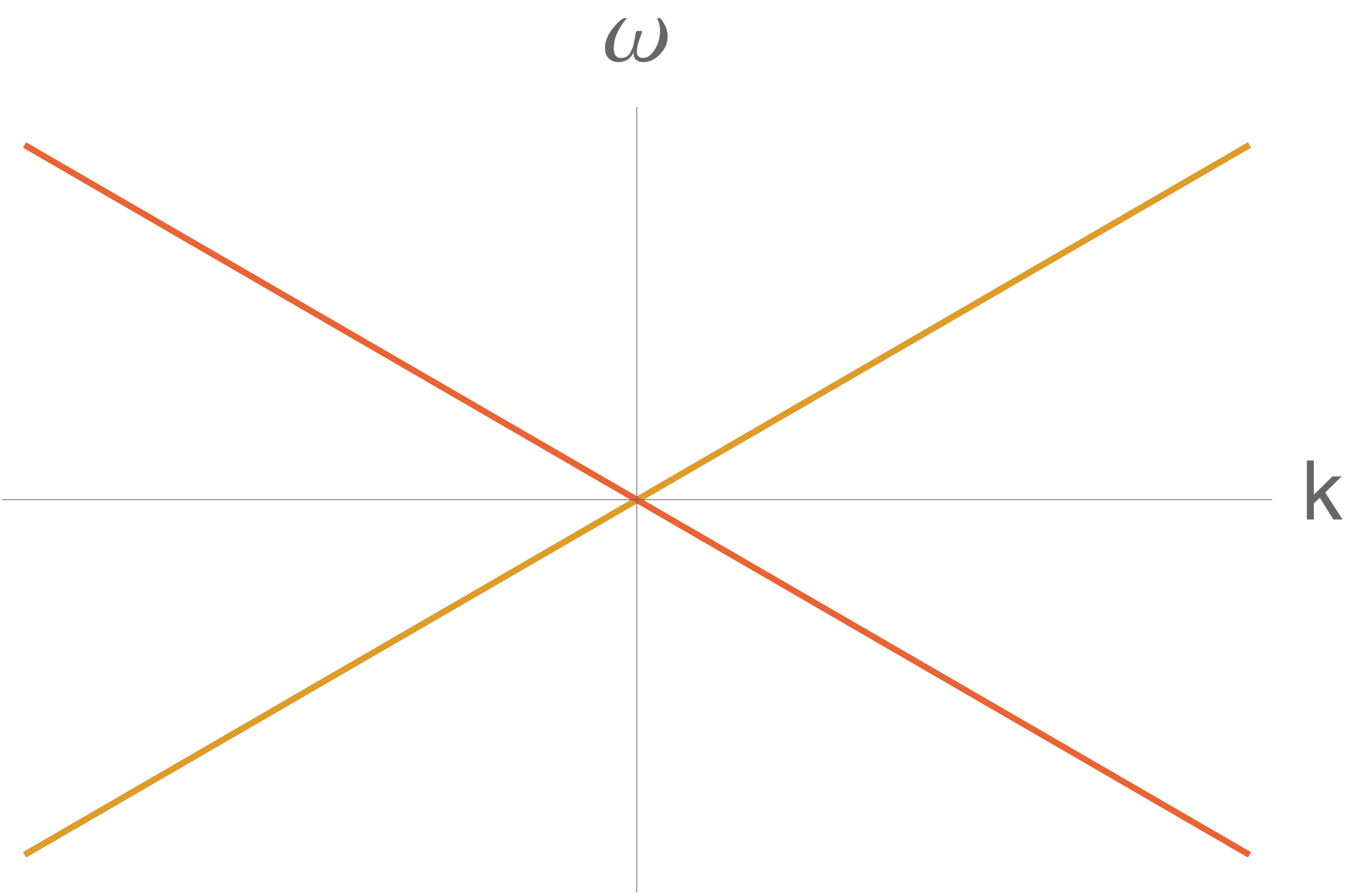}   ~~~ 
      \includegraphics[width=0.270\textwidth]{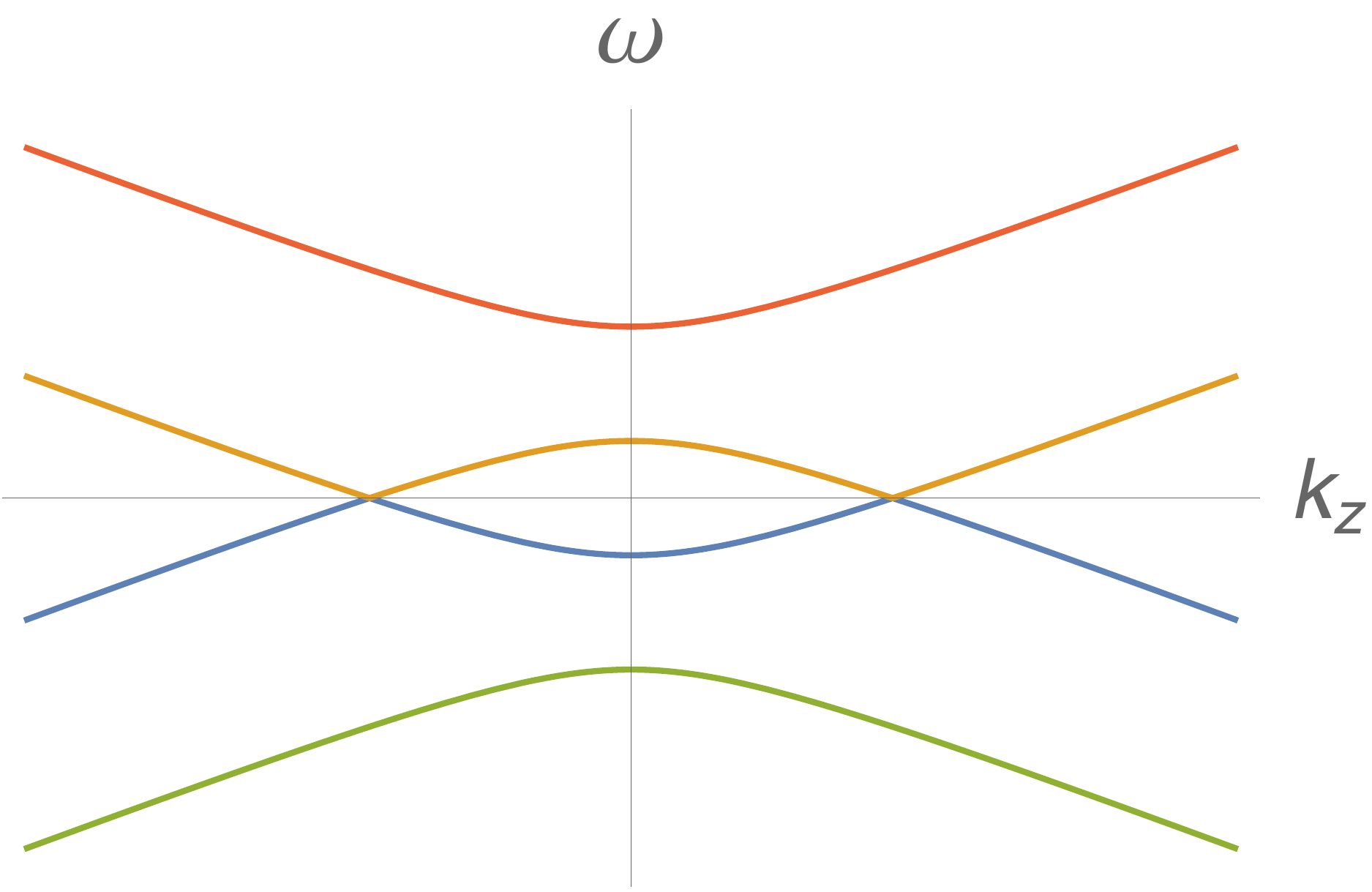}
  \caption{\small The spectrums in relativistic hydrodynamics for $\omega, k\ll T$ ({\it left}), Dirac semimetal ({\it middle}) and Weyl semimetal ({\it right}). }
  \label{fig:spectrums}
\end{figure}

To the first order in $k$, dissipative terms disappear and the spectrums of the four modes are real, which cross each other at $\omega={\bf k}=0$ (see the left plot in Fig. \ref{fig:spectrums}).\footnote{A recent study on the stability issue for the first order hydrodyanmics can be found in \cite{Kovtun:2019hdm}.} This spectrum looks similar to the spectrum of Dirac semimetals (the middle plot in Fig. \ref{fig:spectrums}), except that we have two flat bands here. In the next section we will show that a gap could also open in these hydrodynamic modes after introducing non-conservation terms for the energy momentum tensor and with more non-conservation terms, the hydrodynamic system could develop gapless topologically nontrivial modes similar to the Weyl semimetal states (the right plot in Fig. \ref{fig:spectrums}). 


\section{Topological hydrodynamic modes}
\label{sec:tophm}

The spectrum of the relativistic hydrodynamic modes crosses at $\omega={\bf k}=0$ as a consequence of energy momentum conservation. To deform the spectrum of the hydrodynamic modes and find nontrivial topological structure in the hydrodynamic modes, we introduce non-conservation terms for the energy momentum tensor in (\ref{conserve}) and make sure that the non-conservation terms are small enough to stay within the hydrodynamic limit. The non-conservation of energy and momentum could come from a certain external system which couples to the hydrodynamic system that we study.  We assume that the constitutive equations for perturbations of the hydrodynamic system being considered do not change or the change does not affect the final spectrum and the change will be discussed in section \ref{sec:origin}, while the conservation equations would change due to interchange of energy momentum with external systems. In this section, we consider dissipativeless non-conservation terms that deform the spectrum of the hydrodynamic modes. Before that we will first introduce some basic knowledge about topological states of matter and calculations of topological invariants.

\subsection{A simple introduction of topological states of matter}

In this subsection, we introduce very briefly some basic knowledge about topological states of matter, especially gapless ones. Topological states of matter have attracted a lot of interest among condensed matter physicists during the last years. Topological states of matter are states that have topologically nontrivial band structure in the momentum space. Here we use the terminology ``bands" as in electronic systems and in fact ``bands" could refer to any structure of the spectrum from the eigenvalues of a Hamiltonian. Various kinds of gapped and gapless topological states of matter have been found. Gapped topological states of matter include topological insulators, anomalous Hall effect, topological superconductors, etc., and gapless topological states of matter include Weyl semimetals, nodal line semimetals, etc..
 
In gapped topological states of matter, the bands are separated by a nonzero energy interval and because of the nontrivial topology of the band structure the system cannot be adiabatically deformed to a trivial vacuum state without closing the gap.  Gapless topological states of matter have connected conduction and valence bands in the spectrum, and they cannot be adiabatically deformed to a trivial gapless state, which is a state whose gap could open due to arbitrarily small perturbations of the system. Gapless topological states could not be gapped by an infinitesimal perturbation of the system. 

\begin{figure}[h!]
  \centering
  \includegraphics[width=0.40\textwidth]{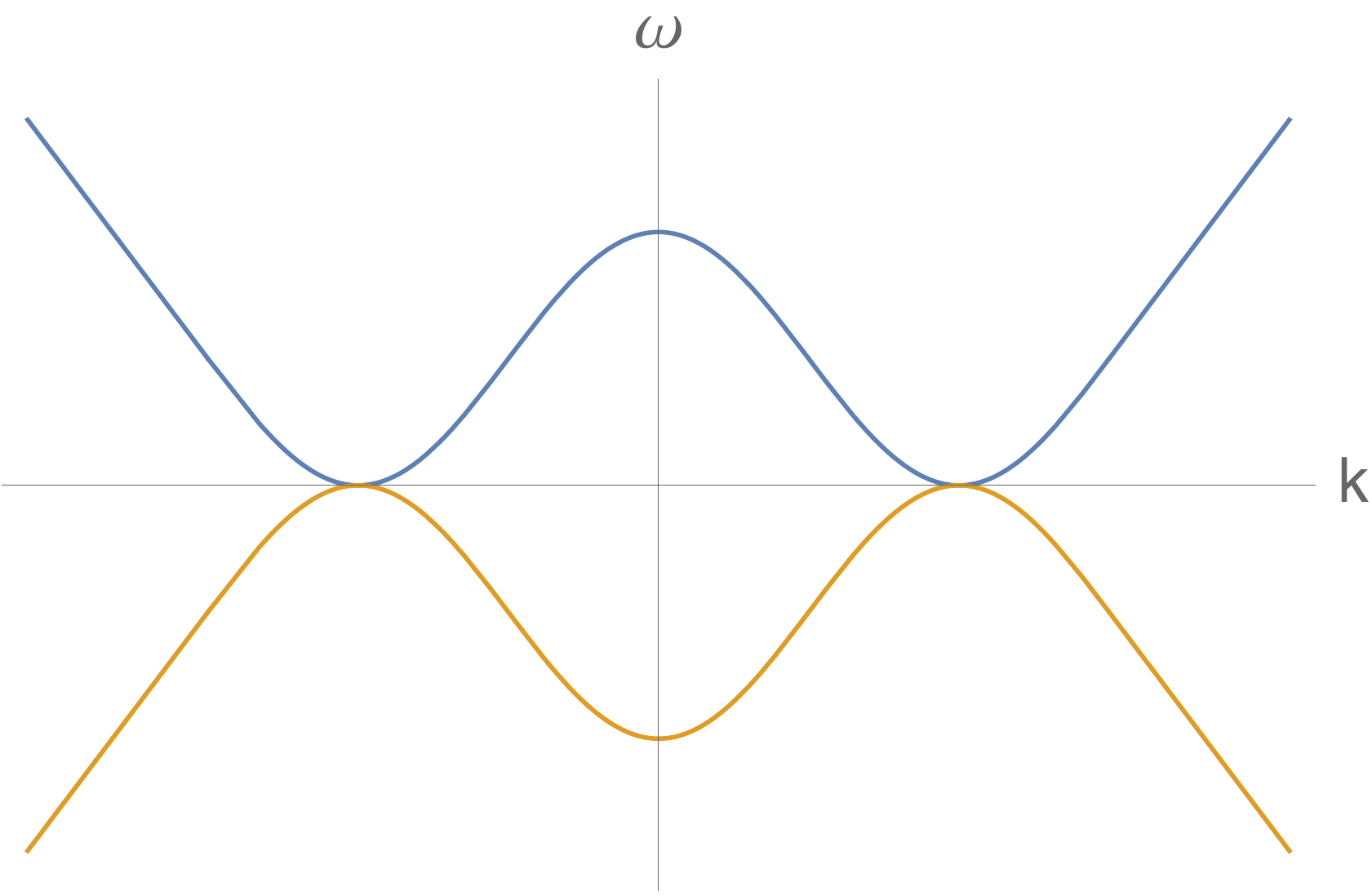}
  \,~~~~~~
  \includegraphics[width=0.40\textwidth]{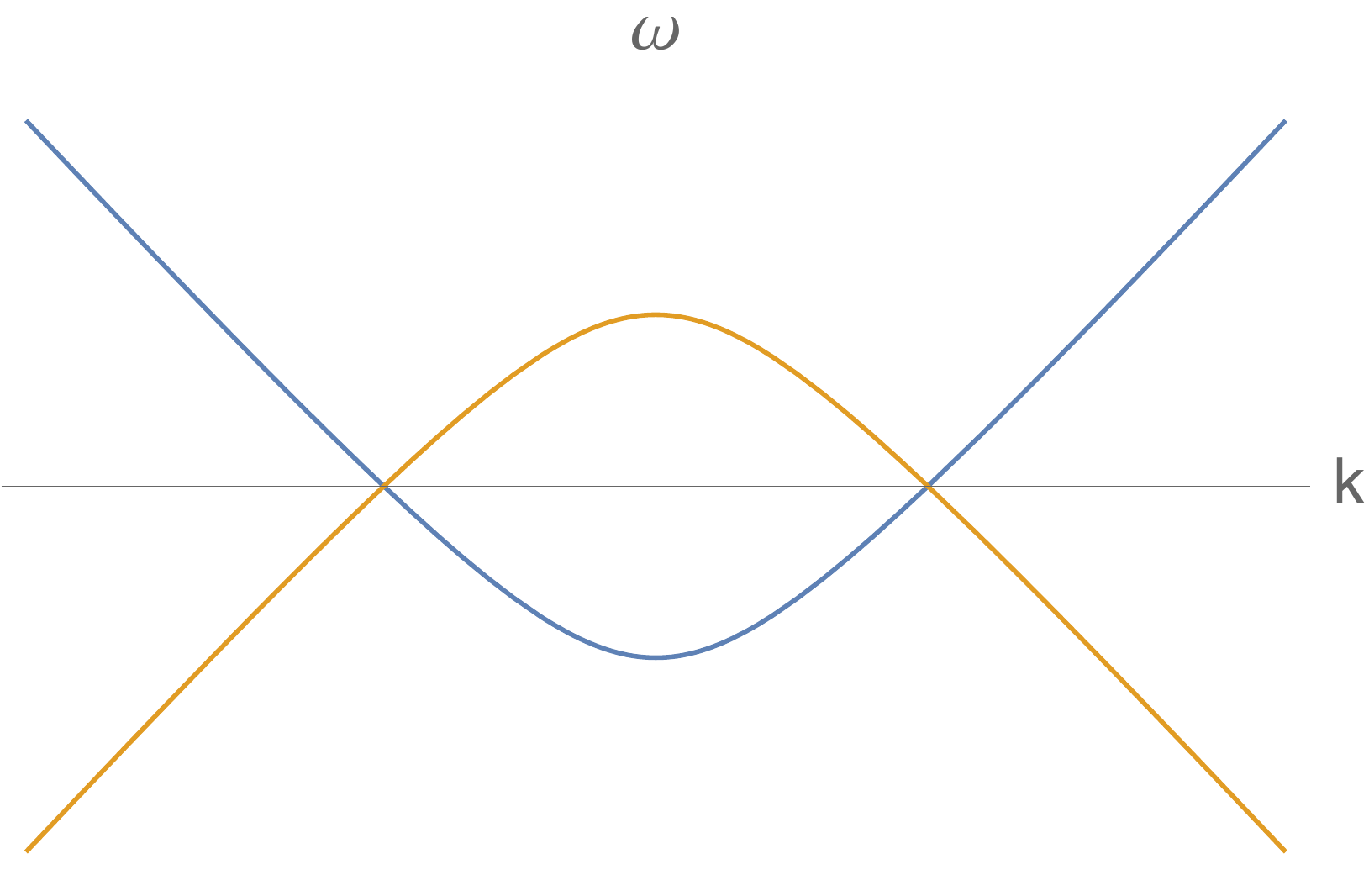}
  \caption{\small An accidental band crossing ({\em left}) and a topologically nontrivial band crossing ({\em right}).}
  \label{fig:ill}
\end{figure}

Fig. \ref{fig:ill} gives an illustration of band crossings that behave differently in their topological properties. The left plot in Fig. \ref{fig:ill} shows an accidental band crossing which would be gapped under an arbitrarily small perturbation and the right plot in Fig. \ref{fig:ill}  shows a topologically nontrivial band  crossing which is still gapless under small perturbations. 

In the figure each curve with the same color denotes the same set of eigenstates with an eigenvalue depending on $k$ as a continuous function $E_{1,2} (k)$. The same set of eigenstates and their eigenvalues would deform continuously under parameter change of the system. The two curves in each figure indicate two sets of eigenstates with two different eigenvalues $E_{1,2} (k)$. At certain values of $k=k_0$, the two eigenvalues coincide with $E_{1} (k_0)=E_{2} (k_0)$ and we have band crossings at these $k_0$. Then at each of these band crossing points, there are two degenerate eigenstates. Then in the left figure of Fig. \ref{fig:ill}, under a small perturbation that moves the upper band up while the lower band down, the system would immediately develop a gap, while in the right figure, under a small perturbation that modifies the curves slightly, the band crossings could not vanish no matter how the two curves move. The right figure represents a topologically nontrivial gapless state while the left one gives a trivial gapless state.

For topologically nontrivial gapless states, the behavior in the previous paragraph could be explained by the fact that band crossing nodes are protected by a nontrivial topological structure and each possesses a nontrivial topological invariant. For example, a Weyl node is protected by a nontrivial topological monopole charge, which could be calculated from the surrounding first Chern number in the three dimensional momentum space. Thus to know if a gapless state is topologically nontrivial or not, there are two ways which should be consistent with each other: one is to see if arbitrarily small perturbations could gap the nodes and the other is to calculate the topological invariant and see if it is different from the value obtained for a trivial vacuum state. More details about calculation of topological invariants for gapless systems in various dimensions would be found in Section \ref{sec:ti}.

\subsection{Effective Hamiltonian}

Before calculating the hydrodynamic modes in different situations, we first review our definition of the effective Hamiltonian in \cite{Liu:2020ksx}, which is a physical quantity parallel to the Hamiltonian matrix of topological electronic systems \cite{shenbook}. In the case of electronic systems, Dirac fermions are four component fields, thus the Hamiltonian from $i\partial_t\Psi=H\Psi$ is a matrix whose eigenvalues give the spectrum of the electronic system. In hydrodynamics, the modes that we consider are perturbations of the energy and momentum densities, thus there are also multiple components of fields in the equation. We could in principle define a similar matrix of effective Hamiltonian whose eigenvalues give the spectrum of the hydrodynamic modes.

We start from the conservation equation of energy momentum tensor which determines the dynamics of hydrodynamic modes.  Substituting the constitutive equations for the perturbations (\ref{eq:constitutive}) into  $
\partial_\mu \delta T^{\mu\nu}=0,$ we could rewrite the equations for the perturbations of energy and momentum densities into
\be
i\partial_t\Psi=H\Psi
\ee
where we have defined
\be\label{effH}
\Psi=\begin{pmatrix} 
\delta \epsilon \\
\delta \pi^x \\
\delta \pi^y  \\
\delta \pi^z
\end{pmatrix}\,,~~~~~~
H=\begin{pmatrix} 
0 & ~~k_x & ~~k_y & ~~ k_z \\
k_x v_{s}^2 & ~~0 & ~~ 0 & ~~ 0 \\
k_yv_{s}^2 & ~~ 0 & ~~ 0 & ~~0 \\
k_z v_{s}^2 & ~~0 & ~~ 0 & ~~ 0 
\end{pmatrix}\,
\ee at leading order in $k$, i.e. omitting dissipative terms at $\mathcal{O}(k^2)$.

In analogy to the electronic systems, in this way we have defined an effective Hamiltonian matrix $H$ whose eigenvalues give the 
spectrum of hydrodynamic modes.\footnote{Note that this effective Hamiltonian is different from the hydrodynamic Hamiltonian, e.g. in section 2.4 of \cite{Kovtun:2012rj} in the sense that this effective Hamiltonian matrix $\Psi^T (-k) H \Psi(k)$ gives the ``kinetic" part of the Hamiltonian while the latter only counts in the potential part associated with the source terms.}  The four eigenvalues of the matrix Hamiltonian above give the sound modes $\omega=\pm v_s \sqrt{k_x^2+k_y^2+k_z^2}$ and double copies of transverse modes $\omega=0$. The form (\ref{effH}) is the ``free" Hamiltonian matrix for a conserved energy momentum tensor.



\subsection{Gapping the system } 

With exact energy momentum conservation the hydrodynamic modes are ``band crossed" at $\omega={\bf k}=0$. However, when the energy momentum tensor is not conserved, the modes might become gapped. To introduce nontrivial topological structure into the spectrum of hydrodynamic modes, the first step is to generate gaps in the modes as in electronic systems. In this section we will directly modify the effective Hamiltonian to find terms that we need and later in section \ref{sec:origin} we will show how these terms could arise from external fields.



We modify the conservation equation for $T^{\mu\nu}$ to 
\bea\label{eq:mod1} 
\begin{split}
\partial_\mu \delta T^{\mu t}&=m_1 \delta T^{tx}\,,\\
\partial_\mu \delta T^{\mu x}&=-m_2 v_s^2 \delta T^{tt}
\,,\\
\partial_\mu \delta T^{\mu y}&=0\,,\\
\partial_\mu \delta T^{\mu z} &=0\,,
\end{split}
\eea
where $m_{1,2}$ are parameters leading to a gap in the spectrum of the hydrodynamic modes. To stay in the hydrodynamic limit, $m_{1,2}$ has to be small compared to $T$. Here we have introduced non-conservation of energy and momentum in the $x$ direction and this non-conservation has to be in a specific form as indicated above. The amount of non-conserved energy/$x$-direction momentum density is proportional to the $x$-direction momentum/energy density and the sign of $m_{1,2}$ has to be the same. This sign could be either positive or negative indicating that the modes could either gain or lose energy. For simplicity, in the following we will choose $m_1=m_2=m$.

\begin{figure}[h!]
  \centering
  \includegraphics[width=0.450\textwidth]{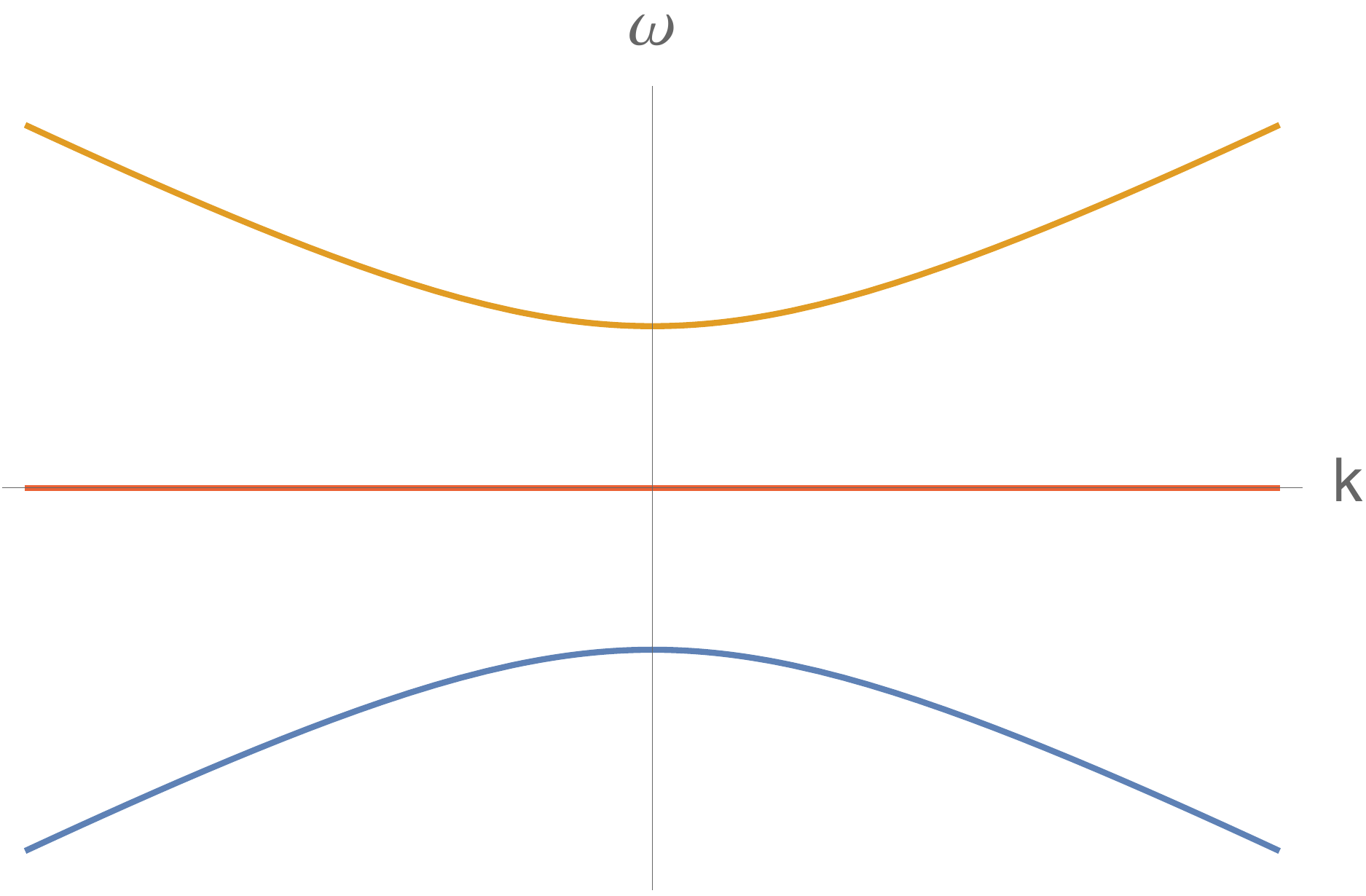}
  \caption{\small The spectrum of the modified hydrodynamics with dynamical equation (\ref{eq:mod1}). Here $m_1=m_2=m$ and it plays the role to gap the spectrum of sound modes. Here and in all the following pictures $\omega, k, m\ll T$. }
  \label{fig:gap}
\end{figure}

At leading order in $k$ the effective Hamiltonian becomes
\be
H=\begin{pmatrix} 
0 & ~~k_x-i m & ~~k_y & ~~ k_z \\
(k_x+i m) v_{s}^2 & ~~0 & ~~ 0 & ~~ 0 \\
k_yv_{s}^2 & ~~ 0 & ~~ 0 & ~~0 \\
k_z v_{s}^2 & ~~0 & ~~ 0 & ~~ 0 
\end{pmatrix}\,
\ee and $k^2$ effects will be studied in section \ref{sec:2nd}. The four eigenvalues of the matrix $H$ are now \be\omega=\pm v_s\sqrt{k_x^2
+k_y^2+k_z^2+m^2}\ee and double sectors of $ \omega=0 $ . This shows that $m$ terms indeed gap the sound modes while do not change the transverse modes. Fig. \ref{fig:gap} shows the ``band structure" of the system at nonzero $m$. We can see that sound modes only exist above/below a certain frequency while transverse modes are not affected.

Effects in the spectrum here would be the same if the momentum $m$ term is in the $y$ or $z$ direction instead of the $x$ direction, or if $m$ terms in all $x$, $y$ and $z$ directions exist. This is important later when we need to check carefully if the topological structure of the modes would not be destroyed by these $m$ terms in all directions or only in certain directions. Here we also emphasize that gapping the system is a first step to realize topologically nontrivial gapless states as we need to show that these mass terms which could gap the standard hydrodynamic system will not gap the topologically nontrivial gapless state that we introduce later confirming that they are indeed topologically nontrivial. However we are not saying that the gapped state in this section is topologically nontrivial. 





\subsection{Topologically nontrivial nodes}

After gapping the system, we could continue to search for possible modifications to the effective Hamiltonian that change the topologically trivial crossing nodes of the hydrodynamic modes into nontrivial ones with more complicated ``band structures". This procedure is similar to introducing time reversal symmetry breaking terms in a Dirac system to obtain a topologically nontrivial Weyl semimetal. Here we need to find a topologically nontrivial spectrum by introducing modifications to the effective Hamiltonian. The possible modifications here are not arbitrary as most deformations of the effective Hamiltonian would give non-meaningful results. We have tried a lot of possible deformations to finally find the ones that could realize the topologically nontrivial spectrum here.

There are more than one ways to realize this and we list several possibilities in this section, including the single 4D hydrodynamic system found in \cite{Liu:2020ksx}, and interacting hydrodynamic systems with two sectors who further exchange energy and momentum with each other.\footnote{Hydrodynamical systems with multiple sectors of energy momentum tensors and currents have also been studied in e.g. \cite{Lucas:2016omy}.} Further possibilities and possible ways to get topologically nontrivial gapped hydrodynamic modes will be left for future work.

\subsubsection{The single 4D hydrodynamic system}
\label{subsec:single4D}

The first possibility is to take a 3+1D hydrodynamic system of (\ref{eq:mod1}) and further break the translational symmetry in the $y$ and $z$ directions, as found in \cite{Liu:2020ksx}. The momentum in the two directions is non-conserved in a specific form so that the resulting spectrum of the hydrodynamic modes would develop crossing nodes.

More precisely, the conservation equation is now
\bea\label{eq:4dhydro}
\begin{split}
\partial_\mu \delta T^{\mu t}&=m_1 \delta T^{tx}\,,\\
\partial_\mu \delta T^{\mu x}&=-m_2 v_s^2 \delta T^{tt}
\,,\\
\partial_\mu \delta T^{\mu y}&=b_1 v_s \delta T^{tz}\,,\\
\partial_\mu \delta T^{\mu z} &=-b_2 v_s \delta T^{ty}\,,
\end{split}
\eea
where $m_{1,2}$ terms are mass terms while $b_{1,2}$ terms change the momentum position of the crossing nodes in the spectrum. Here we have used different values of mass and $b$ terms in equations for generality and in the following we will take $m_1=m_2=m$ and $b_1=b_2=b$ for simplicity.

Note that in the attempt to modify the effective Hamiltonian to get topologically nontrivial hydrodynamic modes, we have to keep the modified effective Hamiltonian in a Hermitian form or similar to a Hermitian matrix at order $\mathcal{O}(k)$. The eigenvalues of $H$ would then be real and there would be no dissipative effects at order $\mathcal{O}(k)$.

After substituting the constitutive equation (\ref{eq:constitutive}) into 
the conservation equation above, the equation (\ref{eq:4dhydro}) can be written as 
\be
i\partial_t\Psi=H\Psi
\ee
where
\be\label{h4d}
\Psi=\begin{pmatrix} 
\delta \epsilon \\
\delta \pi^x \\
\delta \pi^y  \\
\delta \pi^z
\end{pmatrix}\,,~~~~~~
H=\begin{pmatrix} 
0 & ~~k_x+im & ~~k_y & ~~ k_z \\
(k_x-im)v_{s}^2 & ~~0 & ~~ 0 & ~~ 0 \\
k_yv_{s}^2 & ~~ 0 & ~~ 0 & ~~ib v_s \\
k_z v_{s}^2 & ~~0 & ~~ -ib v_s & ~~ 0 
\end{pmatrix}\,.
\ee
Redefine $\delta \epsilon\to \frac{1}{v_s}\delta \epsilon$, it is easy to see that $H$ is similar to a Hermitian matrix where $v_s$ becomes an overall scaling factor, thus this effective $H$ has real eigenvalues. In the following for simplicity we will ignore the factor of $v_s$ and when necessary it can be taken back by an inverse redefinition of $\delta \epsilon\to \frac{1}{v_s}\delta \epsilon$.

The spectrum of the hydrodynamic modes are now
\be
\omega=\pm \frac{1}{\sqrt{2}}\sqrt{b^2 + k^2  + m^2\pm\sqrt{(k_x^2 + m^2 - b^2)^2+(k_y^2 + k_z^2)^2 + 2 (k_y^2 + k_z^2) (k_x^2 + m^2 + b^2)}}\,,\nn
\ee 
where $k=\sqrt{k_x^2+k_y^2 + k_z^2}$.

\begin{figure}[h!]
  \centering
  \includegraphics[width=0.240\textwidth]{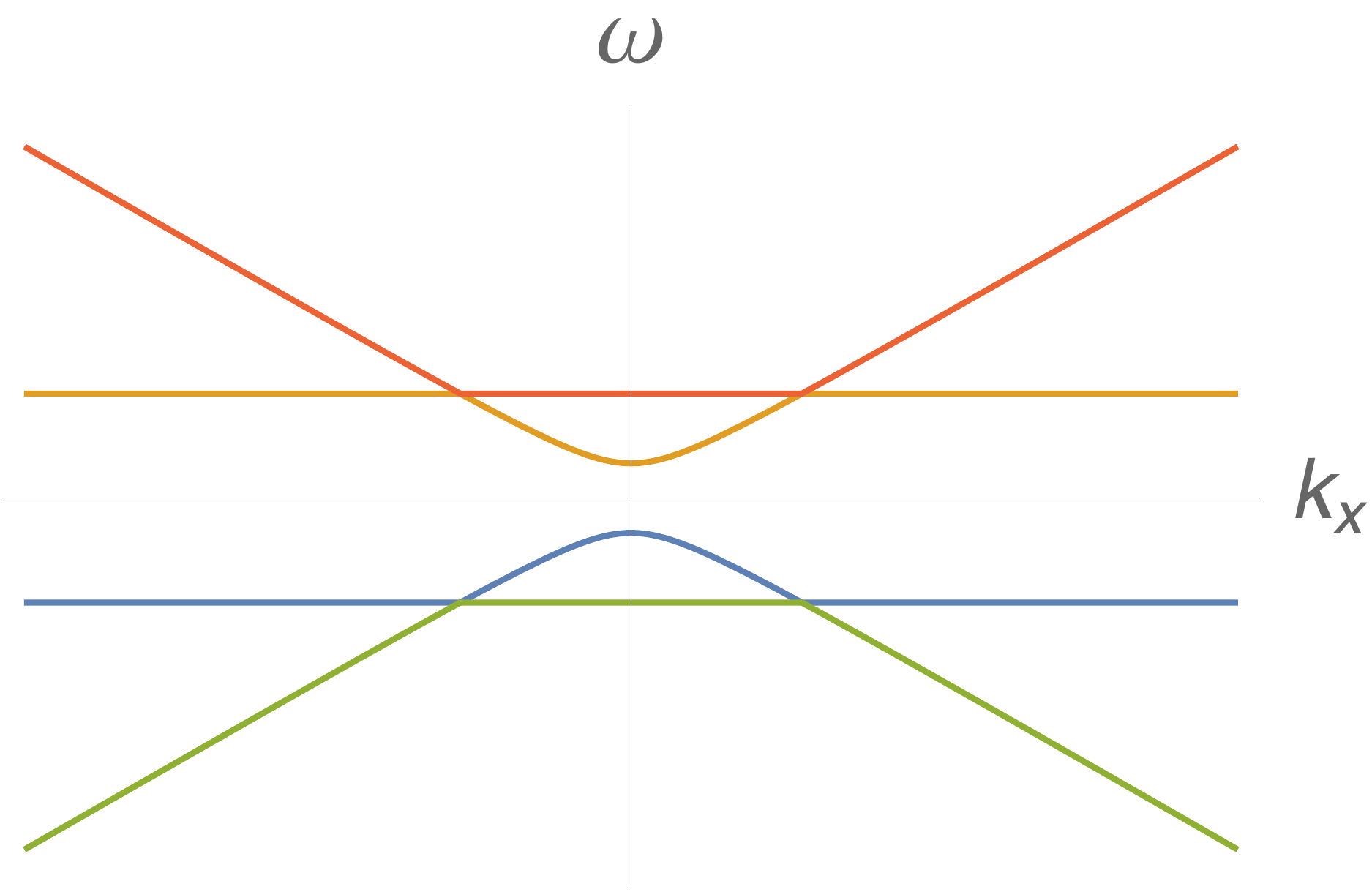}~~
    \includegraphics[width=0.240\textwidth]{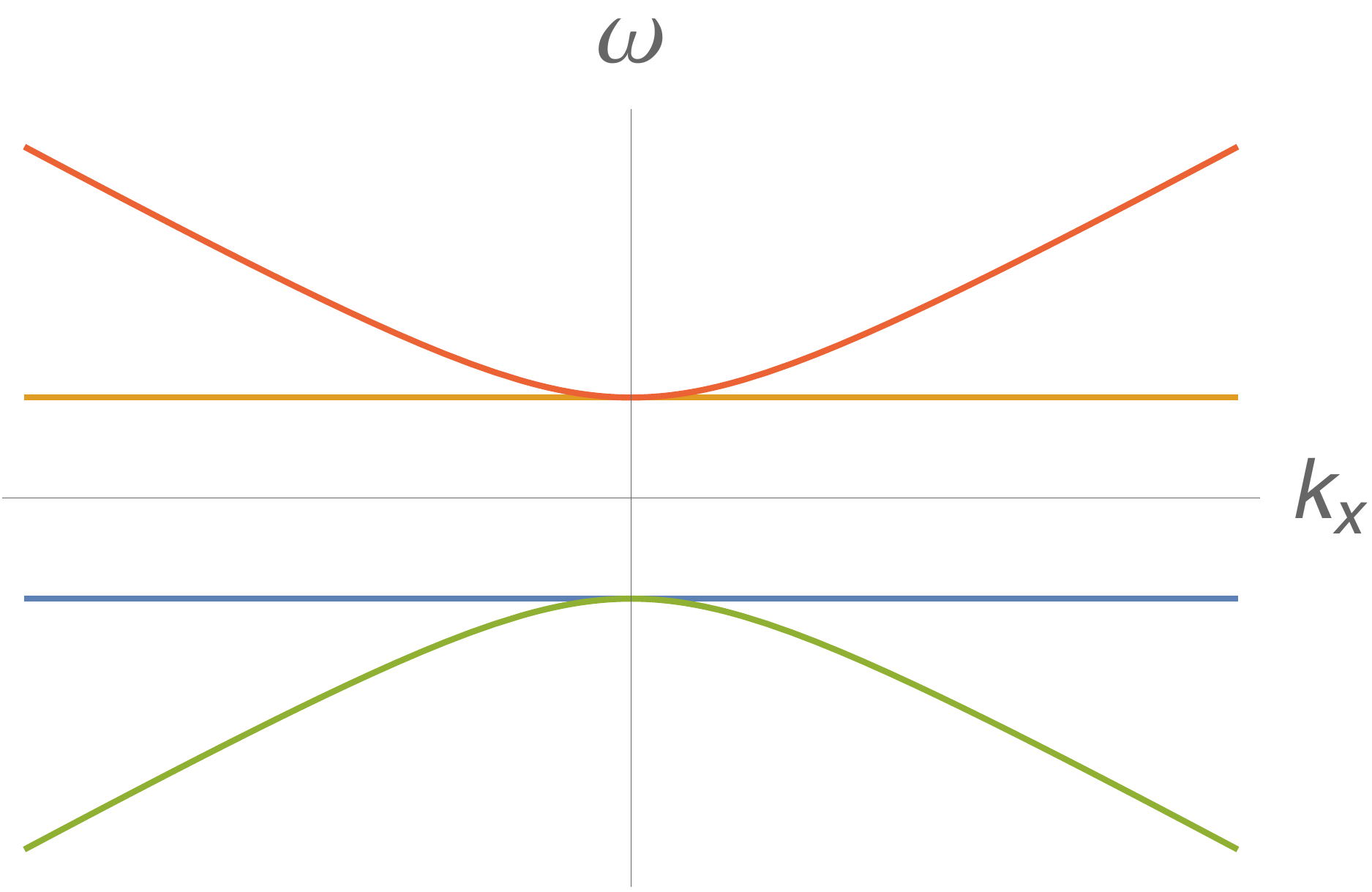}~~
      \includegraphics[width=0.240\textwidth]{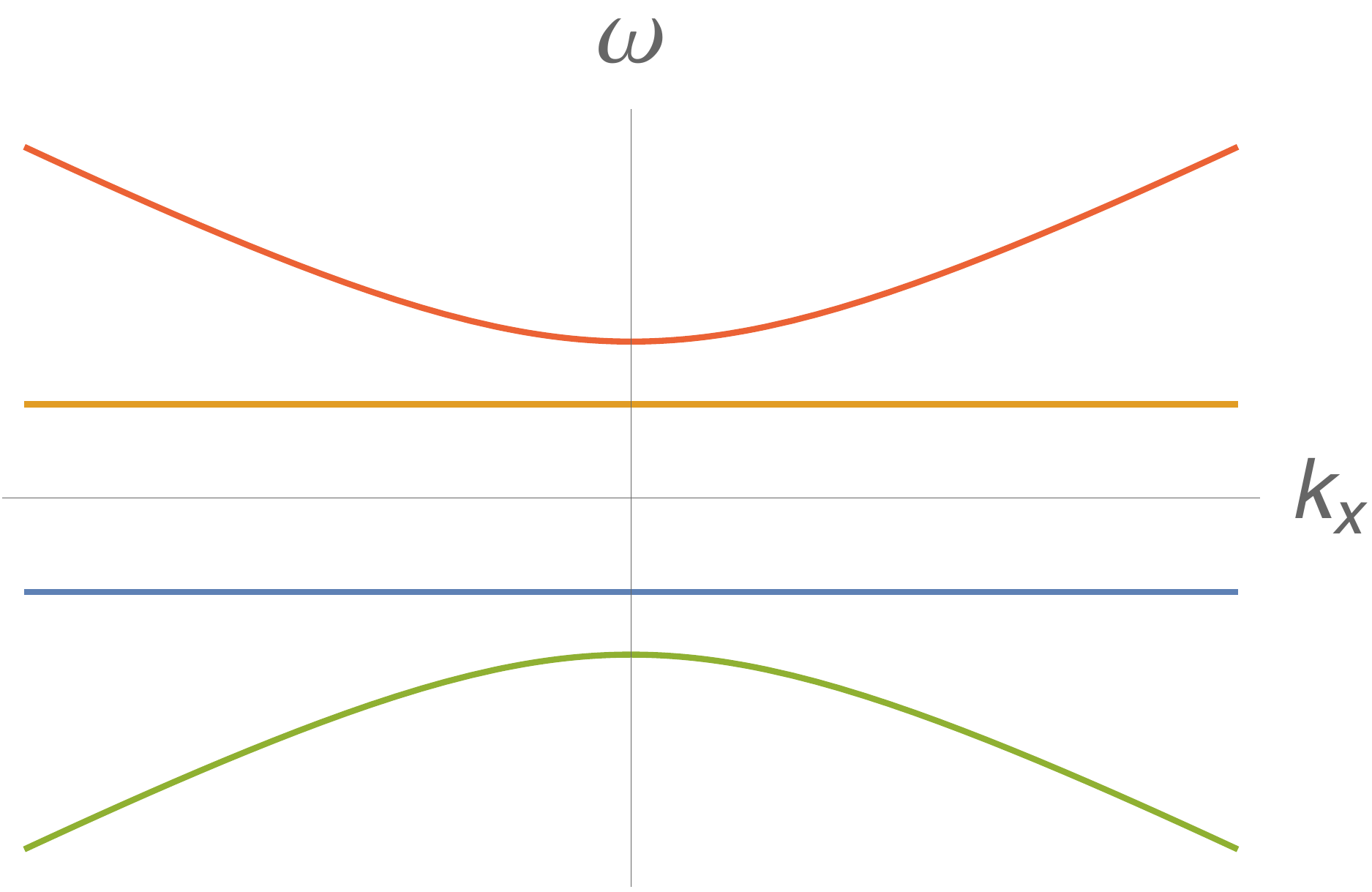}~~
        \includegraphics[width=0.240\textwidth]{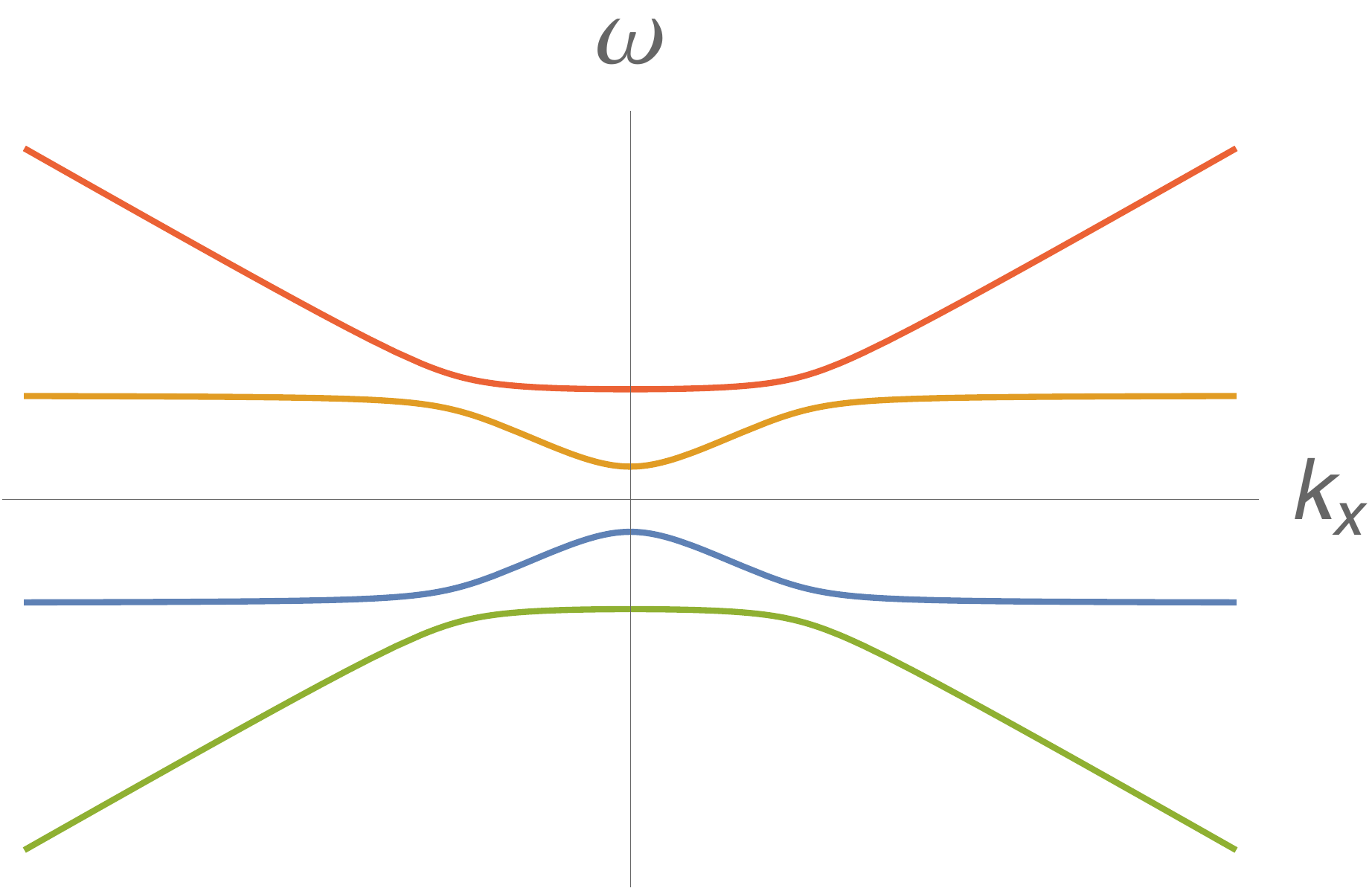}
  \caption{\small The spectrum of the modified hydrodynamics with dynamical equation (\ref{eq:4dhydro}). From left to right: in the first three plots we have $m<b$, $m=b$ and $m>b$ respectively and $k_y=k_z=0$. The fourth plot is for $k_y> 0, k_z=0$ and  $m<b$. }
  \label{fig:4D1}
\end{figure}

Figure \ref{fig:4D1} shows this spectrum as functions of $k_x$ for $k_y=k_z=0$ in three different situations: $m<b$, $m=b$ and $m>b$ as well as for $k_y> 0, k_z=0$ at  $m<b$. The effect of $b$ terms is to lift and lower the two transverse flat bands to symmetric positions of opposite sides of the $k$ axis. The effect of $m$ terms is still to gap the two sound modes. Now the sound and transverse modes are mixed with each other. In this way, the modes have band crossings at nonzero values of $k$ for $m<b$.

From figure \ref{fig:4D1}, we could see that for $m<b$ now we have four band crossing nodes at $k_y=k_z=0$ while $k_x \neq 0$, though $\omega\neq 0$ for these modes. Note that in condensed matter terminology, this kind of system is called gapless as there are band crossings in the spectrum. However, this is different from the usual standard ``hydrodynamic" definition of gapless, which means that $\omega(k=0)=0$. Thus here we do not follow the condensed matter terminology to call this kind of system gapless. Instead, we call them band crossings from here on. Similar to the Weyl semimetal case, these four nodes are still points in the expanded space of $\omega$, $k_x$, $k_y$ and $k_z$ as can be seen from the fourth plot in figure \ref{fig:4D1}.  
For $m=b>0$, the system becomes critical with 2 nodes and for $m>b$ the band crossings in the system disappear again. This behavior is qualitatively similar to the topological phase transition of a Weyl semimetal \cite{Landsteiner:2015pdh}. For $m<b$, now there could still be band crossings though small perturbations of these $m$ terms could previously gap the system and the gap is still there at $\omega=0$. This property is a defining feature of a topologically nontrivial gapless state, which we call the topologically nontrivial band crossing state here to avoid the conflict with terminology in hydrodynamics. We will show later that this is in fact a symmetry protected topological band crossing state.

 Note that different from the Weyl semimetal case, here the band crossings are not at $\omega=0$. Similar to what happens in condensed matter physics, the nodes do not need to be at $\omega=0$ for them to be topologically protected nodes and the important ingredient is that the nodes that we study need to be real band crossings instead of accidental band touchings. Topologically nontrivial band crossing states are those states that possess a spectrum with no gap at any $\omega$ and at the same time the spectrum would not develop a gap under small perturbations. In this sense, we need two bands to cross so there would be band crossings instead of all bands being fully separated in the spectrum (again at any $\omega$) and for the band crossings to be topologically nontrivial we need the band crossings to be stable under small perturbations. Also another reason that the nodes do not need to be at $\omega=0$ is that the absolute value of energy is not important and we usually study effective excitations near each of the band crossing points so the energy at the band crossings could each be set to zero effectively. 

We also need to emphasize here that the definition of a topologically band crossing state is a state whose band crossings would not disappear under small perturbations. This is because as long as a band crossing could disappear under small perturbations, this would mean that the band crossed state is topologically equivalent to the trivial vacuum. Thus for a topologically nontrivial band crossing state, by being topologically protected, we refer to the fact that the state remains band crossed under small perturbations of the system which usually could gap a system. This indicates that the band crossing points should be singular points in the momentum space. Being topologically inequivalent to the trivial vacuum, this topologically nontrivial band crossing state should possess a nontrivial topological invariant which takes a different value compared to the trivial vacuum, e.g. the two nodes of a Weyl semimetal possess nontrivial chiral charges which are different from the trivial value of zero thus the two nodes cannot disappear under small perturbations of the system. The topology of states depends on the Hamiltonian or equivalently the wave functions and is an intrinsic property of the Hamiltonian.

\begin{figure}[h!]
  \centering
  \includegraphics[width=0.440\textwidth]{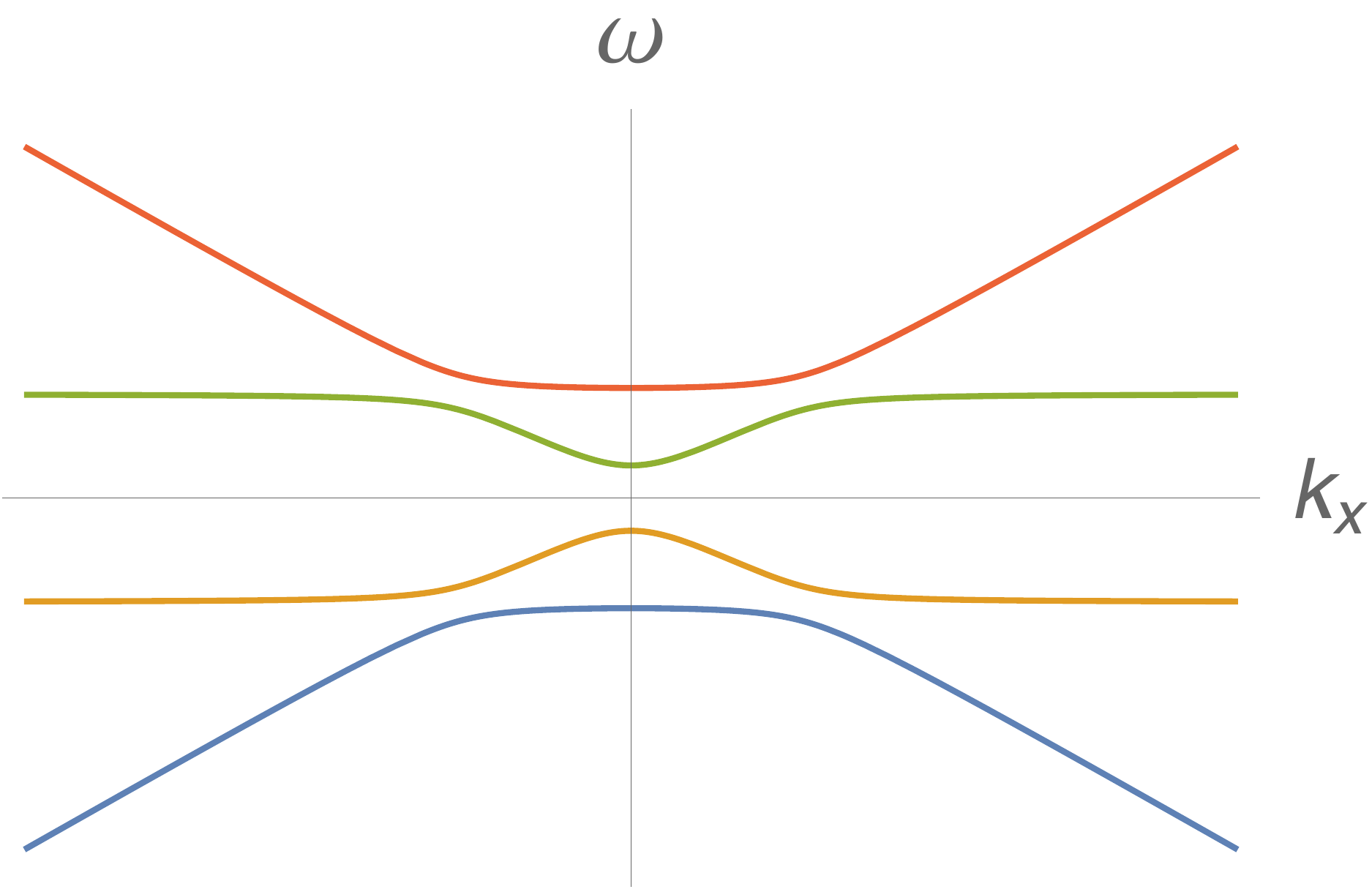}
  \caption{\small The spectrum of the hydrodynamic equation (\ref{eq:4dhydro}) modified by an extra mass term in the $y$ or $z$ direction or both these two directions. This plot is for $k_y=k_z=0$ and $m<b$. The four nodes disappear. }
  \label{fig:4D-hydro2mass}
\end{figure}

The $m$ terms in (\ref{h4d}) do not eliminate the four band crossing nodes in the $m<b$ case, however, if we have $m$ terms in the $y$ or $z$ directions, the band crossing will disappear no matter how small the $y$ or $z$ mass parameters are as shown in figure  \ref{fig:4D-hydro2mass}. In this situation, there are no crossing nodes anymore. According to the definition above, the nodes are not topologically protected as they could be  destroyed by $m$ terms in the $y$ or $z$ directions. These in fact belong to symmetry protected topologically band crossing states, which could only endure perturbations that respect the symmetry. Here the nodes should be topologically nontrivial under the protection of the symmetry that forbids the $m$ terms in the $y$ and $z$ directions as will be further confirmed in section \ref{sec:ti}, i.e. the symmetry that would be broken by $m$ terms in the $y$ and $z$ directions. As a rough estimate, the symmetry that we need here should be related to the translational symmetry in the $y$ and $z$ directions. This symmetry is a restriction to the system while not the solutions. The explicit form of this symmetry will be shown in section \ref{sec:symmetry} {from a more accurate calculation.} In this sense, the system (\ref{eq:4dhydro}) experiences a symmetry protected topological phase transition which happens at the critical point $m=b$. 

Another thing that we need to emphasize here is that the fact that $m$ terms in the $y$ and $z$ directions would eliminate the band crossings in the system has nothing to do with the fact that the system has no band crossing at nonzero $k_y$, $k_z$ and also $k_x$ away from the node values. The notion of band crossing only requires a band crossing in the whole momentum space so though the system has no band crossing at nonzero $k_y$, $k_z$ and also $k_x$ away from the node values, the system is still a band crossed state because of the band crossing at $k_y=k_z=0$. This latter fact is used to show that the band crossings in this case are nodal points in the momentum space while not nodal lines or surfaces and this property is the same as Weyl semimetals.

Several remarks are in order:
\begin{enumerate}
\item For the hydrodynamical modes $\omega(k=0)$ is not zero anymore due to the non-conservation of energy, i.e. energy is continuously pumped into or out of the system; 
\item In the case $m<b$, the crossing nodes have nonzero $\omega$ and $k$. This indicates that two degenerate copies of modes would suddenly arise at a special value of $\omega$ and $k$, leading to a sudden rise in the amplitude of sound. Besides this possible observational effect, the effect of these energy momentum non-conservation terms in transport coefficients will be shown in section \ref{sec:tran4d}. 
\item These crossing nodes at $m<b$ are dissipative when order $k^2$ terms are taken into account as we will see in  section \ref{sec:2nd}. This is different from the $\omega=k=0$ nodes which are real poles in hydrodynamics with unbroken translational symmetries.
\end{enumerate}

There is one extra point that we need to emphasize, the new $m$ and $b$ terms above are not dissipative so they only change the shape of the spectrum while do not introduce any imaginary parts in the dispersion relation. Readers who are interested in the difference between the two systems could find more details in \ref{sec:app}.

\subsubsection{Two 2D hydrodynamic systems}
\label{sec:2d2d}

The second choice is to start from two separately conserved hydrodynamic systems. After introducing weak interchange of energy and momentum between the two systems, there will be more possibilities for the existence of  topological band crossing  modes.

We start from the simplest case, where we have two 1+1D hydrodynamic systems exchanging energy and momentum with each other weakly. In this case, we have two variables to solve ($T^{00}$ and $T^{0x}$) in each sector and four in total. 
The conservation equation under consideration has the form of 
\bea
\label{eq:2dhydro}
\begin{split}
\partial_\mu \delta T_L^{\mu t}&=m_1 \delta T_L^{tx}+b_1 \delta T_R^{t t}\,,\\
\partial_\mu \delta T_L^{\mu x}&=-m_1 v_{sL}^2 \delta T_L^{tt}+b_1 \delta T_R^{t x}\,,\\
\partial_\mu \delta T_R^{\mu t} &=m_2 \delta T_R^{tx}-b_2 \delta T_L^{t t}\,,\\
\partial_\mu \delta T_R^{\mu x} &=-m_2 v_{sR}^2 \delta T_R^{tt}-b_2 \delta T_L^{t x}\,,
\end{split}
\eea
in which there are two different sectors of matter and they can exchange both energy and momentum as well as gain or lose energy and momentum from or to the environment. Furthermore, we assume that the parameters $m_{1,2} , b_{1,2}\ll T$ to be in the hydrodynamic limit. The $m$ terms are the same as in the single 4D case, except that now there is only one spatial direction. Both $m$ and $b$ terms involve exchange of energy and momentum within the two sectors as well as with the environment. For simplicity we shall focus on the case $m_1=m_2=m, b_1=b_2=b, v_{sL}=v_{sR}=v_s,$ and $v_s$ could be set to be $1$ as shown in section \ref{subsec:single4D}.

In two dimensions, the first order corrections to the ideal fluid components of energy momentum tensors are zero. 
The constitutive equations for the perturbations of energy momentum tensors are 
\bea
\begin{split}
\delta T^{00} &=\delta \epsilon\,,\\
\delta T^{x0} &=\delta T^{0x}=(\epsilon+p) \delta u^x=\delta\pi \,,\\
\delta T^{xx}  &=\frac{\partial P}{\partial \epsilon} \delta\epsilon=v_s^2\delta\epsilon\,,
\end{split}
\eea
where we have omitted the subscript $L$ and $R$ for simplicity. 

In momentum space, the equation (\ref{eq:2dhydro}) can be written as 
\be
i\partial_t\Psi=H\Psi
\ee
where
\be\label{eq:2D2DH}
\Psi=\begin{pmatrix} 
\delta \epsilon_L  \\
\delta \pi_L \\
\delta \epsilon_R  \\
\delta \pi_R
\end{pmatrix}\,,~~~~~~
H=\begin{pmatrix} 
0 & ~~k_x+im & ~~ ib & ~~ 0 \\
(k_x-im) v_s^2 & ~~0 & ~~ 0 & ~~ ib \\
-ib & ~~ 0 & ~~ 0 & ~~ k_x+im \\
0 & ~~-ib & ~~ (k_x-im) v_{s}^2 & ~~ 0 
\end{pmatrix}\,.
\ee

\begin{figure}[h!]
  \centering
  \includegraphics[width=0.300\textwidth]{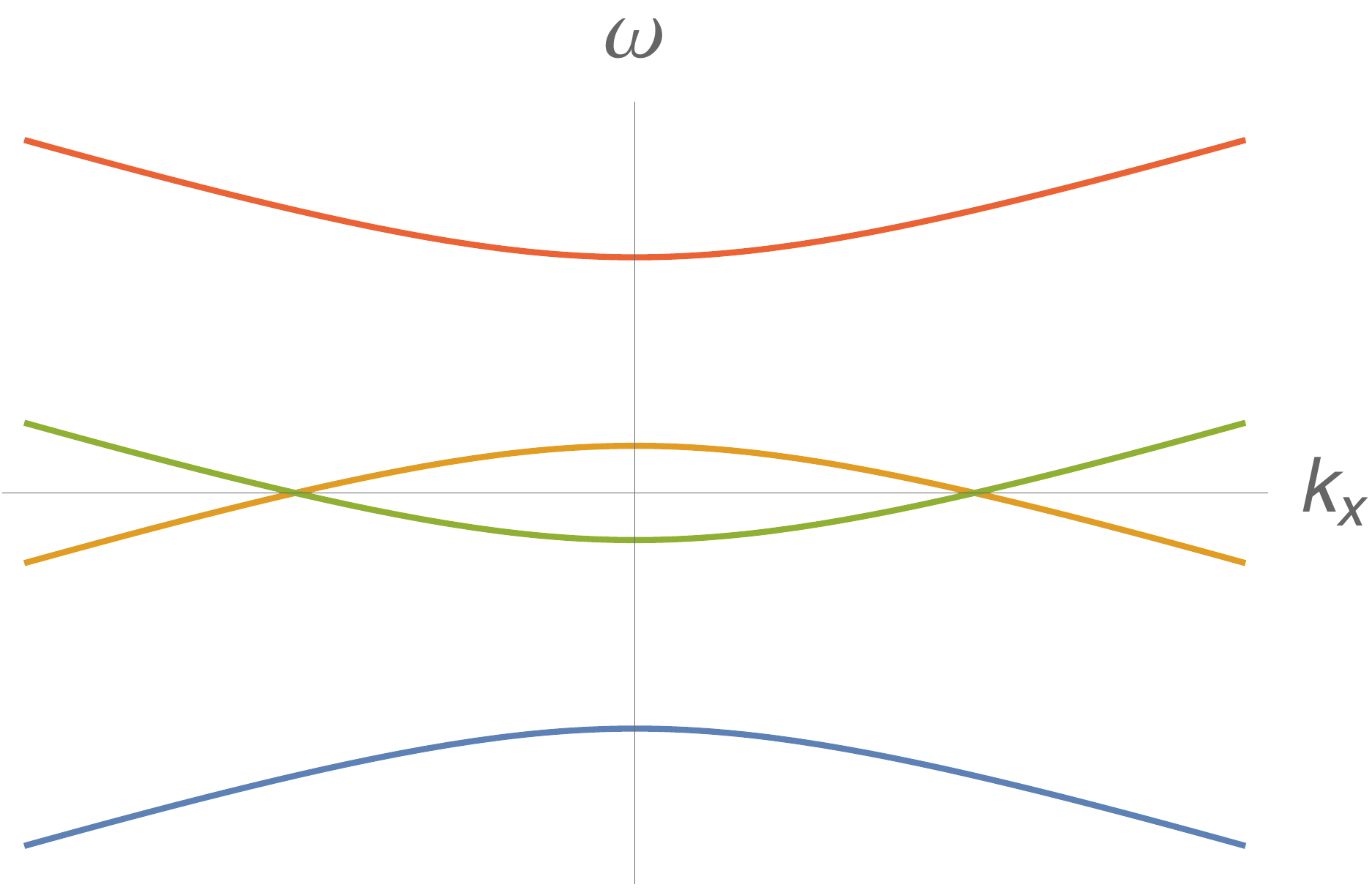}~~~~~
    \includegraphics[width=0.300\textwidth]{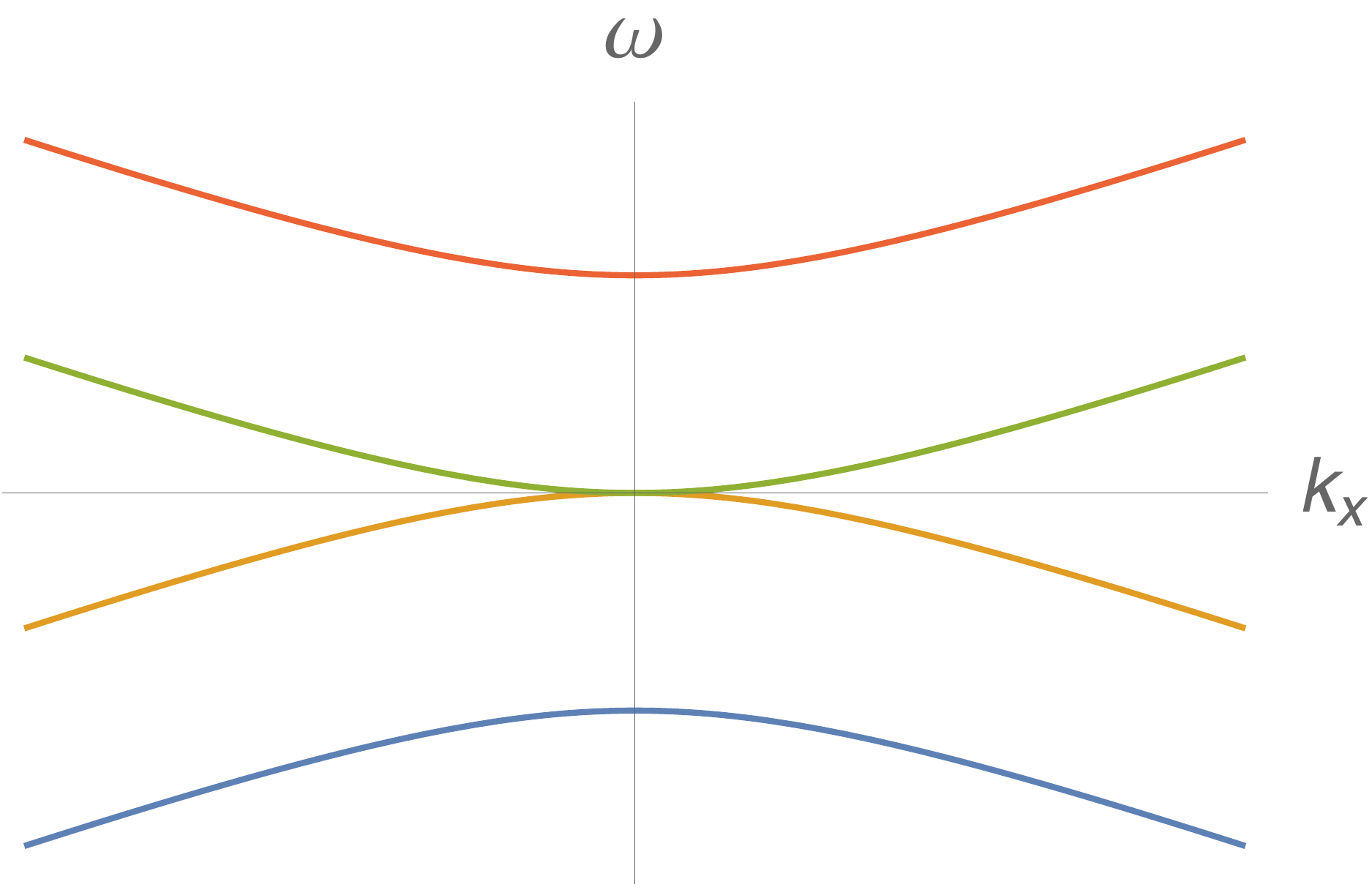}~~~~~
      \includegraphics[width=0.300\textwidth]{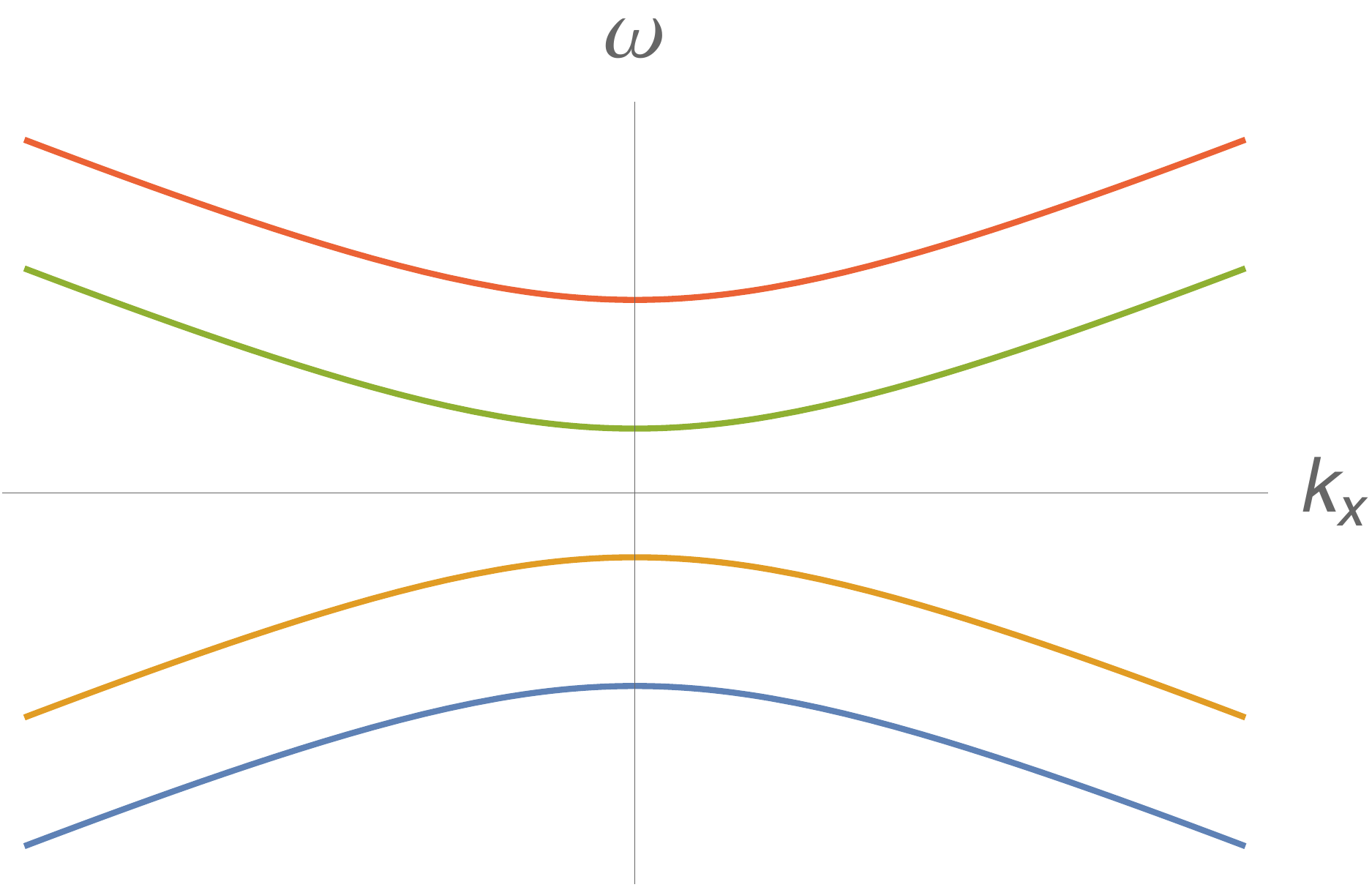}
 \caption{\small The spectrum of the modified hydrodynamics with dynamical equation (\ref{eq:2dhydro}). From left to right, $m<b$, $m=b$ and $m>b$ respectively.  }
  \label{fig:2D2D}
\end{figure}

The eigenvalues of $H$ give 
\be\label{eq:2d2dspec}
\omega=\pm b \pm \sqrt{m^2+k_x^2} v_s\,.
\ee  
Without $m$ and $b$ terms there are two sets of two sound modes and these modes will mix together when $m$ and $b$ become nonzero. $m$ is to gap the hydrodynamical modes and $b$ is to separate the crossing nodes of the hydrodynamic modes, thus generating non-trivial topological structure in the spectrum. The structure of the spectrum here looks very similar to the band structure of the Weyl semimetal. From figure \ref{fig:2D2D} we can see that for $m<b$, the system is in the topological phase with two crossing nodes. For $m=b$, the system is in the critical point where the two nodes merge into one. As effects of $m$ and $b$ terms cannot cancel each other even when $m=b$, in this case the spectrum is not the same as when $T^{\mu\nu}$ is conserved. For $m>b$, the system is in the trivial phase where no band crossing exists. More evidence of the nontrivial topological structure from the calculation of Berry phase will be shown in section \ref{sec:ti}.

Note that in this case there is only one spatial direction, so these nodes in the topological phase will not disappear due to new mass terms. Also as we mentioned above, there are no dissipative terms in 2D, so that the nodes in this case are non-dissipative at order $\mathcal{O}(k^2)$, which is different from the single 4D case. The same as the single 4D case, there should be a sudden rise in amplitude of sounds at a certain momentum for $m<b$, however, different from the 4D case, now the frequency is zero. 

\subsubsection{More possibilities with topological hydrodynamic modes}
\label{subsec:moreposs}
We can generalize the 2D+2D system above to higher spatial dimensions to get more possible topological hydrodynamic modes. To generalize the $b$ terms, there are several choices. We may generalize the $b$ terms to all spatial dimensions or only to some of them. In the following we show the generalization to 3D+3D and 4D+4D systems with $b$ interaction terms in some or all of the dimensions. The spectrum is much more complicated in these cases and we will mainly list the qualitative features of these systems in this subsection. 

\vspace{0.3cm}
\noindent {\bf I. 3D+3D/4D+4D with $b$ interaction terms not in all directions}



For 3D+3D systems, to have $b$ interaction terms not in all directions, the only possible effective Hamiltonian matrix is 
\be\label{eq:3D3D}
H_{3D+3D,I}=\begin{pmatrix} 
0 & ~~k_x+im &~~ k_y &~~ ib & ~~ 0& ~~ 0 \\
k_x-im & ~~0& ~~0 & ~~ 0 & ~~ ib& ~~0 \\
k_y & ~~0 &~~ 0 &~~ 0 & ~~ 0& ~~ 0 \\
-ib & ~~ 0 & ~~ 0 & ~~ 0 & ~~ k_x+im& ~~ k_y \\
0 & ~~-ib& ~~ 0  & ~~ k_x-im & ~~ 0 & ~~ 0 \\
0&~~ 0 & ~~ 0& ~~ k_y & ~~ 0& ~~ 0\\
\end{pmatrix}\,.
\ee 
For 4D+4D there are two possibilities for the $b$ terms
\be\label{eq:4D4D}
H_{4D+4D,I1}=\begin{pmatrix} 
0 & ~~k_x+im &~~ k_y&~~ k_z &~~ ib & ~~ 0& ~~ 0& ~~ 0 \\
k_x-im & ~~0& ~~0 & ~~ 0& ~~ 0 & ~~ ib& ~~0& ~~ 0 \\
k_y & ~~0 &~~ 0 &~~ 0 & ~~ 0& ~~ 0& ~~ 0& ~~ 0 \\
k_z & ~~0 &~~ 0 &~~ 0 & ~~ 0& ~~ 0& ~~ 0& ~~ 0 \\
-ib & ~~ 0 & ~~ 0 & ~~ 0 & ~~ 0  & ~~ k_x+im& ~~ k_y& ~~ k_z \\
0 & ~~-ib& ~~ 0 & ~~ 0 & ~~ k_x-im & ~~ 0 & ~~ 0 & ~~ 0 \\
0&~~ 0 & ~~ 0& ~~ 0& ~~ k_y & ~~ 0& ~~ 0& ~~ 0\\
0&~~ 0 & ~~ 0& ~~ 0& ~~ k_z & ~~ 0& ~~ 0& ~~ 0\\
\end{pmatrix}\,,
\ee and 
\be
\label{eq:4D4Dhy2}
\begin{split}
H_{4D+4D,I2}=\begin{pmatrix} 
0 & ~~k_x+im &~~ k_y+im&~~ k_z &~~ ib & ~~ 0& ~~ 0& ~~ 0 \\
k_x-im & ~~0& ~~0 & ~~ 0& ~~ 0 & ~~ ib& ~~0& ~~ 0 \\
k_y -i m& ~~0 &~~ 0 &~~ 0 & ~~ 0& ~~ 0& ~~ ib& ~~ 0 \\
k_z & ~~0 &~~ 0 &~~ 0 & ~~ 0& ~~ 0& ~~ 0& ~~ 0 \\
-ib & ~~ 0 & ~~ 0 & ~~ 0 & ~~ 0  & ~~ k_x+im& ~~ k_y+im& ~~ k_z \\
0 & ~~-ib& ~~ 0 & ~~ 0 & ~~ k_x-im & ~~ 0 & ~~ 0 & ~~ 0 \\
0&~~ 0 & ~~ -ib& ~~ 0& ~~ k_y-im & ~~ 0& ~~ 0& ~~ 0\\
0&~~ 0 & ~~ 0& ~~ 0& ~~ k_z & ~~ 0& ~~ 0& ~~ 0\\
\end{pmatrix}\,.
\end{split}
\ee 
Here $ \Psi=\begin{pmatrix} 
\Psi_1  \\
\Psi_2
\end{pmatrix}\,,$ with $\Psi_{1,2}$ denoting the 3 or 4 modes for the first and the second systems separately and we have ignored $v_s$ factors. For these systems, there are no simple analytical expressions for the spectrum.

\begin{figure}[h!]
  \centering
  \includegraphics[width=0.400\textwidth]{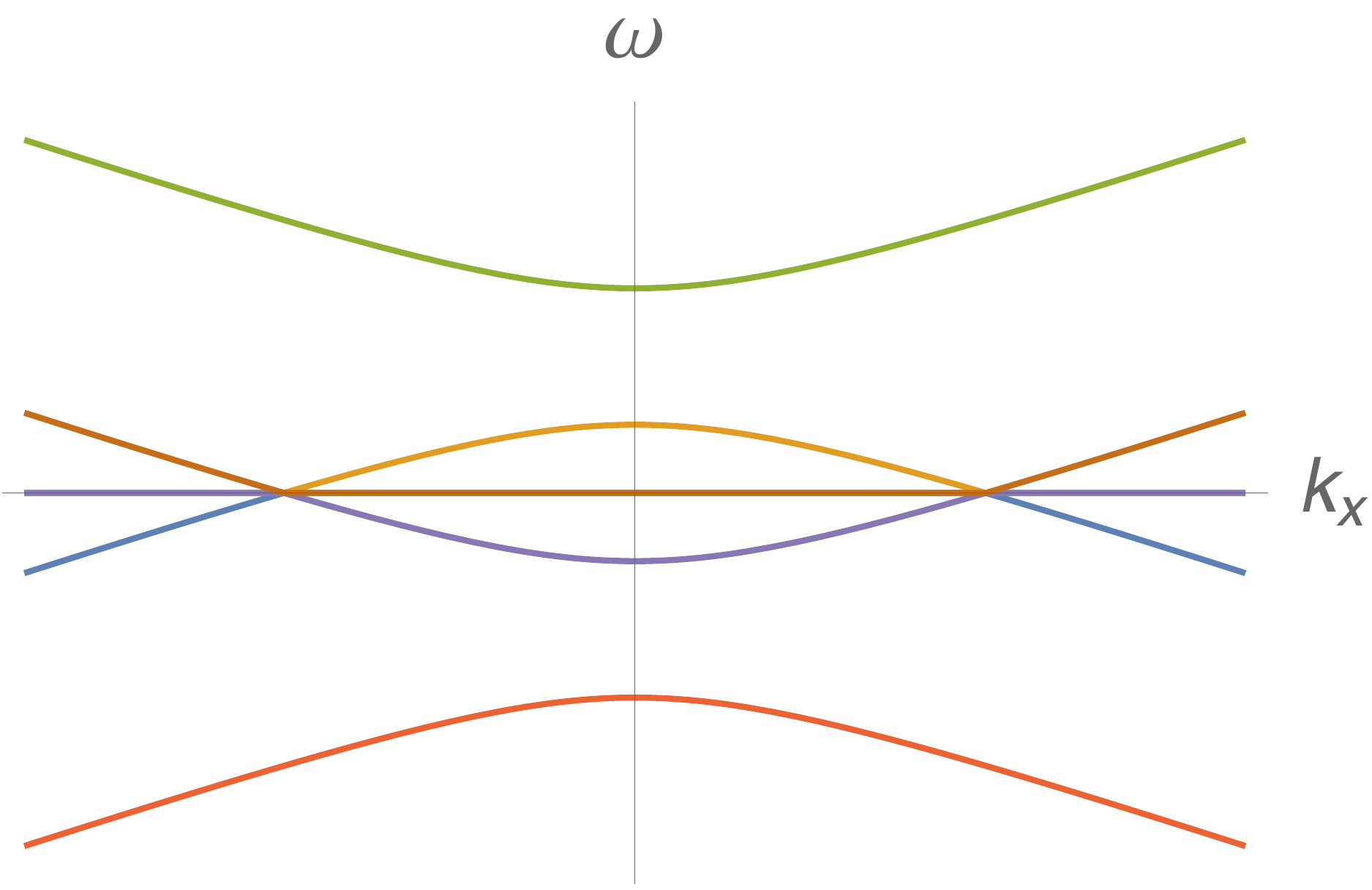}~~~~~
    \includegraphics[width=0.400\textwidth]{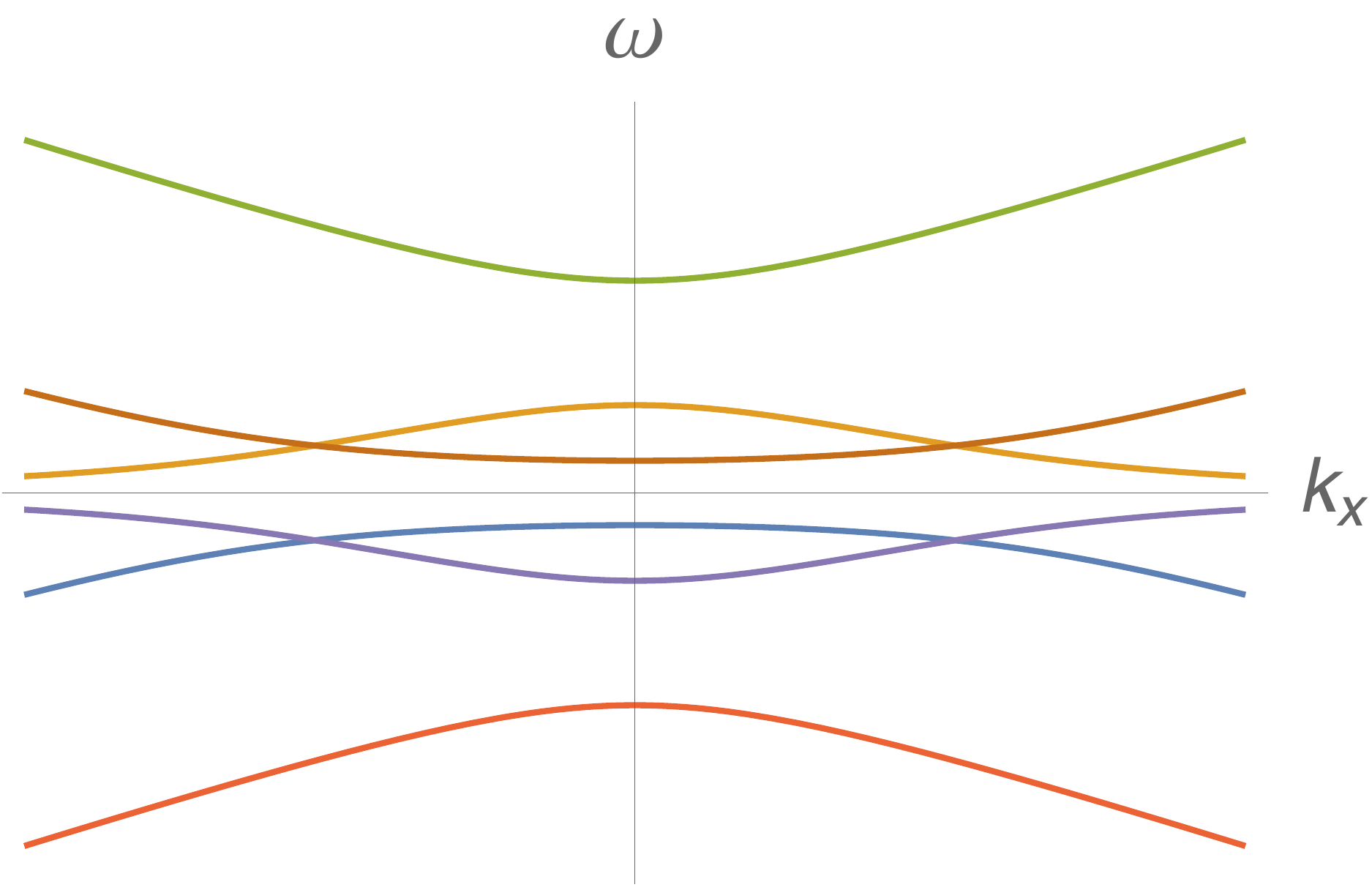}
  \caption{\small The spectrum of the modified hydrodynamics of (\ref{eq:3D3D}) $H_{3D+3D,I}$ with $m<b$ and $k_y=0$ ({\it left}) and $k_y>0$ ({\it right}). The spectrum of  $H_{3D+3D,I}$ with an extra $m$ term in the $y$ direction at $k_y=0$ is the same as the right plot.}
  \label{fig:3D3D1}
\end{figure}

\begin{figure}[h!]
  \centering
  \includegraphics[width=0.400\textwidth]{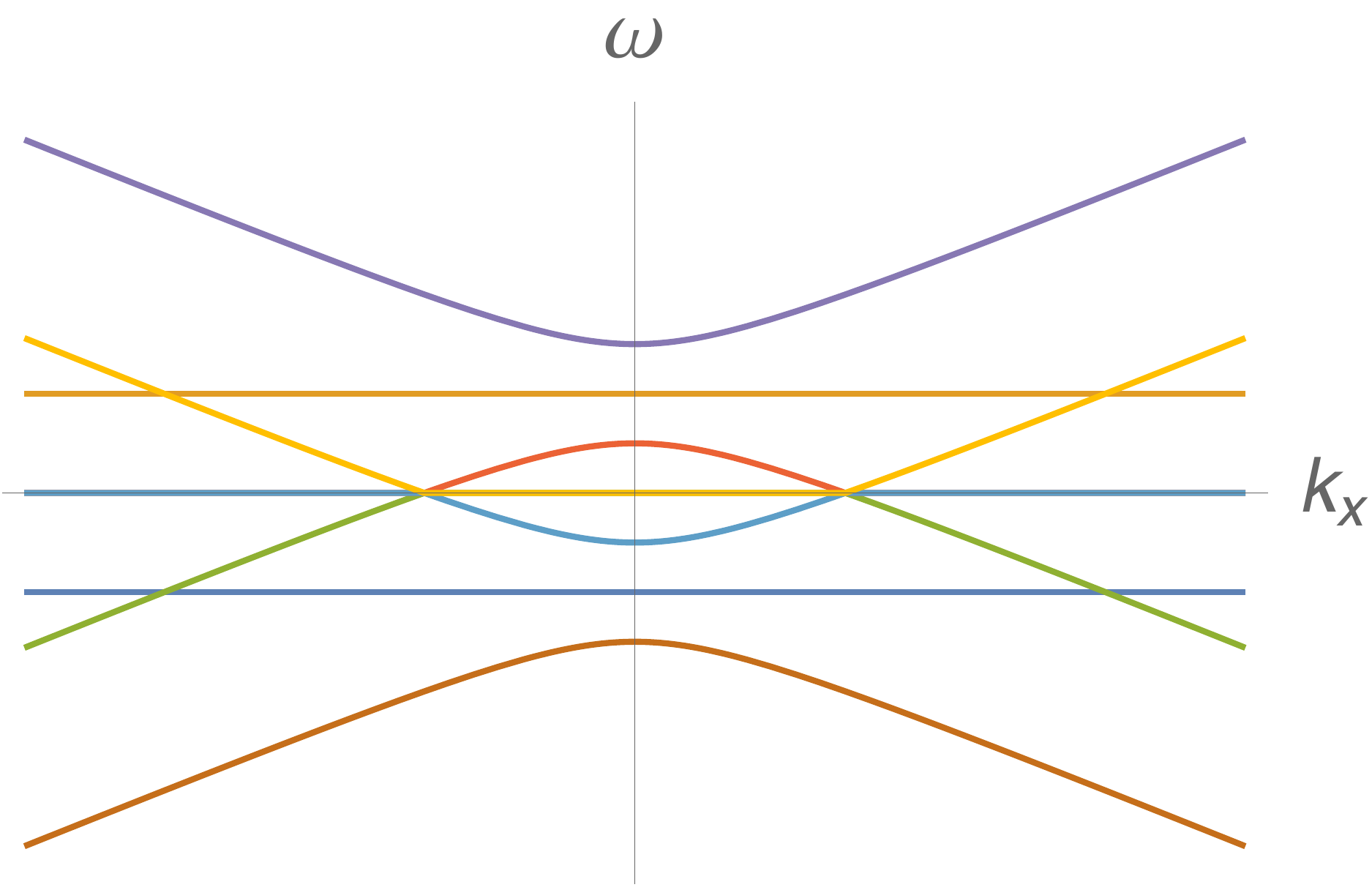}~~~~~
    \includegraphics[width=0.400\textwidth]{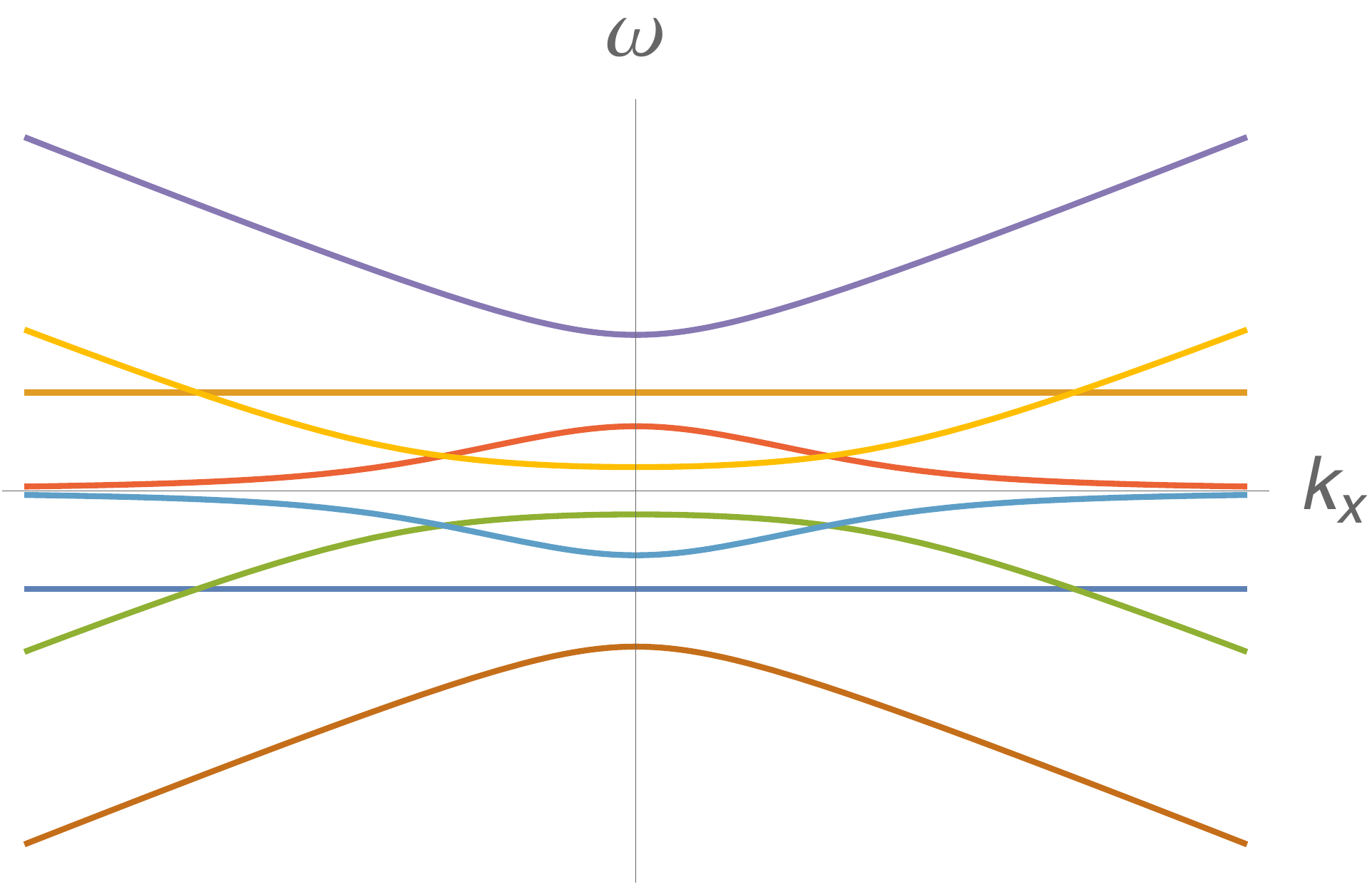}
  \caption{\small The spectrum of the modified hydrodynamics of (\ref{eq:4D4D}) $H_{4D+4D,I1}$ with $m<b$ and $k_1=\sqrt{k_y^2+k_z^2}=0$ ({\it left}) and $k_1>0$ ({\it right}). The spectrum of  $H_{4D+4D,I1}$ with an extra $m$ term in the $y$ direction at $k_1=0$ is the same as the right plot. }
  \label{fig:4D4D1}
\end{figure}

Numerically we could see that the spectrum looks similar to that in 2D+2D with slight differences. The spectrums of $H_{3D+3D, I}$ and $H_{4D+4D,I}$ are almost the same with the latter having two more flat bands corresponding to two transverse modes (see Figs. \ref{fig:3D3D1} and \ref{fig:4D4D1}). In the following we list the qualitative behavior for the spectrum in all three cases above.

\begin{figure}[h!]
  \centering
  \includegraphics[width=0.450\textwidth]{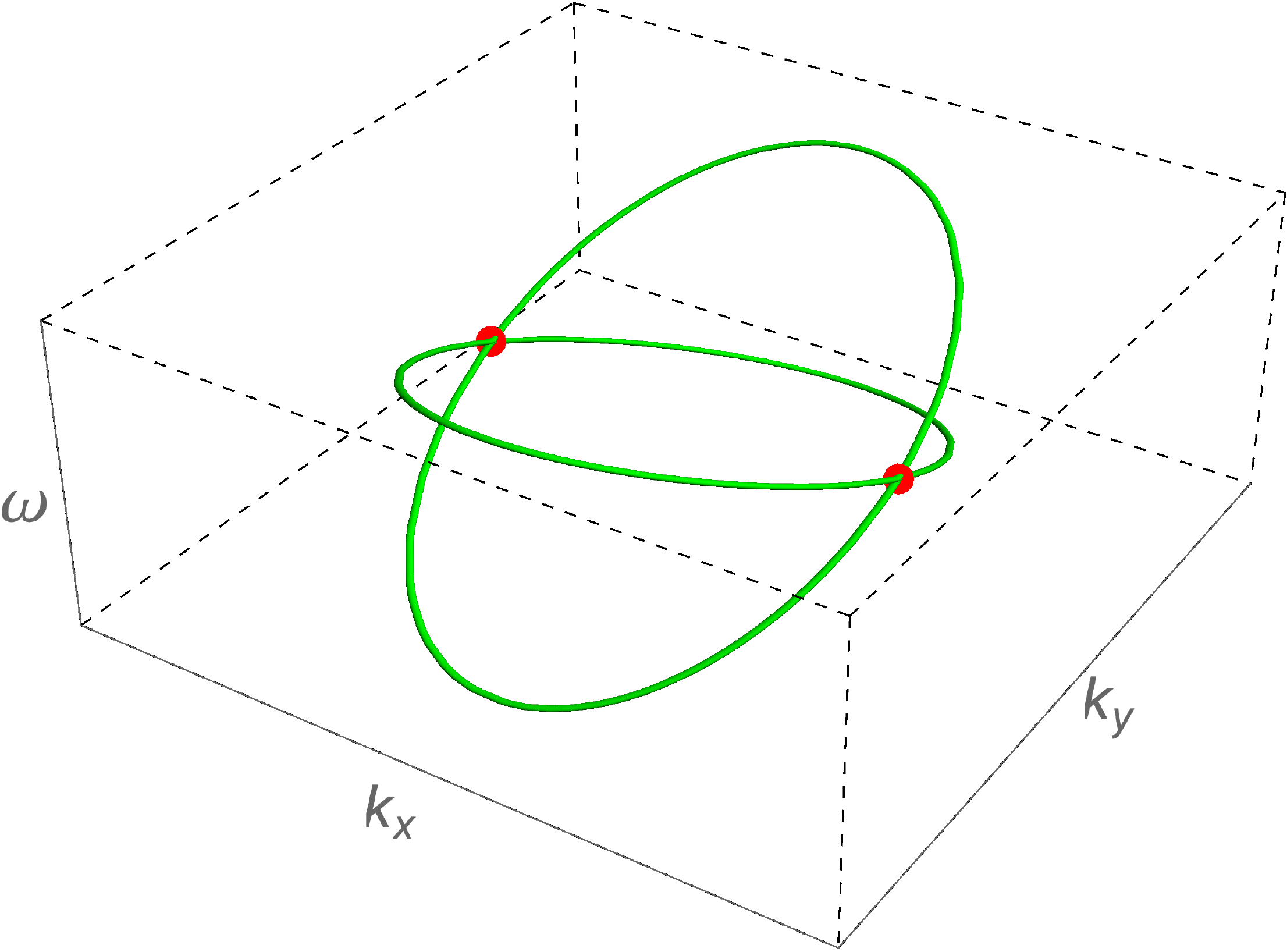}~~~~~
    \includegraphics[width=0.450\textwidth]{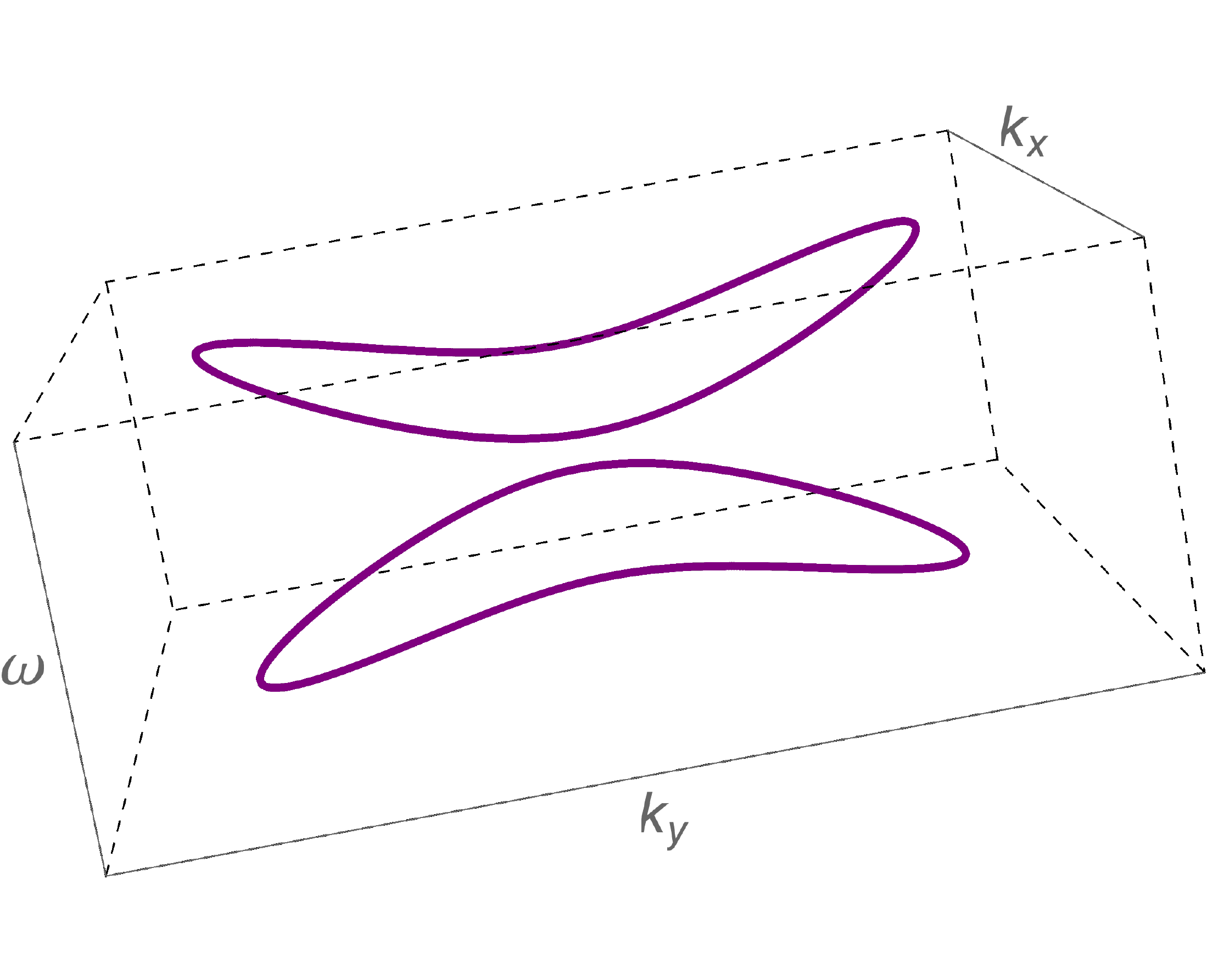}
  \caption{\small The illustration plot in the $\omega$, $k_x$, $k_y$ space for the crossing nodes of $3D+3D$ hydrodynamical systems (\ref{eq:3D3D}) ({\it left}) and the crossing nodes of its generalization with an extra $m$ term in the $y$ direction ({\it right}). }
  \label{fig:3d3dill}
\end{figure}

\noindent  $H_{3D+3D,I}$: As shown in figure \ref{fig:3D3D1}, we could see that in the 3D+3D case at $m<b$ the crossing nodes form two circles pinned together at two opposite points in $\omega$, $k_x$, $k_y$ space. With an extra $m$ term in the $y$ direction, the two pinned nodes will disappear but the two circles of crossing nodes remain. 
Figure \ref{fig:3d3dill} is an illustration in the $\omega$, $k_x$, $k_y$ space for the crossing nodes only, which can be viewed as an extended space version of the left two and right four crossing points in figure \ref{fig:3D3D1}. In the left figure the two circles of crossing nodes are pinned by two opposite points without $m$ term in the $y$ direction, while in the right plot the two pinned points disappear with $m$ term in the $y$ direction and now two circles do not intersect with each other. 
This indicates that the two pinned nodes are topologically nontrivial protected by the symmetry that forbids the $m$ term in the $y$ direction while the two circles of crossing nodes are topologically nontrivial without the need of symmetry protection. In section \ref{sec:ti} we will confirm the topological structure of these nodes from the calculation of the Berry phase.  In this case, the nodes at nonzero frequency are supposed to be dissipative with higher order terms considered. 

\noindent $H_{4D+4D,I1}$: As now there are three spatial dimensions, there are two extra flat bands corresponding to two transverse modes and the spectrum looks more complicated. 
With all spatial dimensions taken into account, the circles in the $H_{3D+3D, I}$ case now become spheres in the $\omega, k_x, k_y, k_z$ space. As there is a rotational symmetry in the $y$-$z$ plane, we can use $k_{1}=\sqrt{k_y^2+k_z^2}$ in the consideration of the spectrum. In the $\omega$, $k_x$, $k_{1}$ space,  the behavior of the spectrum is the similar for $H_{4D+4D, I}$ to the $H_{3D+3D, I}$ case as shown in figure \ref{fig:4D4D1}. Compared to the $H_{3D+3D, I}$ case, there are four additional crossing nodes in the $\omega, k_x$ space due to two extra flat bands in $H_{4D+4D, I}$ case, i.e. four additional circles of crossing nodes in the $\omega, k_x, k_1$ space.  With an extra $m$ term in the $y$ direction, the pinned circles will become two non-intersecting circles, indicating the two spheric nodes in the $\omega, k_x, k_y, k_z$ space are topologically nontrivial without the need of symmetry protection.

\begin{figure}[h!]
  \centering
  \includegraphics[width=0.245\textwidth]{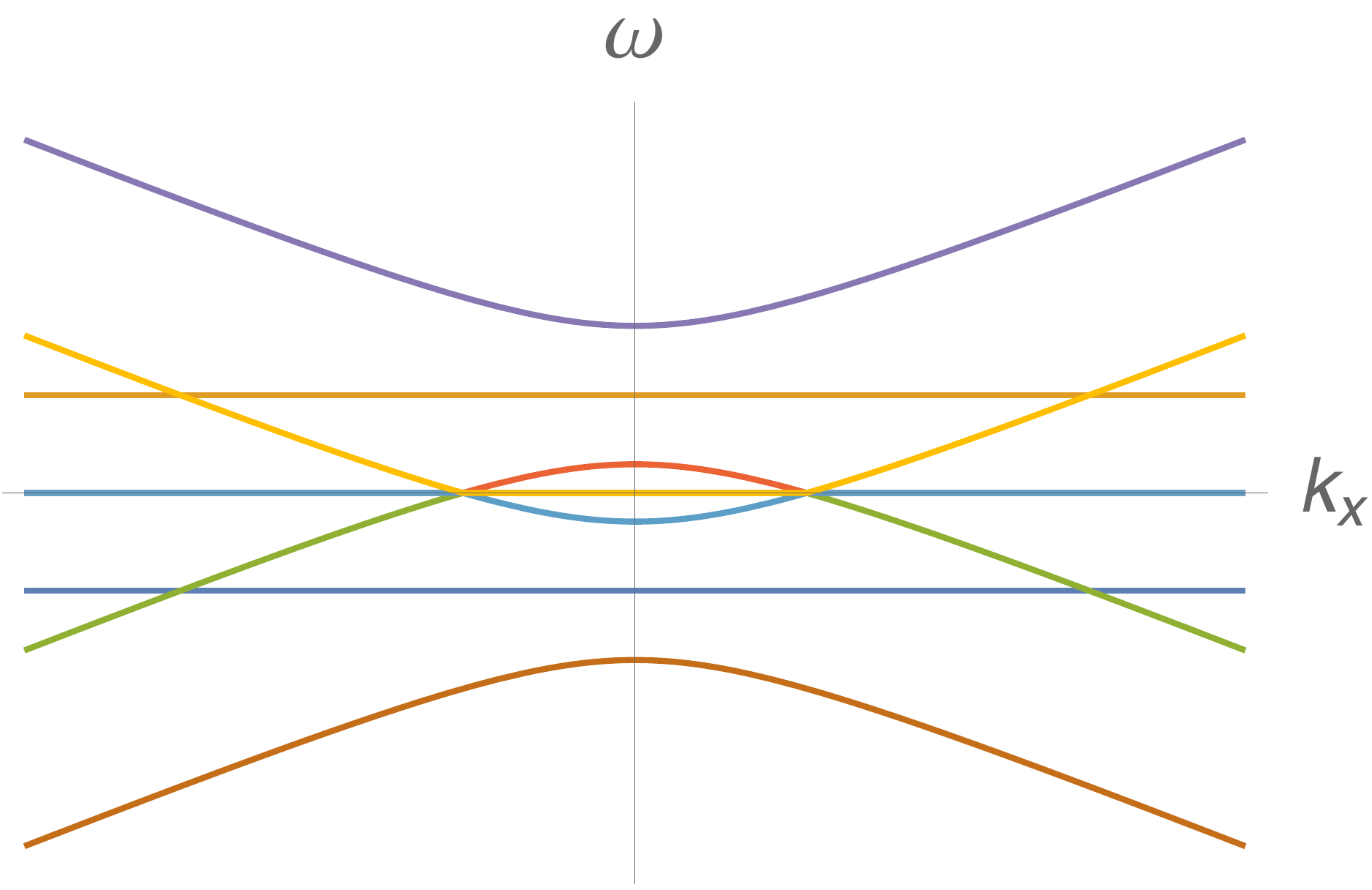}~
    \includegraphics[width=0.245\textwidth]{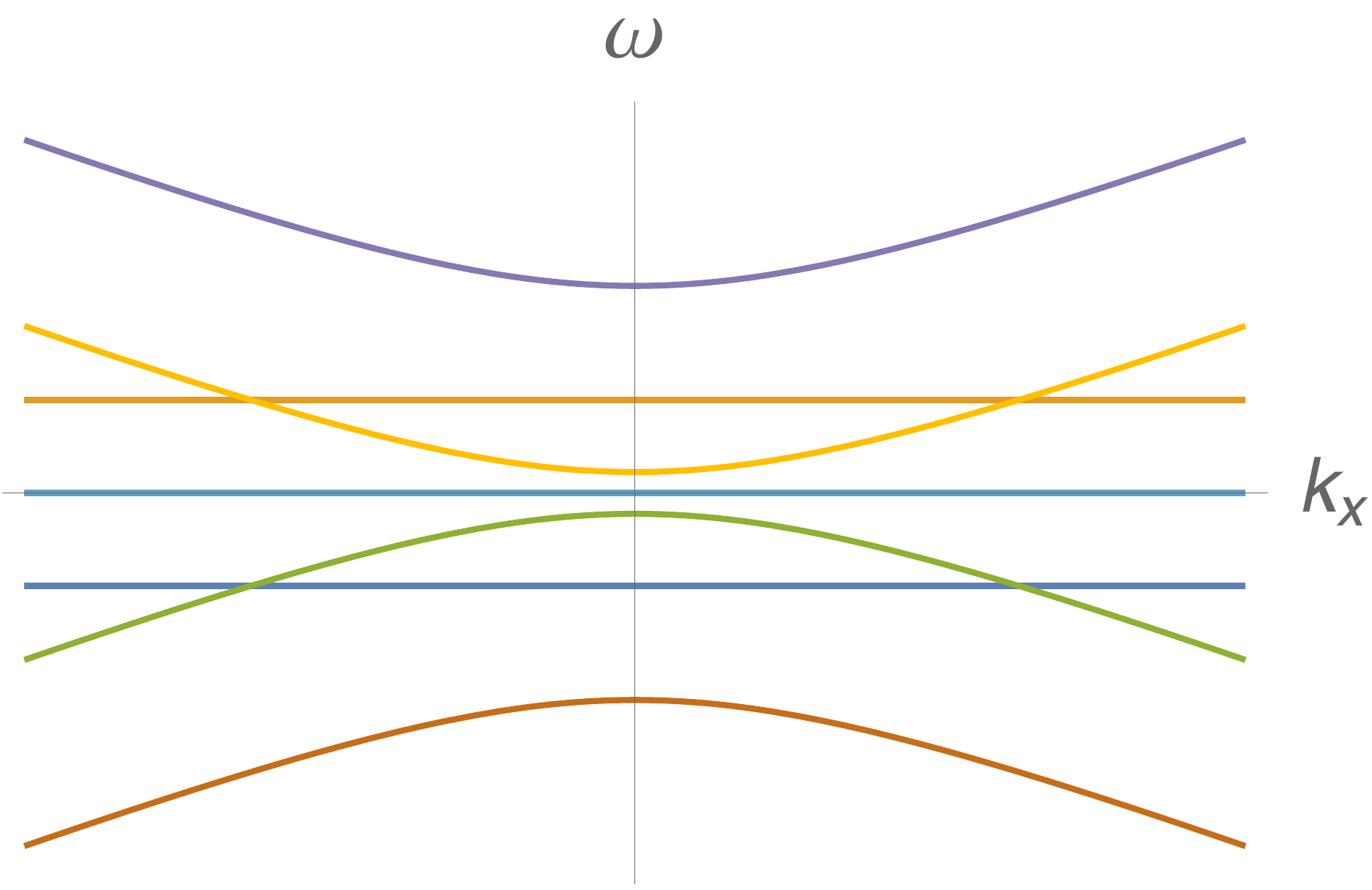}~
      \includegraphics[width=0.245\textwidth]{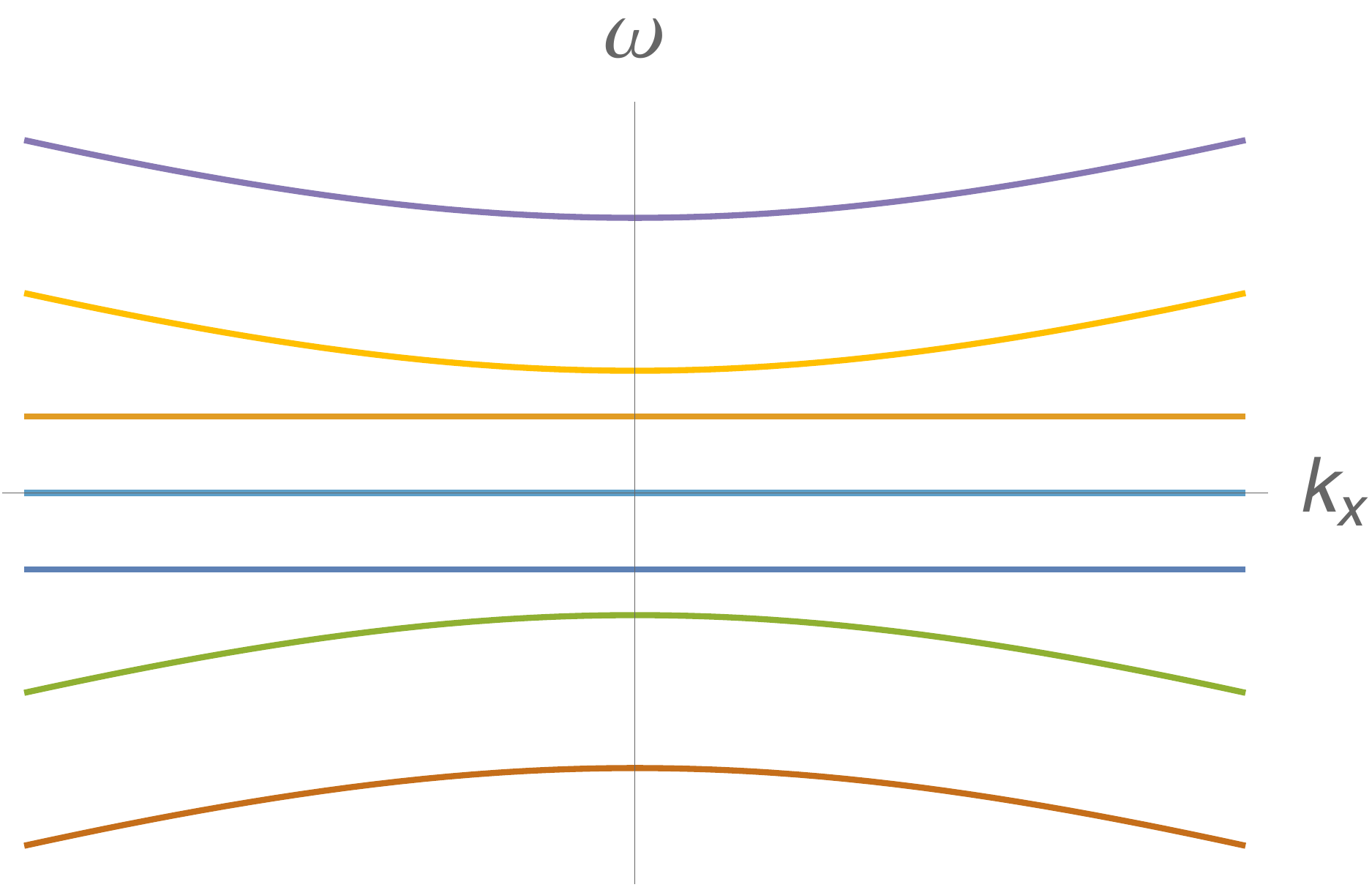}~
    \includegraphics[width=0.245\textwidth]{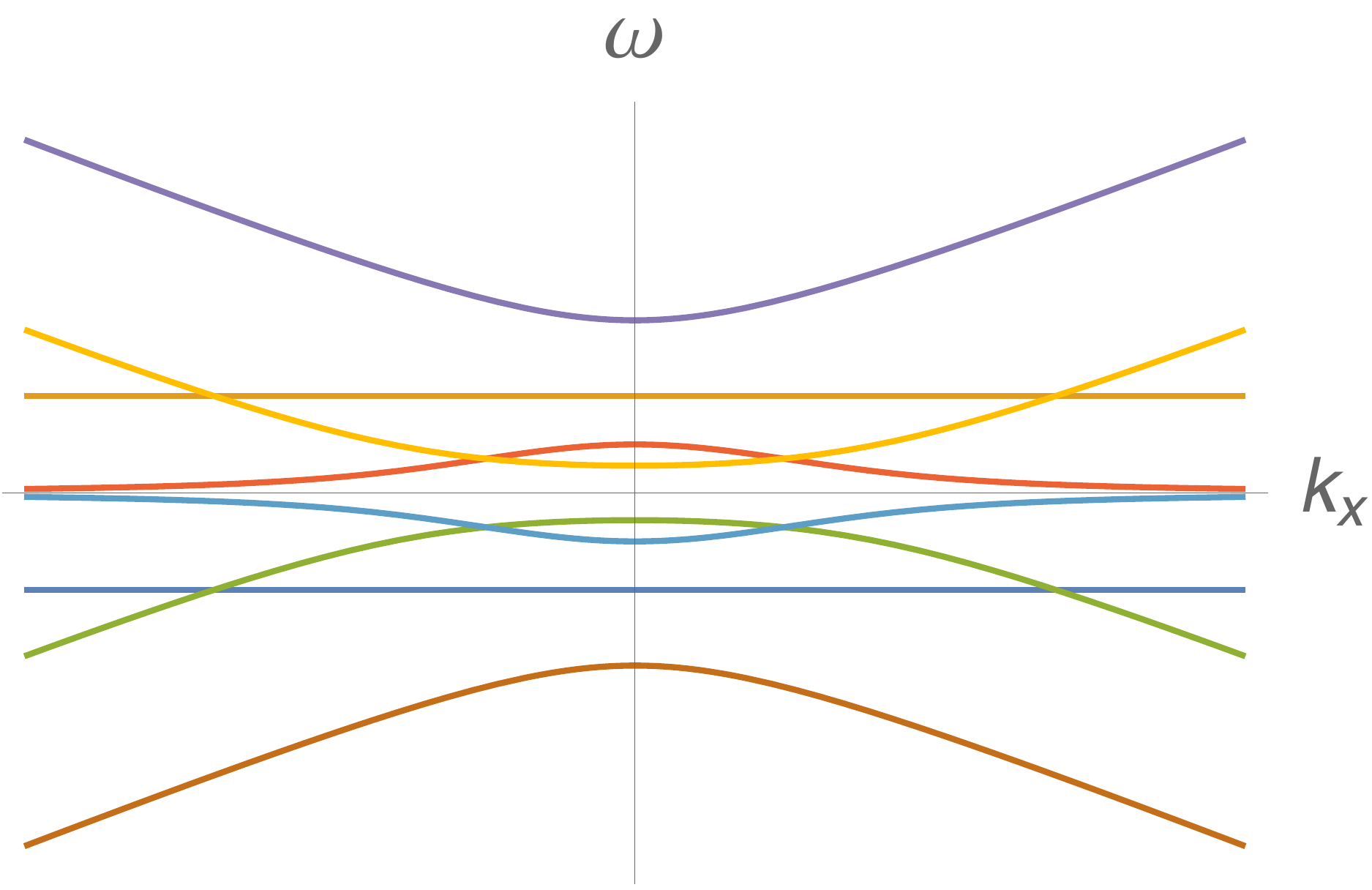}
  \caption{\small The spectrum of the modified hydrodynamics (\ref{eq:4D4Dhy2}) of $H_{4D+4D,I2}$ with $m<b$. We set $k_z=0$ and increase $k_y$ from the left to right for the first three plots. In the right plot $k_y=0$ while a nonzero $k_z$. 
 }
  \label{fig:4D4D2}
\end{figure}

\noindent $H_{4D+4D,I2}$: This system is much more complicated and we only describe the qualitative behavior briefly. As could be seen from figure \ref{fig:4D4D2}, in this case the nodes form two large and non-intersecting circles and one small circle in the middle in the $\omega,k_x,k_y$ space at $k_z=0$. For nonzero $k_z$, the one small circle becomes two non-intersecting circles at opposite $\omega$, meaning that in the $\omega,k_x,k_z$ space at $k_y=0$, the nodes are similar to the $3D+3D$ case above, where the two small circles are pinned at two opposite points of $k_x$ at $k_z=0$ as in figure \ref{fig:4D4D2}. In the $k_x,k_y,k_z$ space, the nodes become two large spheres and two small spheres.  This behavior will not change as long as there is no mass term in the $z$ direction, i.e. $m$ could appear in either $k_x$ or $k_y$ or both $k_x$ and $k_y$ directions. With a mass term in the $z$ direction, the same as in the $3D+3D$ case with the $m$ term in the $y$ direction, in the $\omega,k_x,k_z$ space, the two small circles in the middle which are pinned at two points at opposite values of $k_x$ would become two separate circles.

\vspace{0.3cm}
\noindent {\bf II. 3D+3D/4D+4D with $b$ interaction terms in maximal dimensions}

The second choice is to generalize the $b$ terms to all the spatial directions. For the 3D+3D case, there are two choices for the $m$ terms: to have them in both spatial directions or only one spatial direction. 

\begin{figure}[h!]
  \centering
  \includegraphics[width=0.300\textwidth]{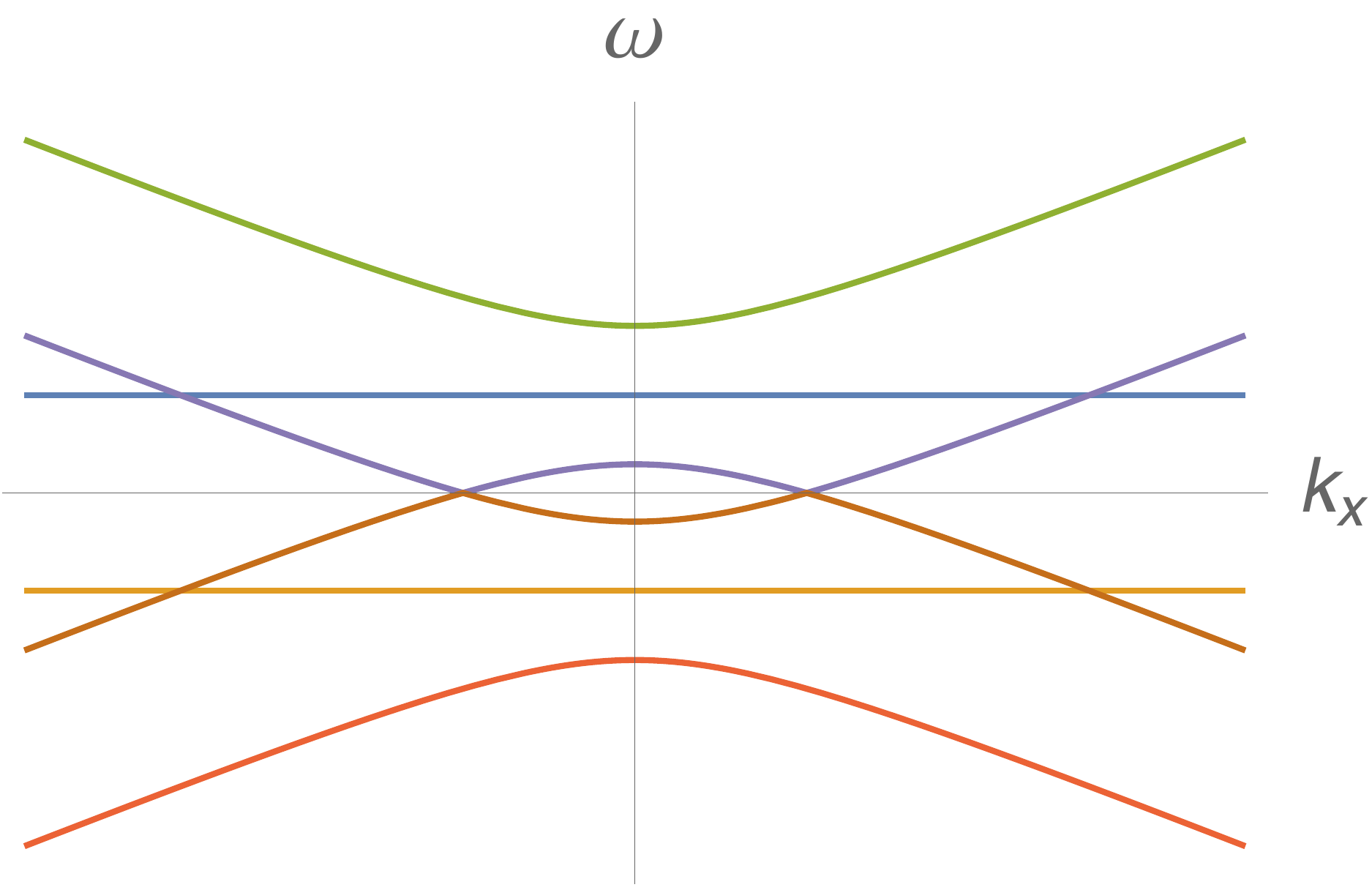}~~
   \includegraphics[width=0.300\textwidth]{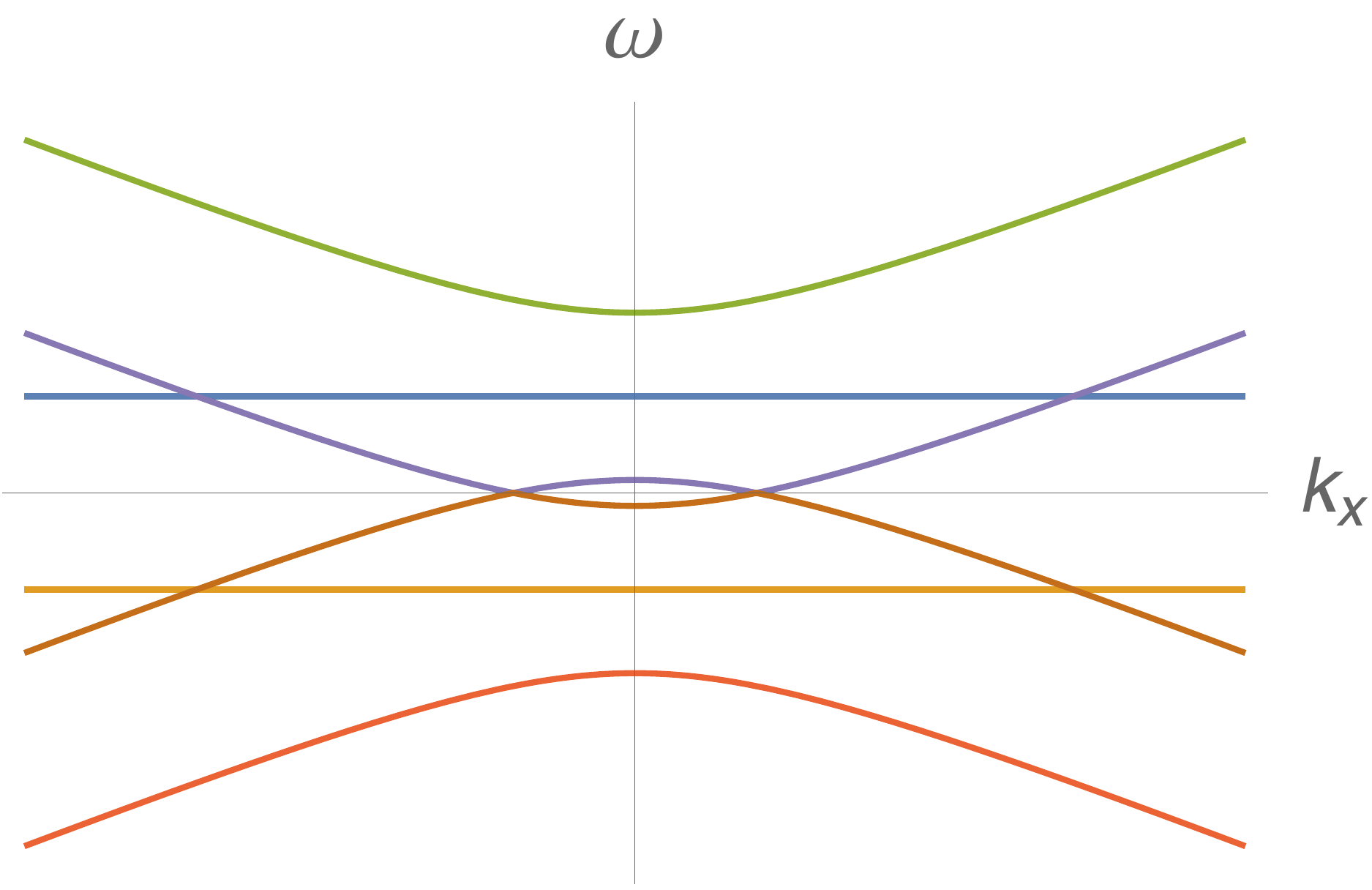}~~
    \includegraphics[width=0.300\textwidth]{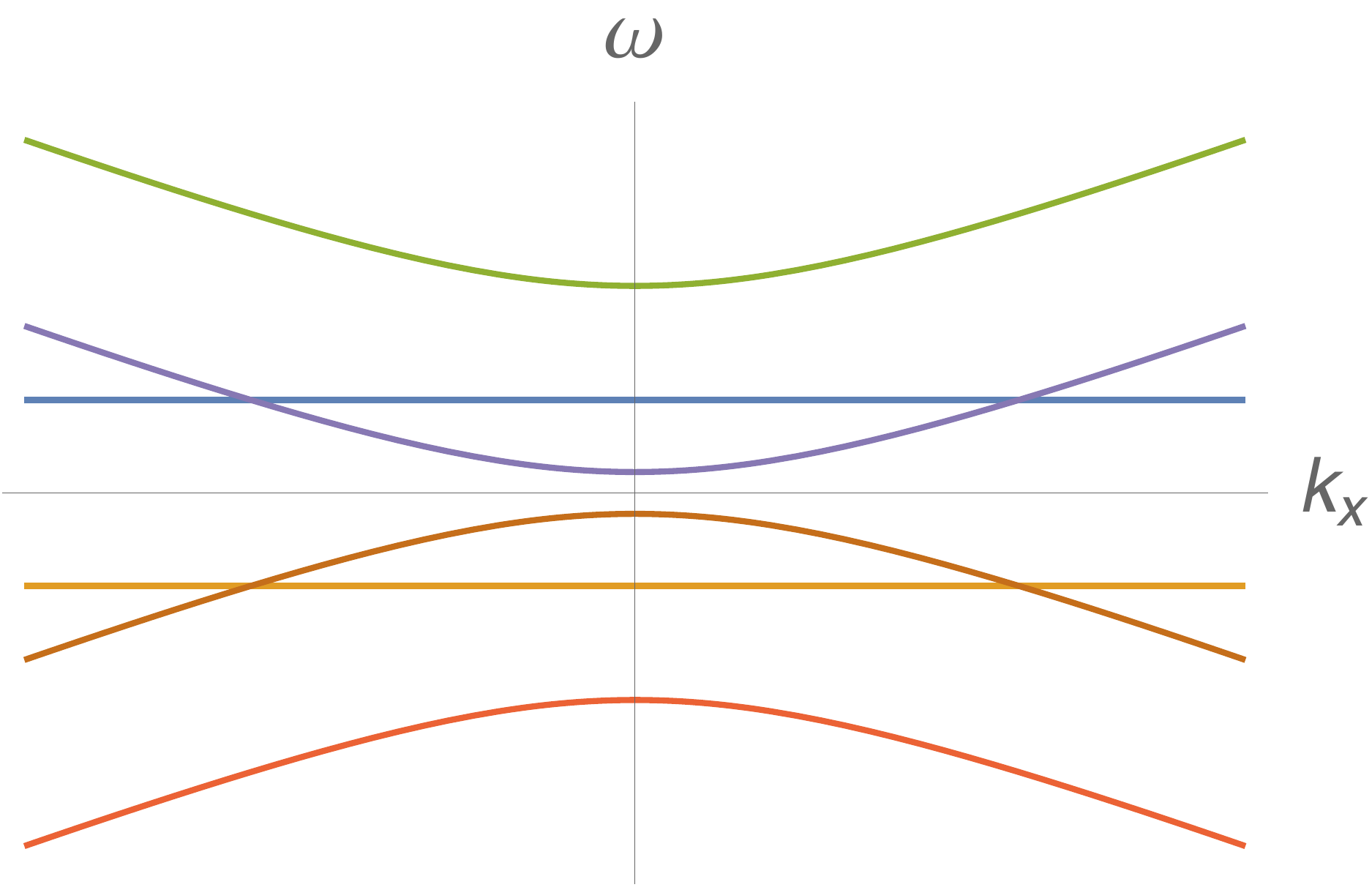}
  \caption{\small The spectrum of the modified hydrodynamics with (\ref{eq:3d3dhr2}) $H_{3D+3D,II}$  at $m<b$ and increasing from $k_y=0$ to larger values from left to right. }
  \label{fig:3d3d-hy}
\end{figure}

First we have $m$ terms in both $x$ and $y$ directions. The effective Hamiltonian is
\be
\label{eq:3d3dhr2}
H_{3D+3D,II}=\begin{pmatrix} 
0 & ~~k_x+im &~~ k_y+im &~~ ib & ~~ 0& ~~ 0 \\
k_x-im & ~~0& ~~0 & ~~ 0 & ~~ ib& ~~0 \\
k_y-im & ~~0 &~~ 0 &~~ 0 & ~~ 0& ~~ ib \\
-ib & ~~ 0 & ~~ 0 & ~~ 0 & ~~ k_x+im& ~~ k_y +im\\
0 & ~~-ib& ~~ 0  & ~~ k_x-im & ~~ 0 & ~~ 0 \\
0&~~ 0 & ~~ -ib& ~~ k_y-im & ~~ 0& ~~ 0\\
\end{pmatrix}\,.
\ee  
We could also have the $m$ term only in one direction, the $k_x$ direction, and the qualitative behavior for the spectrum does not change at all. In figure \ref{fig:3d3d-hy} we show the spectrum for the $H_{3D+3D,II}$ case in the $\omega, k_x$ space at $k_y=0$ and $k_y\neq 0$. We could see that in the $\omega, k_x, k_y$ space there are three circles of band crossing nodes with two large ones and one small one. All these band crossing nodes will not disappear due to the $m$ term in any direction. When $m$ terms in both directions disappear, the middle circle becomes a point.

We continue to the 4D+4D case. From the 3D+3D case we have seen that with maximal $b$ terms there will still be crossing band nodes even when $m$ terms in all spatial directions are turned on.  Thus here for simplicity we have $m$ terms in all spatial directions and for cases with only one or two $m$ terms the behavior would be qualitatively the same. 

The effective Hamiltonian is 
\bea
H_{4D+4D,II}=\begin{pmatrix} 
0 & ~~k_x+im &~~ k_y+im&~~ k_z+im &~~ ib & ~~ 0& ~~ 0& ~~ 0 \\
k_x-im & ~~0& ~~0 & ~~ 0& ~~ 0 & ~~ ib& ~~0& ~~ 0 \\
k_y -i m& ~~0 &~~ 0 &~~ 0 & ~~ 0& ~~ 0& ~~ ib& ~~ 0 \\
k_z-im & ~~0 &~~ 0 &~~ 0 & ~~ 0& ~~ 0& ~~ 0& ~~ ib \\
-ib & ~~ 0 & ~~ 0 & ~~ 0 & ~~ 0  & ~~ k_x+im& ~~ k_y+im& ~~ k_z+im \\
0 & ~~-ib& ~~ 0 & ~~ 0 & ~~ k_x-im & ~~ 0 & ~~ 0 & ~~ 0 \\
0&~~ 0 & ~~ -ib& ~~ 0& ~~ k_y-im & ~~ 0& ~~ 0& ~~ 0\\
0&~~ 0 & ~~ 0& ~~ -ib& ~~ k_z-im & ~~ 0& ~~ 0& ~~ 0\\
\end{pmatrix}\,.\nn
\eea

\begin{figure}[h!]
  \centering
  \includegraphics[width=0.300\textwidth]{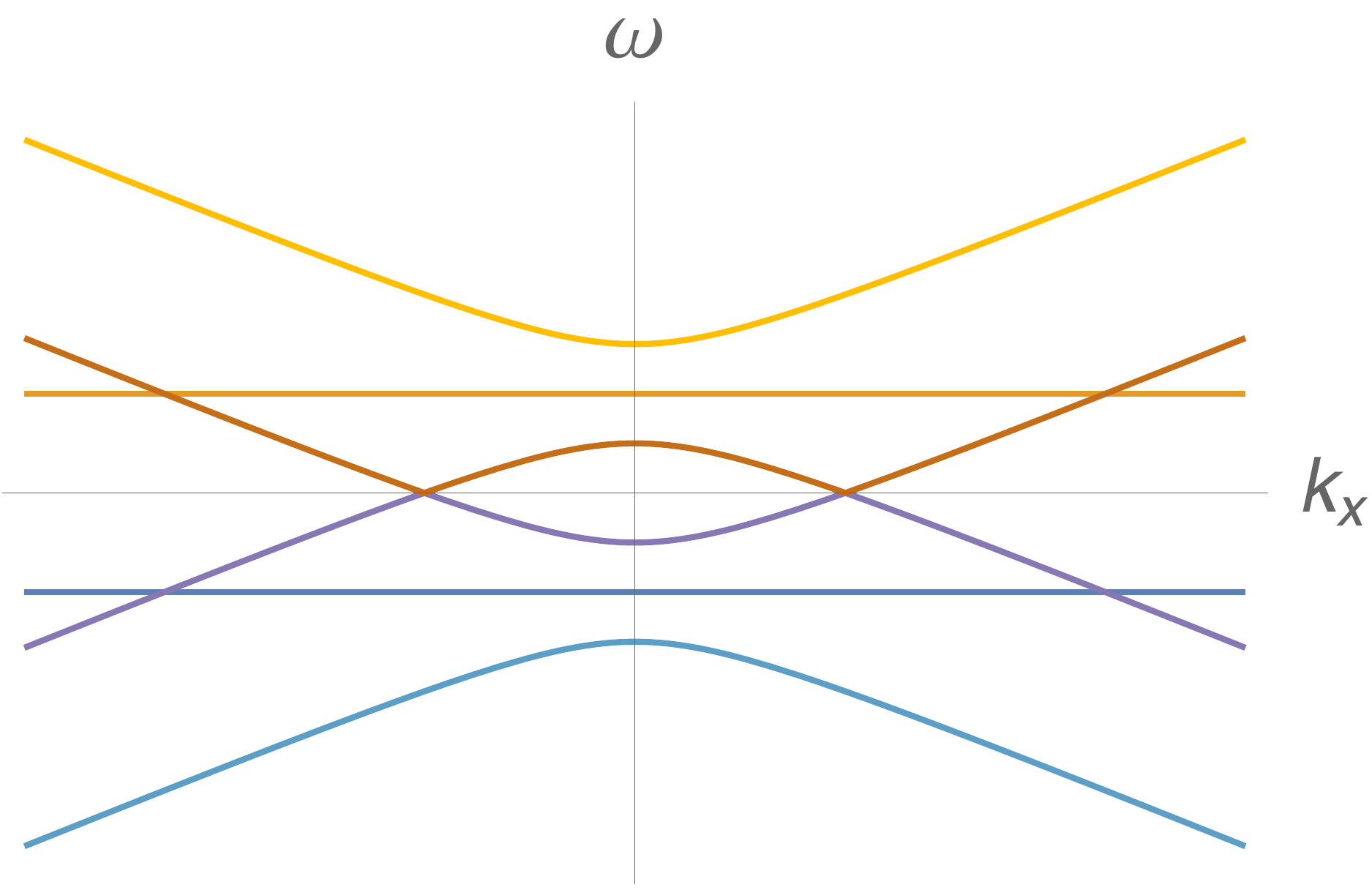}~~~
    \includegraphics[width=0.300\textwidth]{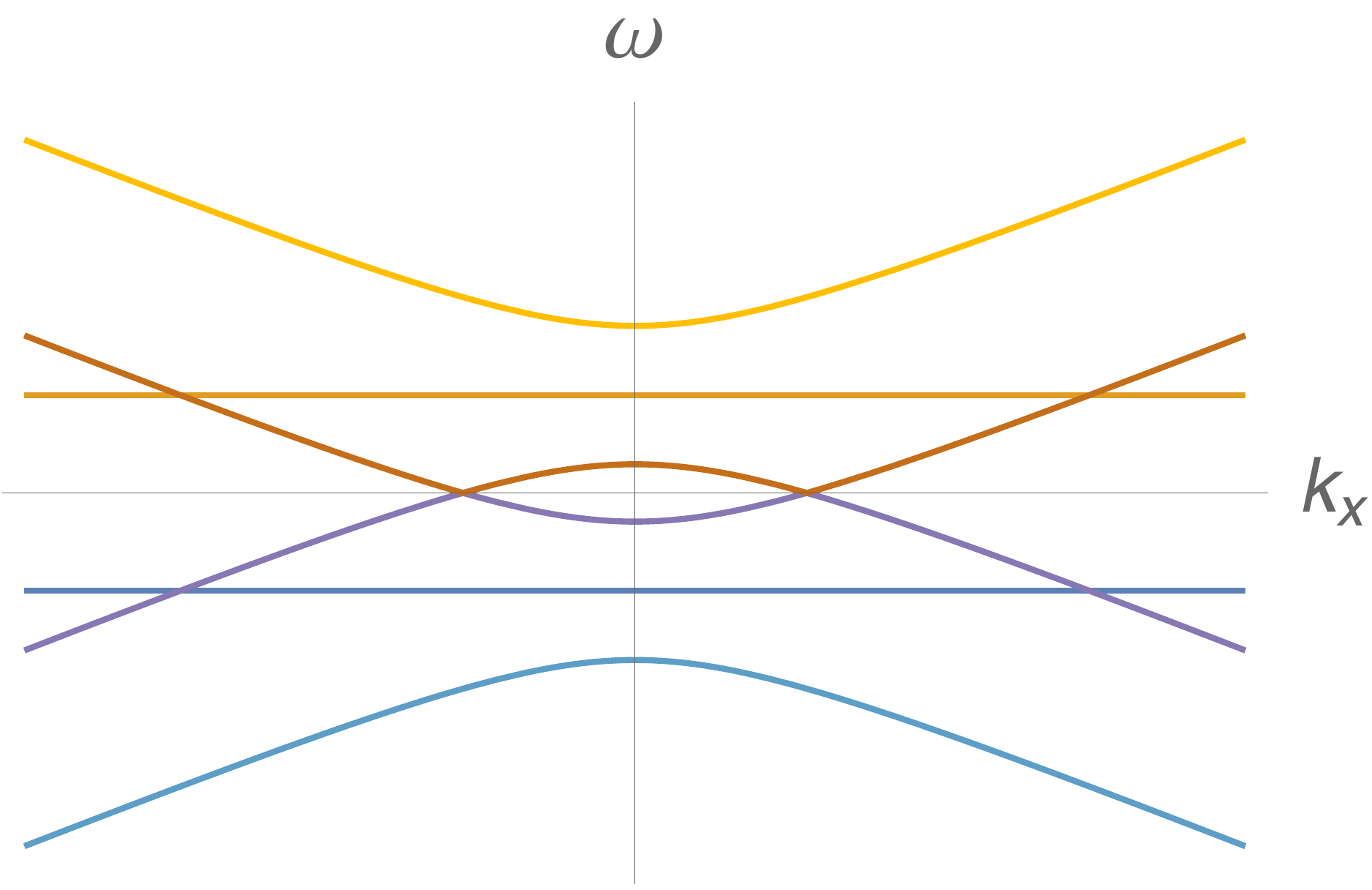}~~~
      \includegraphics[width=0.300\textwidth]{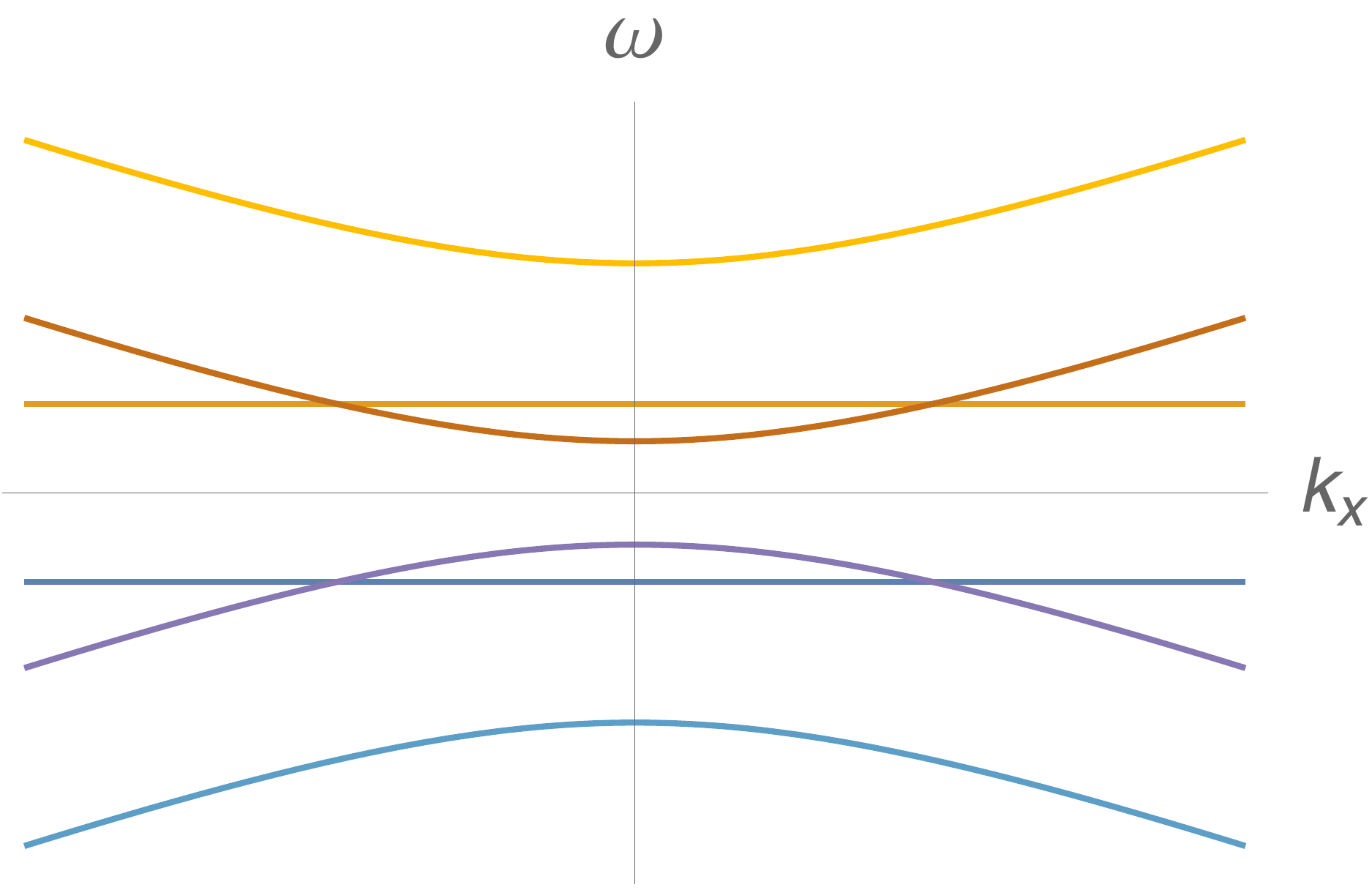}
  \caption{\small The spectrum of the modified hydrodynamics with $H_{4D+4D,II}$ at $m<b, k_z=0$ and increasing from $k_y=0$ to larger values from left plot to right plot.}
  \label{fig:4d4dhy3}
\end{figure}

Now there is an SO(3) symmetry in the $x,y,z$ directions. Again there is no analytic result for the spectrum, and numerically we could see from figure \ref{fig:4d4dhy3} that for $m<b$, in the $\omega,k_x,k_y$ space at $k_z=0$, the crossing nodes form two large circles at opposite values of $\omega$ and one small circle at $\omega=0$.  In the $k_x, k_y, k_z$ space, the nodes  become three spheres, which are codimension $1$ surfaces in the momentum space. 

Though we have not written out explicitly, in all the cases above when $m$ and $b$ values change, there will be phase transitions between fully trivial spectrum (for large enough $m/b$) and spectrum with crossing nodes which are topologically nontrivial (protected by special spacetime symmetries in certain directions in some cases). Also as may have been noticed, all topologically nontrivial nodes (without symmetry protection) are codimension one manifolds while topologically nontrivial nodes that require the protection of spacetime symmetries in $n$ dimensions are codimension $n+1$ manifolds in the $\omega$, ${\bf k}$ space.

\section{Possible origin for non-conservation terms of $T^{\mu\nu}$ }
\label{sec:origin}

We have added {a specific set of} non-conservation terms in the effective Hamiltonian to get topologically nontrivial hydrodynamic modes. In this section, we consider possible origin for these $m$ and $b$ terms. The non-conservation of $T^{\mu\nu}$ could come from external fields that couple to the system we are studying, i.e. the system constantly loses or gains energy or momentum to the external fields.

In the case of momentum dissipation terms that have been studied extensively \cite{Hartnoll:2016apf},  the external fields could be scalar fields \cite{Andrade:2013gsa, Donos:2013eha} or vector fields \cite{Donos:2012js} that break translational symmetry. However, for the $m$ and $b$ terms in section \ref{subsec:single4D}, it has been explicitly checked that scalar or vector external fields cannot give this form of energy momentum non-conservation. The simplest way to realize such extra terms is to introduce an external rank two symmetric tensor field $f_{\mu\nu}$. This external tensor field could either be an effective external tensor matter field that couples to the energy momentum tensor of the system in a specific way or could be viewed as an external  gravitational field. In the latter case, the non-conservation of $T^{\mu\nu}$ could even  be the effect of observing in a non-inertial reference frame. In the following we show more details implementing the two possibilities. In this section, we will mainly focus on the 4D case of section \ref{subsec:single4D}. More general cases will be studied in future work.

\subsection {Origin I: Effective external tensor matter field}
\label{sec:real1}

We consider an external symmetric matter field $f_{\mu\nu}$ to be the source that breaks energy momentum conservation and it couples to the operator $O^{\mu\nu}$ in the Lagrangian.  This contributes an extra $f_{\mu\nu}O^{\mu\nu}$ term in the Lagrangian of the system effectively. With this extra term the energy momentum of the system will not be conserved as it can be transferred to the external system whose energy momentum will not be counted into the system that we study.

 We could calculate the non-conservation equation for $T^{\mu\nu}$ using the method introduced in section 5 of \cite{Landsteiner:2016led}. The Lagrangian of the system is now 
 \be \mathcal{L}=\mathcal{L}_0+f_{\mu\nu}O^{\mu\nu}\,.\ee 
 We assume that $\mathcal{O}(f_{\mu\nu})\sim \mathcal{O}(k)$ so that the conservation of energy momentum tensor is weakly broken. The whole action with restored metric $S=\int d^4 x\sqrt{-g}  \mathcal{L}$ is diffeomorphism invariant and under infinitesimal coordinate transformations 
 $ x_{\mu}'=x_{\mu}+\epsilon_{\mu}$, the change in the action is 
 \be\label{acS} \delta S=\int d^4 x  \bigg(  \frac{\delta (\sqrt{-g}\mathcal{L})}{\delta g_{\mu\nu}} \delta g_{\mu\nu}+\sqrt{-g}\frac{\delta (f_{\alpha\beta}O^{\alpha\beta}) }{\delta f_{\mu\nu}} \delta f_{\mu\nu}\bigg)=0\,,\ee 
 where 
 \be\label{gfmunu} 
 \delta g_{\mu\nu}=\nabla_{\mu}\epsilon_{\nu}+\nabla_{\nu}\epsilon_{\mu}\,,~~~~~~ \delta f_{\mu\nu}=\epsilon^{\rho} \partial_{\rho}f_{\mu\nu}+f_{\rho\nu}\partial_{\mu}\epsilon^{\rho}+f_{\rho\mu}\partial_{\nu}\epsilon^{\rho}\,.\ee 
 Note that 
 we have used the fact that $\mathcal{L}_0$ does not depend on $f_{\mu\nu}$.  
Here only $g_{\mu\nu}$ and $f_{\mu\nu}$ are external fields while $T_{\mu\nu}$ defined as   \be\label{eq:emtensor}
   T^{\mu\nu}=\frac{2}{\sqrt{-g}}\frac{\delta (\sqrt{-g}\mathcal{L})}{\delta g_{\mu\nu}}\,.\ee
    could be composed of internal fields whose variations vanish due to their equations of motion. We have also assumed that $\epsilon^{\mu}$ vanishes at the boundary so all surface terms are ignored.
  
 Substituting (\ref{gfmunu}) into (\ref{acS}) and omit terms at order of $\mathcal{O}(k^2)$ or higher, we get the non-conservation equation for $T^{\mu\nu}$
 \be\label{fmunu} \partial_{\mu}T^{\mu\nu}=O^{\rho\mu}(\partial_{\nu} f_{\rho\mu}-2 \partial_{\mu} f_{\rho\nu})\,. \ee  To get the non-conservation terms that we need, this $O^{\mu\nu}$ in the formula above should be $T^{\mu\nu}$ and the nonzero components of $f_{\mu\nu}$ are 
\bea
&&f_{tt}=f_{xx}=m x\,,~~f_{tx}=f_{xt}=\frac{1}{2}m t(v_s^2+1)\,,~~\\
&&f_{ty}=f_{yt}=-\frac{1}{2}b v_s z\,,~~f_{tz}=f_{zt}=\frac{1}{2}b v_s y\,.
\eea

Note that here when $O^{\mu\nu}=T^{\mu\nu}$, it seems that we need some kind of ``fine-tuning" to make sure that $T^{\mu\nu}$ defined from (\ref{eq:emtensor}) is the same as the operator $T^{\mu\nu}$ in $f_{\mu\nu} T^{\mu\nu}$ which itself also determines the forms of the operator $T^{\mu\nu}$ from (\ref{eq:emtensor}). In fact here as $\mathcal{O}(\partial f_{\mu\nu})\sim \mathcal{O}(k)$, the term $f_{\mu\nu}T^{\mu\nu}$ in the Lagrangian only contributes a higher order term in the definition of $T^{\mu\nu}$ thus does not contribute to the leading order equation of the non-conservation equation for $T^{\mu\nu}$. Therefore we could directly take $O^{\mu\nu}=T^{\mu\nu}_{(0)}=\frac{2}{\sqrt{-g}}\frac{\delta (\sqrt{-g}\mathcal{L}_0)}{\delta g_{\mu\nu}}$, which is more natural, and $T^{\mu\nu}_{(0)}=T^{\mu\nu}$ at order $\mathcal{O}(k^0)$. At the fundamental level, the only possibility for $f_{\mu\nu}$ to have this kind of coupling is the metric field of the next subsection, however, we could not rule out the possibility of an effective coupling in this form in an effective theory, e.g. in elastic theories, though we do not have any concrete examples at hand.

\subsection{Origin II: Gravitational field, non-inertial reference frame}
\label{sec:rel2}
This is a more natural and at the same time very interesting possibility.\footnote{We thank Karl Landsteiner for pointing this out.} As we need an external symmetric tensor field which couples to the energy momentum tensor in the form of $f_{\mu\nu}T^{\mu\nu}$, it is natural to consider if this external field could be a metric field. With this possibility the spacetime would be nontrivial and the energy momentum tensor would not be conserved in the form of $\partial_\mu  T^{\mu\nu}=0$ but as $\nabla_\mu  T^{\mu\nu}=0$. In this way, when we have a nontrivial metric field, $\partial_\mu  T^{\mu\nu}=0$ will not hold anymore.  

We start from the following covariant form of energy momentum conservation equation
\be
\label{eq:curvecon}
\nabla_\mu \delta T^{\mu\nu}=0\,.
\ee We assume that the new spacetime metric is $g_{\mu\nu}=\eta_{\mu\nu}+h_{\mu\nu}$, thus we get
\be
\partial_\mu\delta T^{\mu\nu}=-\frac{1}{2}\partial_\alpha h \delta T^{\alpha\nu}-\frac{1}{2}\eta^{\nu\beta}(2\partial_{\mu}h_{\alpha\beta}-\partial_\beta h_{\mu\alpha})\delta T^{\mu\alpha}\,.
\ee Again, we  have assumed that $\mathcal{O}(\partial h_{\mu\nu})\sim \mathcal{O}(k)$ and only kept leading order in $k$ terms.  To get the exact $m$ and $b$ terms in the effective Hamiltonian of the single 4D system, we could choose the components of $h_{\mu\nu}$ to be \bea
&&h_{tt}=h_{xx}=m x\,,~~h_{tx}=h_{xt}=\frac{1}{2}m t(v_s^2+1)\,,~~\\
&&h_{ty}=h_{yt}=-\frac{1}{2}b v_s z\,,~~h_{tz}=h_{zt}=\frac{1}{2}b v_s y
\eea
while all the other components are zero. With this choice, the conservation equation (\ref{eq:curvecon}) reduces to (\ref{eq:4dhydro}).\footnote{We assume that we are working in a large but finite volume of spacetime and $m$, $b$ are so small that $m x, b y, b z, mt\ll 1$ in the finite volume. In fact this requires $m\ll k$ so we assume that $\mathcal{O}(k^2)<\mathcal{O}(m)< \mathcal{O}(k)$. In this way we still keep all terms at order $m, ~b$ and order $k$ so all the calculations are not affected. We ignore the boundary effects as the volume is large.  }

\subsubsection{Symmetry of the single 4D system}
\label{sec:symmetry}

{In this subsection, we obtain the symmetry of this single 4D symmetry from the Killing vector of the new metric and then introduce the symmetry that we need to protect the topology of the crossing nodes, i.e. the symmetry that would be broken by $m$ terms in the $y$ and $z$ directions.}

In the system of (\ref{eq:4dhydro}), original Poincare symmetry is broken and new spacetime symmetries of the system could be found from the new metric $g_{\mu\nu}=\eta_{\mu\nu}+h_{\mu\nu}$. From the covariant conservation equation $\nabla_\mu  T^{\mu\nu}=0$, the isometry of $g_{\mu\nu}$ could keep the equation (\ref{eq:4dhydro}) unchanged. The Killing vectors generating the isometry of the new spacetime metric $g_{\mu\nu}=\eta_{\mu\nu}+h_{\mu\nu}$ are
\be
K_\mu=
\sum_{i=0}^3 a_i \chi_i+ \sum_{i=1}^6 c_i \theta_i
\ee
where $a_0,..., a_3, c_1, ..., c_6$ are constants with
\bea\begin{split}
\chi_0&=\left(1-\frac{m x}{2},~ -\frac{mt}{2}, ~0, ~0\right)\,,~~~~\chi_1=\left(\frac{mtv_s^2}{2}, ~1+\frac{mx}{2},~ 0, ~0\right)\,,~~\\
\chi_2&=\left(-\frac{b z v_s}{4}, ~0, ~1, ~-\frac{b t v_s}{4}\right)\,,~~~~\chi_3=\left(\frac{b  yv_s}{4}, ~0, ~\frac{b tv_s}{4}, ~1\right)\,,~~
\end{split}\eea
and
\bea\begin{split}
\theta_1&=\left(-\frac{m (x^2+ v_s^2 t^2)}{4}+x,~ -t-\frac{mt x}{2}, ~0, ~0\right)\,,~~\\
\theta_2&=\left(\big(1-\frac{mx}{2}\big)y,~ -\frac{mt y}{2}, ~\big(-1+\frac{mx}{2}\big)t, ~\frac{bt^2v_s}{4}\right)\,,~~\\
\theta_3&=\left(\big(1-\frac{mx}{2}\big)z,~ -\frac{mt z}{2}, ~-\frac{bt^2v_s}{4},~ \big(-1+\frac{mx}{2}\big)t\right)\,,~~\\
\theta_4&=\left(\frac{mt yv_s^2}{2}+\frac{bxz v_s}{4},~-\frac{btz v_s}{4}+y\big(1+\frac{mx}{2}\big), ~-\frac{m (x^2+ v_s^2 t^2)}{4}-x,~\frac{btxv_s}{4}\right)\,,~~\\
\theta_5&=\left(\frac{mt zv_s^2}{2}-\frac{bxyv_s}{4},~\frac{bty v_s}{4}+z\big(1+\frac{mx}{2}\big), ~-\frac{btxv_s}{4}, ~-\frac{m (x^2+ v_s^2 t^2)}{4}-x\right)\,,~~\\
\theta_6&=\left(-\frac{b v_s (y^2+ z^2)}{4},~ 0, ~z, ~-y\right)\,.~~
\end{split}
\eea  These Killing vectors generate the isometry of the metric $g_{\mu\nu}$, i.e. with a coordinate transformation $x^{\mu}\to x^{\mu}+s K_\mu$ where $s$ is an infinitesimal constant, the metric $g_{\mu\nu}$ would not change.  These coordinate transformations can be viewed as the spacetime symmetries of the hydrodynamic system with dynamical equation (\ref{eq:4dhydro}).  We have ten independent coordinate transformations which keep the system invariant. $\chi_0$ is a combination of translation in the $t$ direction and a boost in the $t$-$x$ directions. $\chi_1$ is a combination of translation in the $x$ direction as well as a boost in the $t$-$x$ directions. 

Among these ten Killing vectors, the ones that forbid $m$ terms in the $y$ and $z$ directions could be chosen to be $\chi_2$ and $\chi_3$ which are combinations of translations in $y$ and $z$ directions and boosts in the $t$-$y$ and $t$-$z$ directions. Thus, in the single 4D case, the symmetries that protect the topologically nontrivial band crossing modes could be the two spacetime symmetries generated by $\chi_2$ and $\chi_3$. 

We have solved the infinitesimal spacetime symmetry of the system and besides this infinitesimal coordinate transformation, there are other possible large symmetries of the system that could forbid the $m$ terms in the two directions, e.g. reflection symmetries in the two directions. Also in symmetry protected topological states of matter, the symmetry that is needed to protect the topological structure does not need to be the whole symmetry that forbids the $m$ term. It could be that only a subgroup of the symmetry is at work to prevent the band crossings from being destroyed. Thus, the symmetry that is required for the protection of the topological structure could be a special subgroup of the whole symmetry. Here the symmetry that we need for the protection of the topological structure in the single 4D case is in fact the reflection symmetry $M$ in both $y$ and $z$  directions, i.e. $M: y\to -y, z\to -z$. This is not an infinitesimal coordinate transformation of the spacetime, but we could see that the $M$ symmetry forbids both $m$ terms in the $y$ and $z$ directions as these terms violate the reflection symmetry in the $y$ or $z$ directions. This could be seen from the fact that $T^{yt}\to-T^{yt}$ under the symmetry transformation and the mass term in the $y$ or $z$  directions would violate the reflection symmetry in these two directions. We will see in the next section how this protecting symmetry could be used to calculate the topological invariants.
 
Note that not all systems that were studied in section \ref{sec:tophm} require a protecting symmetry and some of the systems are topologically band crossing states without requiring any symmetry.

\subsubsection{non-inertial reference frame}
This graviton field $h_{\mu\nu}$ could come from sources of massive matter and more intriguingly it could also come from a coordinate transformation from the flat Minkowski metric, where $x_{\mu}'=x_{\mu}+\xi_{\mu}$ with \be\xi_\mu=\bigg(\frac{mxt}{2}, ~~\frac{mx^2}{4}+\frac{mt^2}{4}v_s^2,~~-\frac{b}{4}v_s z t,~~\frac{b}{4}v_s y t\bigg)\,.\ee This is an intriguing result as usually a nontrivial gravitational field could not be transformed to a flat spacetime globally but only locally. It could be checked that this new metric field has all components of the Riemann tensor vanishing at leading order, thus could be transformed to the flat spacetime. Though equivalent to a flat spacetime, $h_{\mu\nu}$ could still be viewed as a non-trivial gravitational field according to the equivalence principle.
 
This result suggests that in a specific non-inertial frame, we could observe hydrodynamic modes that are topologically protected even when they are topologically trivial in the original inertial frame. A third possible circumstance to have this nonzero $h_{\mu\nu}$ could be in analog gravity systems, where certain materials could give rise to effective hydrodynamic equations as if there exists a nontrivial gravitational field. 


Note that in this case, with nonzero components of $h_{\mu\nu}$ the constitutive equations for $T_{\mu\nu}$ could also be written into a covariant form thus leading to extra terms compared to the original constitutive equations. However, it can be explicitly checked that these extra terms do not change the spectrum at leading $k\sim m, b$ order up to a rescaling of parameters $m$, $b$ and $v_s$. More details could be found in the appendix of \cite{Liu:2020ksx}.

\section{Topological invariant}
\label{sec:ti}

The band crossing behavior in the hydrodynamic modes in this paper is very similar to that of the Weyl semimetal or other types of gapless semimetals, raising the question if these modes are indeed topologically protected modes. In section \ref{sec:tophm} we have shown explicitly that these band crossing nodes would not disappear by small perturbations thus have already shown that they should be topologically protected. Note that in several systems we studied in section \ref{sec:tophm}, the perturbations need to obey certain required symmetry and in other systems we do not require any protecting symmetry. In this section we further confirm that the band crossing  hydrodynamical modes are indeed topologically nontrivial modes or symmetry protected topological modes by the existence of nontrivial topological invariants for different cases. Before the detailed calculation of topological invariants in our system, we will first introduce some basic ingredients in the calculation of topological invariants for topological band crossing states of matter.
\vspace{.4cm}\\
\noindent {\bf Topological invariants for band crossing topological states}

A topological invariant is a quantity that does not change unless there is a topological phase transition from a topologically nontrivial phase to a trivial one and the quantity has to have different values for the topologically nontrivial and trivial phases. This quantity could be a number or a property which cannot be identified with a number. For topological band crossing modes, the crossing node is a singular point in the momentum space which cannot disappear due to  small perturbations, thus could be associated with a nontrivial topological invariant which is different from the value obtained for a trivial vacuum state. For accidental band crossing  modes, the crossing node could be disappear due to an arbitrarily small perturbation, thus should possess trivial topological invariants. 

For a Weyl semimetal, the topological charge is the chirality charge that could be calculated from the integral of the Berry curvature on an infinitesimal sphere enclosing the Weyl point in the three dimensional momentum space. The topological invariant for nontrivial Weyl nodes are either $1$ or $-1$ in contrast to the trivial value of $0$. For a nodal line semimetal, as the nodal line spans in one momentum direction, there are only two spatial directions left and the topological invariant is the Berry phase accumulated along the circle that links the nodal line \cite{nlsm,Liu:2018bye,Liu:2018djq}. A nontrivial nodal line semimetal would have a Berry phase of $\pi$ in contrast to $0$ for an accidental one.

\begin{figure}[h!]
  \centering
  \includegraphics[width=0.30\textwidth]{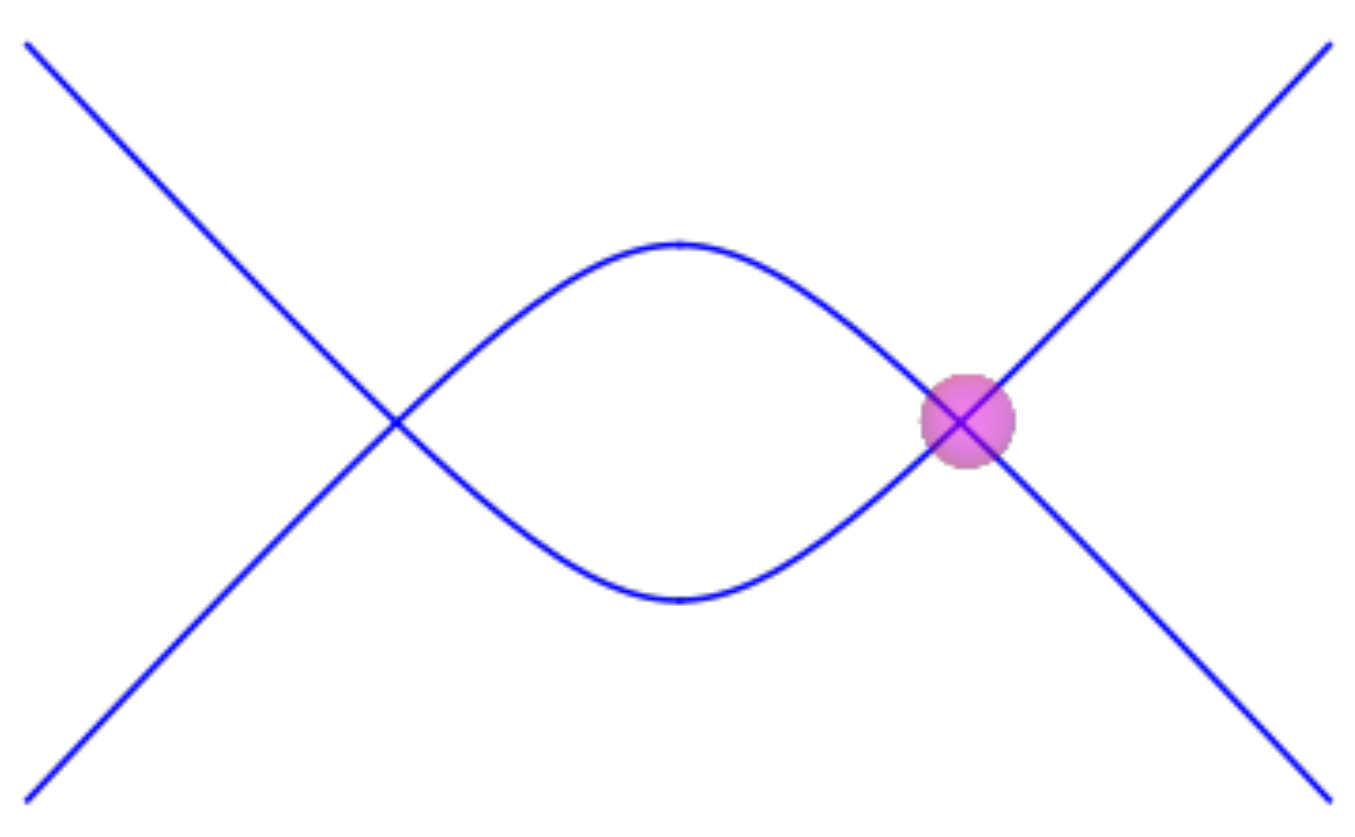}\,~~
    \includegraphics[width=0.30\textwidth]{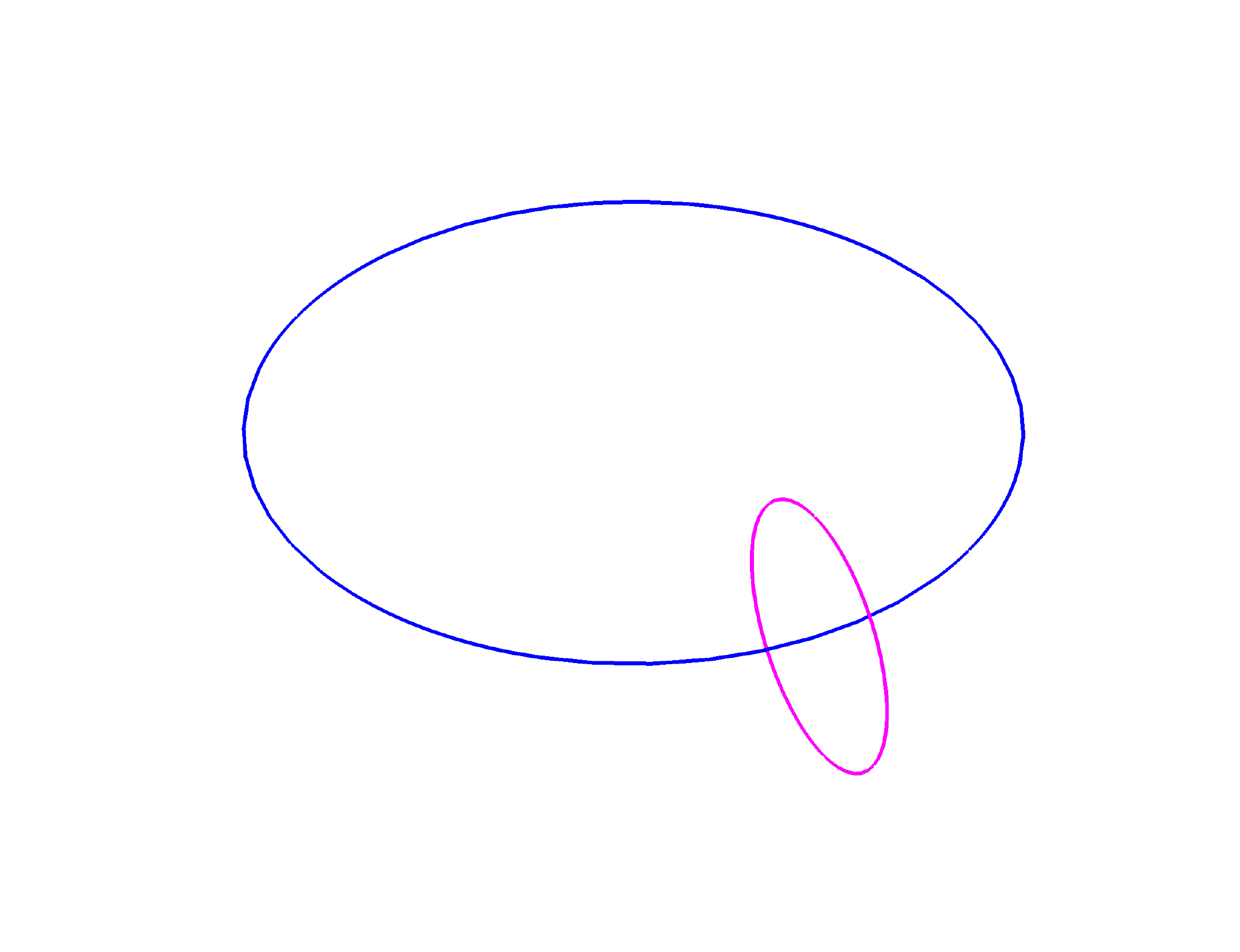}\,~~
      \includegraphics[width=0.30\textwidth]{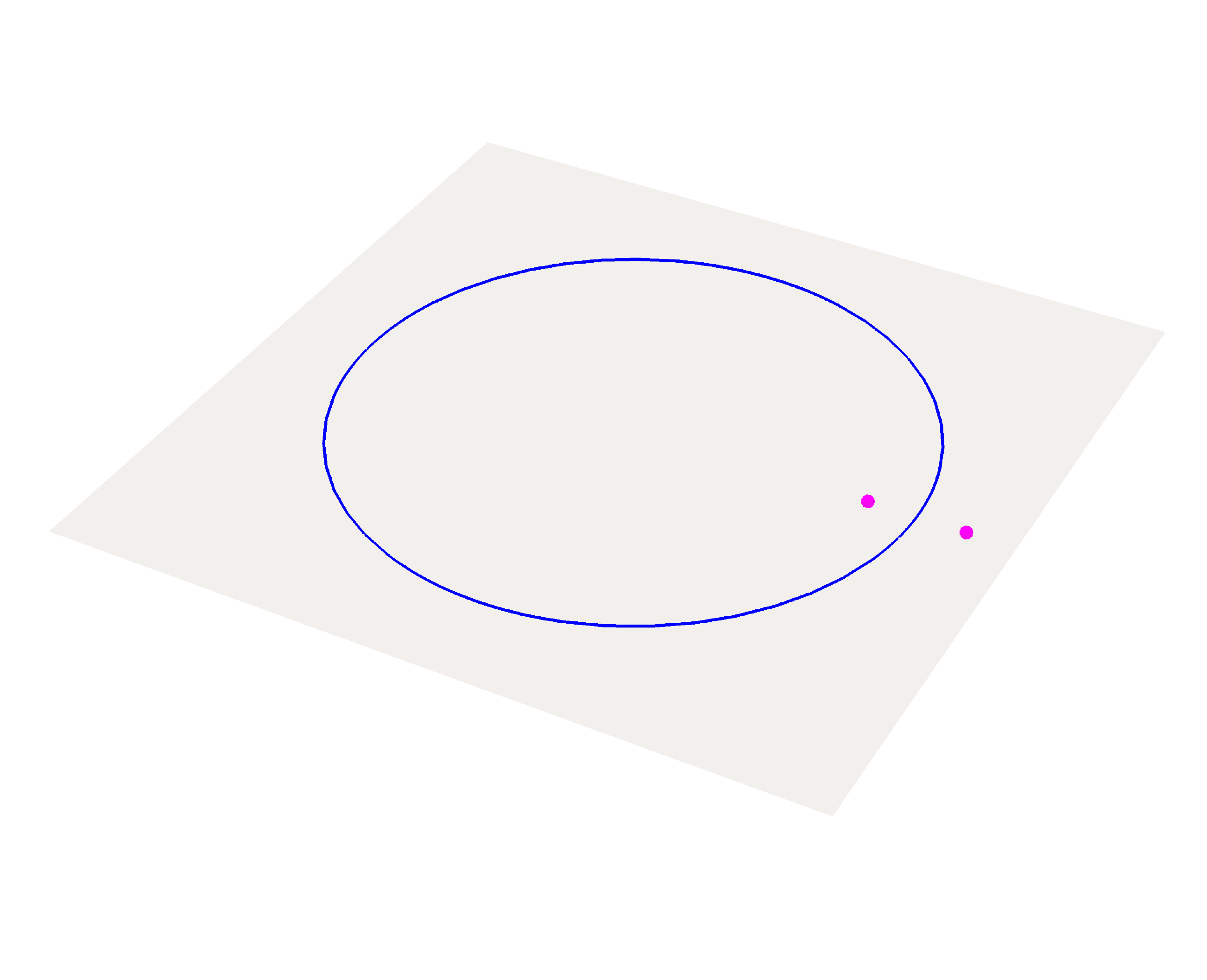}
  \caption{\small Three typical cases where the calculation of the topological charge is on a two/one/zero dimensional manifold respectively. In the left figure, the crossing nodes are two points. In the right two figures blue curves indicate the crossing nodes and the spectrum at other momenta have not been plotted.}
  \label{fig:ill2}
\end{figure}

Note for band crossing states, the calculation of the topological invariant depends crucially on the dimensionality of the system. For band crossing  states, the nodes are momentum space singular points for Green functions and any physical calculation has to avoid passing through this point. Thus for a three spatial dimensional Weyl semimetal, the topological charge is an integration of the Berry curvature on a two dimensional sphere surrounding the node (left figure of Fig. \ref{fig:ill2}), while for a three spatial dimensional nodal line semimetal, as the nodes form a circle, the topological charge is an integration of the Berry phase along a one dimensional manifold: a circle linking the circle of the nodes (middle figure of Fig. \ref{fig:ill2}). 

For some systems, the topological charge has to be calculated on an even lower dimensional manifold, i.e. on zero dimensional manifolds. This could happen for a nodal line semimetal in two spatial dimensions. More commonly, this could come from symmetry protected topological state, where the calculation of the topological invariant has to be on a high symmetric point thus reducing the dimension of the system to effectively zero dimension. An example is the mirror symmetry protected topological nodal line semimetal (right figure of Fig. \ref{fig:ill2}), where the nodes have to be protected by a mirror symmetry, i.e. small perturbations that violate the mirror symmetry could destroy the band crossings of the system while perturbations that do not violate the mirror symmetry would not. In symmetry protected topological states, topological invariants have to be calculated at the high symmetric points in the momentum space, which would give a nontrivial result, and the topological invariants calculated directly (not at the high symmetric points in the momentum space) would give a trivial result. The requirement of the high symmetric points would usually reduce the dimensionality of the system to lower dimensions.

Fig. \ref{fig:ill2} shows the three typical cases where the calculation of the topological charge is on a two/one/zero dimensional manifold respectively. The left plot in Fig. \ref{fig:ill2} is for a typical Weyl node, where the calculation of the topological invariant is an integration of the Berry curvature on a two dimensional sphere (magenta sphere). The middle plot in Fig. \ref{fig:ill2} is for a typical nodal line semimetal, where the calculation of the topological invariant is an integration of the Berry phase on a one dimensional circle (magenta circle). The right plot in Fig. \ref{fig:ill2} is for a mirror symmetry protected topological nodal line semimetal \cite{nlsm}, where the calculation of the topological invariant has to be on the high symmetric plane, which reduces the dimensionality for the calculation to zero.

Here for all the cases in this paper, 
the manifold enclosing the node is zero dimensional effectively, either because the system is symmetry protected so the calculation of the topological invariant needs to be at a high symmetric point or because the nodes form a co-dimension one surface.  Compared to the sphere in three spatial dimensions and the circle in two spatial dimensions surrounding the singular nodes, in the cases here we only have two points surrounding the singular nodes, as illustrated in the right figure of Fig. \ref{fig:ill2}.

In this zero dimensional case, the calculation of the topological invariant is very different. We first briefly review the calculation of the topological invariant of the mirror symmetry protected topological nodal line semimetal in \cite{nlsm} as an example here and then perform our calculation following this calculation later in this section. We will also show that this calculation is in fact equivalent to the requirement of the existence of an undetermined Berry phase. 

\begin{figure}[h!]
  \centering
  \includegraphics[width=0.45\textwidth]{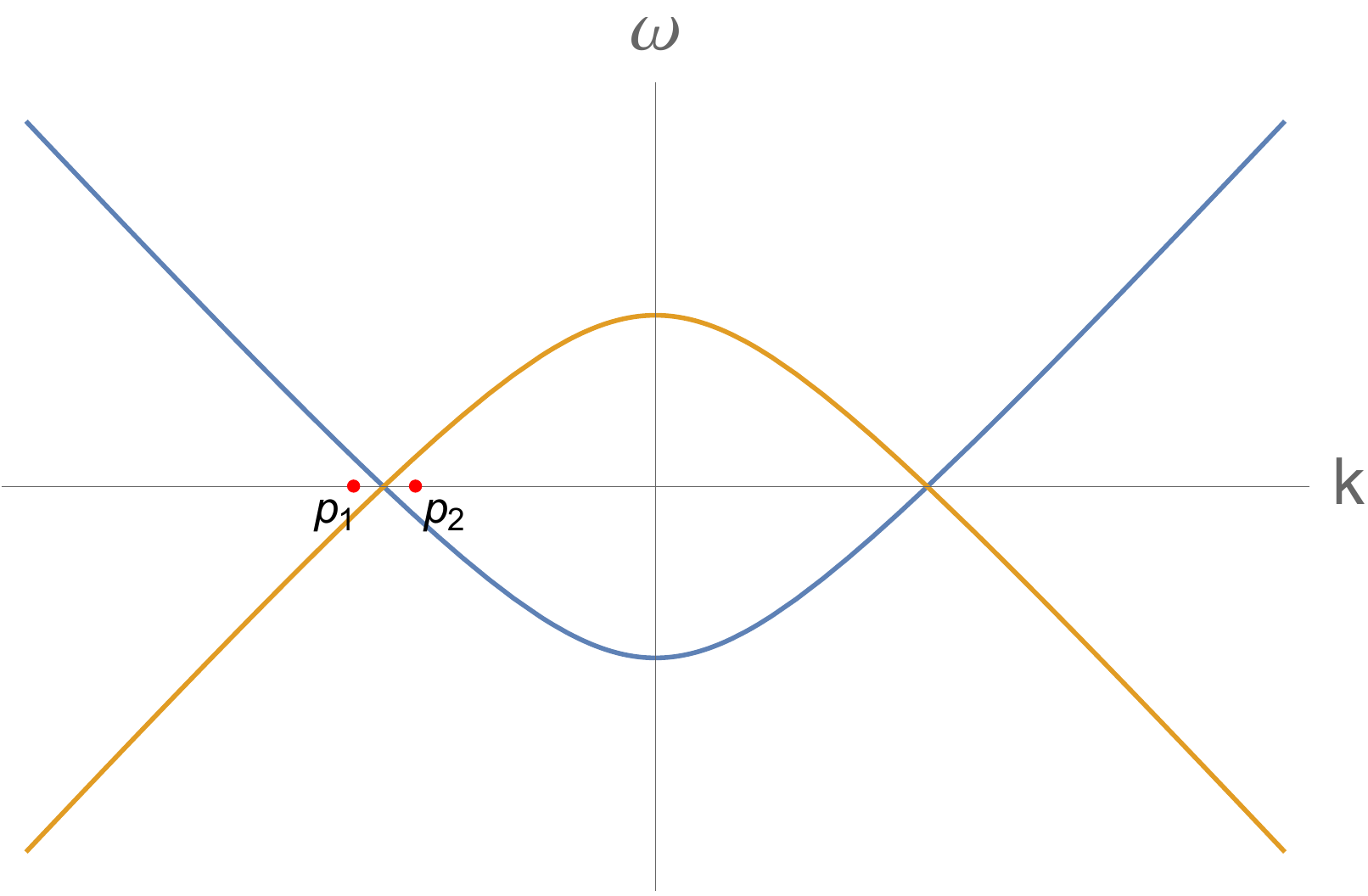}
  \caption{\small 
  $p_1$ and $p_2$ denote the outer and inner points of the nodal loop.}
  \label{fig:p1p2}
\end{figure}

The mirror symmetry protected topological nodal line semimetal is illustrated in the right figure of Fig. \ref{fig:ill2}. To calculate the topological invariant, we have to perform the calculation at the high symmetric points of the system, which is the plane invariant under the reflection, shown in shade in the figure. The nodal line also resides on this plane and we pick two points on the two sides of the nodal loop, denoted by $p_1$ and $p_2$ separately, e.g. we could denote the outer point in the right figure of Fig. \ref{fig:ill2} as $p_1$ and the inner one as $p_2$. As in the right figure of Fig. \ref{fig:ill2} only the nodes are plotted,  at each point $p_i$, there should be a gapped spectrum, i.e. one band above the mirror plane and one band below the mirror plane (occupied band), which has not been plotted out in Fig. \ref{fig:ill2} and is shown in \ref{fig:p1p2}.

Then as explained in Sec.II A of \cite{nlsm}, the topological invariant is defined to be $\xi_0=N_1-N_2$ (equation (6) in \cite{nlsm}), where $N_i$ refers to the number of occupied bands that have mirror symmetry eigenvalue of $+1$ at the point $p_i$.  Here as the mirror symmetry commutes with the Hamiltonian at the high symmetric point, the energy eigenstates carry their Mirror symmetry eigenvalues. Consider Fig. \ref{fig:p1p2} as the $k_y=0$  spectrum of a nodal line semimetal and $p_i$ points are the two inner and outer points of the left node, as shown in Fig. \ref{fig:p1p2}. The two curves with different colors have the mirror symmetry eigenvalue 1 and -1 separately as each curve is for a set of eigenstates with the same eigenvalue $E_i(k)$ which also has the same mirror symmetry eigenvalue. Thus if at $p_1$, the number of occupied $+1$ mirror eigenvalue band is 1 and at the other point $p_2$, the number of occupied $+1$ mirror eigenvalue band would be zero as could seen from Fig. \ref{fig:p1p2}. Thus  the topological invariants could be calculated from the defining formula to be $\pm 1$ for the two nodes. Thus the system is a mirror symmetry protected topologically nontrivial state. Note that in general, the symmetry does not need to be the full symmetry of the system, and one subgroup of the symmetry that is enough for the calculation of the topological invariant.

In fact we could see from the calculation above that this calculation of the topological invariant relies on the fact that the two occupied states (states with energy lower than the crossing nodes) at the two sides of the node belong to two different sets of eigenstates so that they have distinct mirror symmetry eigenvalues. The existence of a nontrivial topological invariant defined this way is also equivalent to the fact that the band crossings cannot be destroyed by small perturbations. This is because the two occupied states at the two sides of the node
 behave differently under small deformations of the system. When the system is perturbed by a small perturbation, each set of eigenstates will get a small deformation and they will still cross at some point as long as the perturbation is small enough. However, if the two lowest energy states at the two sides of the node belong to the same origin of eigenstates (as the left figure of Fig. \ref{fig:ill}), when each band gets a small perturbation, it is possible that the band crossing would disappear, e.g. if the upper band goes up while the lower band goes down.
 
Then the question becomes how do we distinguish whether they belong to the same origin of eigenstates or to two different sets of eigenstates. In the calculation above, different types of eigenstates are characterized by different eigenvalues of the mirror reflection symmetry. In fact, a more direct way is to see if the eigenstates at the two sides the nodes are orthogonal or equal to each other up to a relative phase. If they are orthogonal to each other, they surely belong to two different sets of eigenstates. 

\begin{figure}[h!]
  \centering
  \includegraphics[width=0.55\textwidth]{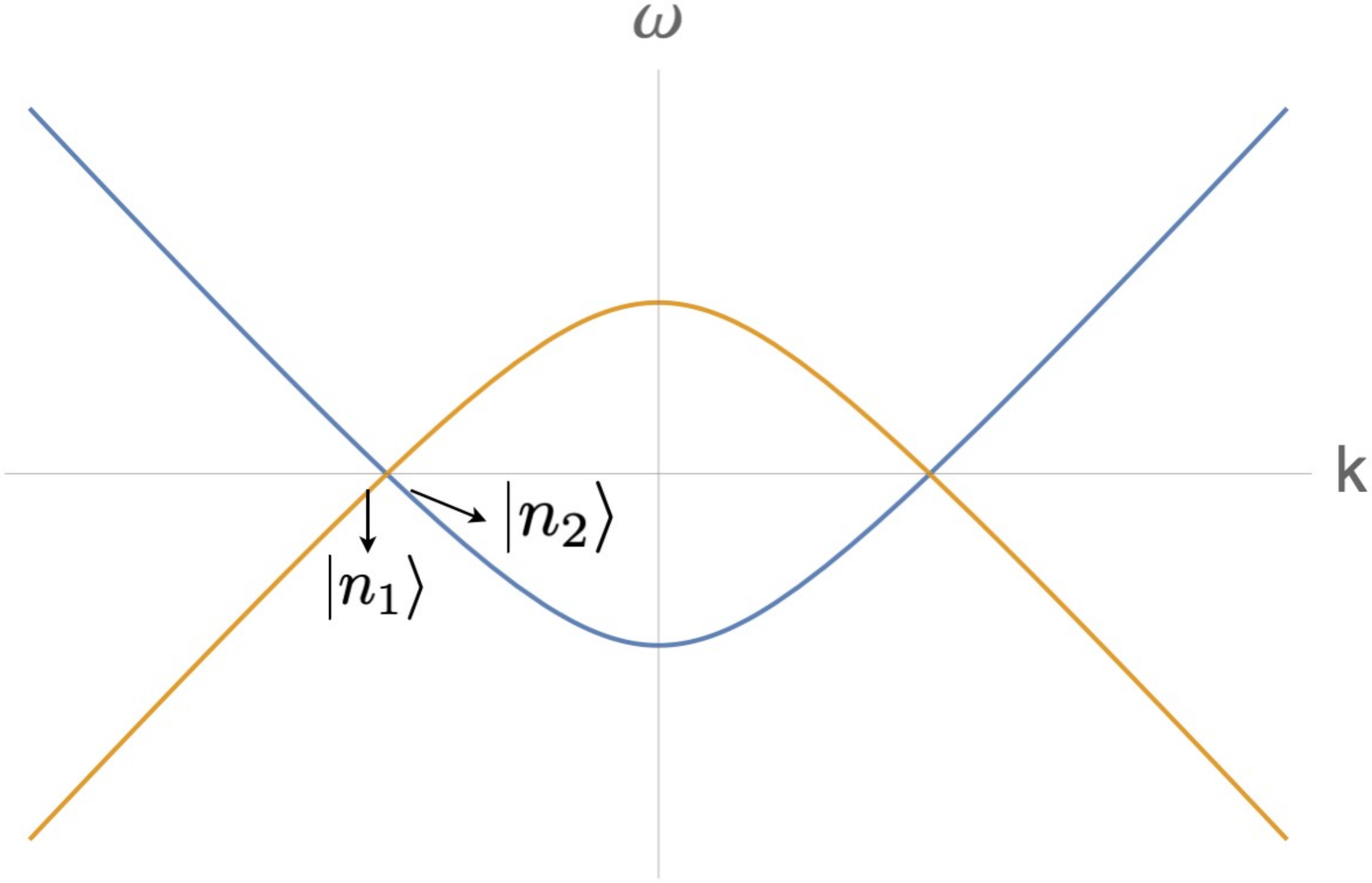}
  \caption{\small 
  $|n_1\rangle$ and $|n2\rangle$ are the left and right limiting states near one node. }
  \label{fig:withn1n2}
\end{figure}

At the singular node, the effective Hamiltonian would have two degenerate orthogonal eigenstates with the same eigenvalue. At the left and right limits of the singular mode in the occupied band (the band with energy lower than the crossing nodes), the two states $|n_1\rangle$ and $|n_2\rangle$ should be extremely close to one of these two degenerate eigenstates as shown in Fig. \ref{fig:withn1n2}. If the left and right limit states of the same {upper or lower} band are two different eigenstates, i.e. if they are orthogonal to each other, then the left and right limits cannot be connected together by performing a small perturbation {as the band crossing point is a singular point which cannot disappear from a small perturbation}. This means that two bands cannot be separated with a small perturbation.

Thus we could also use this fact to define a topological invariant: if the left and right limit states of the node in the same occupied band are orthogonal to each other ($|n_1\rangle$ and $|n_2\rangle$ states in Fig. \ref{fig:withn1n2}), the crossing node is a topologically nontrivial one and for trivial crossing nodes, the two states should be the same to each other up to a relative phase. {The latter case gives a topological invariant the same as the trivial vacuum state indicating a topologically trivial state. As mentioned above, a topological invariant could be a number or could be a quantity/property that is not a number as long as the property distinguishes different topological classes. Here it provides an example of a topological invariant which is not a number.} 

Thus a node would possess a nontrivial topological invariant when an undetermined Berry phase exists, i.e. the two left and right limit states are orthogonal, {which makes the Berry phase an undetermined one}. We name this method the undetermined Berry phase method, however, it in fact refers to the fact of orthogonality of adjacent states on the two sides of the nodes. This method is more useful for systems without a protecting symmetry while the calculation has to be on zero dimensional manifolds. In the following, we will use the first method (a modified version of the formula $\xi_0=N_1-N_2$) to calculate the topological invariant of the single 4D case and show that it is equivalent to the second method, i.e. the undetermined Berry phase method. Then we mainly use the second method to calculate the topological invariants for the more complicated systems. We emphasize again that these two methods are equivalent and they both rely on the fact that the two lower energy states on the two sides of the singular node belong to two different kinds of eigenstates which behave differently under deformations of the system. More details could also be found below for each case.

Before we proceed there is another point that we need to emphasize: the topological structure of the systems that we discussed above is a consequence of the spectrum while not directly related to the modification of the effective Hamiltonians or the modified Ward identities due to the non-conservation terms of the energy momentum tensor. This is similar to what happens to Weyl semimetal: the deformed Hamiltonian with a time reversal symmetry breaking terms produces the spectrum of a topological Weyl semimetal but the topological structure is not directly related to the deformed Hamiltonian.

Finally we need to emphasize that, though the calculation of topological invariants was first done for quantum electronic systems, in principle topological invariants are also properties of the wave phenomenology while not essentially associated to the quantum nature. In classical wave systems, topological invariants could also be defined in the same way as for quantum systems, e.g. Berry phases are thought to be quantum quantities as they are usually associated with quantum mechanical interference, however, it can in principle occur wherever phase interference phenomena exist and are governed by Hermitian eigenvalue problems, as in classical topological systems \cite{Haldane21}.
 
\vspace{.3cm}

\noindent {\bf The single 4D system}


For the single 4D system in section \ref{subsec:single4D}, there are four nodes for $m<b$ when the $y$ and $z$ directions have no $m$ terms. The nodes disappear with $m$ terms in the $y$ and $z$ directions, thus the nodes should be topologically nontrivial ones protected by the special spacetime symmetry in the $y$ and $z$ directions shown in section \ref{sec:rel2}. Here we first review the calculation of the topological invariant for this system as in \cite{Liu:2020ksx}. 




The calculation of the topological invariant for a symmetry protected state is different from the calculation for an ordinary topological state. For systems protected by a certain symmetry, we could calculate the topological invariant at a high symmetric point in the momentum space. The same to the case of mirror symmetry protected nodal line semimetals discussed in \cite{nlsm}, at the high symmetric point the protecting symmetry $M$ commutes with the effective Hamiltonian and the eigenstates would each have an eigenvalue of the reflection symmetry $M$ in our system: $M: y\to -y, z\to -z$. We have to find the eigenvalues of the states under this symmetry. 

The high symmetric point here should be $k_y=k_z=0$, which is invariant under the reflection symmetry transformation $M$. Note that the high symmetric point is the one whose wave function is invariant under the protecting symmetry and here the high symmetric point being $k_y=k_z=0$ is not related to the fact that the system has no band crossing at nonzero $k_y$ and $k_z$. Also the existence of a high symmetric point does not mean that the symmetry of the system requires $k_y=k_z=0$: the symmetry is a restriction to the system while not the solutions and we only need $k_y=k_z=0$ for the calculation of the topological invariant for a symmetry protected topological state.

\begin{figure}[h!]
  \centering
  \includegraphics[width=0.440\textwidth]{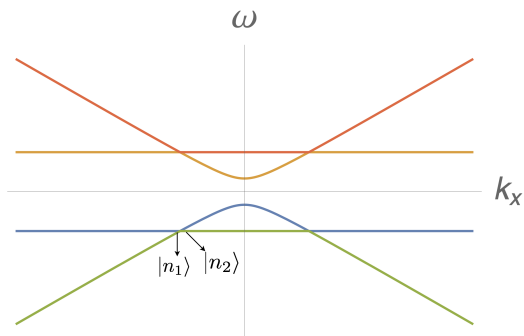}
  \caption{\small The left and right limit of the eigenstates $|n_1\ket$ and $|n_2\ket$ at the crossing point of the modified hydrodynamics with dynamical equation (\ref{eq:4dhydro}). Note that in this figure, different from some of the previous illustration figures, each color denotes  a separate band, which, however, may come from different types of eigenstates of the system as we use the fact wether the value of the energy is larger or lower than the crossing node to distinguish between different bands while we use the continuous functions of eigenvalues to distinguish between different sets of eigenstates of the system.}
  \label{fig:TI4d}
\end{figure}

The solutions are symmetric for $\omega \to -\omega$ and we could focus on the lower two nodes in figure \ref{fig:TI4d}. In the figure, the different two eigenstates of the effective Hamiltonian are denoted in different colors.
In the language of the first method of the topological invariant calculation, the two eigenstates have two different eigenvalues under the symmetry $M$. Now we calculate the topological invariant using this method in detail.

The formula for the topological invariant should be $\xi=N_1-N_2$, where in our case here, $N_i$ is the number of occupied bands at point $p_i$ with eigenvalue of the reflection symmetry $M$ to be $1$. In figure \ref{fig:TI4d}, $|n_i>$ is the occupied state at point $p_i$ and we have $|n_1\ket=\frac{1}{\sqrt{2}}\big(0,0,-i,1\big)$ and $|n_2\ket=\frac{1}{\sqrt{\frac{1}{v_s^2}-1}}\Big(\frac{\sqrt{m^2+k_x^2}}{v_s(m+i k_x)},i,0,0\Big)$. As already shown in section \ref{sec:symmetry}, the protecting symmetry required here is the reflection symmetry in both $y$ and $z$ directions. Under this symmetry, the state $|n_1\ket$ has eigenvalue $-1$  and the state $|n_2\ket$ has eigenvalue $1$. Thus from the formula $\xi=N_1-N_2$, the topological invariant at each node is $\xi=1$ or $\xi=-1$, which is a nontrivial value in contrast to the trivial value of $0$, confirming that the nodes are protected by a nontrivial topological charge.

Now we calculate the topological invariant using the second method, i.e. the fact of  $|n_i\ket$ being orthogonal to each other. For the left node at $k_x=k_1$, the green solutions at the left limit $k_x\to k_{1-}$ and the right limit $k_x\to k_{1+}$ are denoted as $|n_1\ket$ and $|n_2\ket$ separately.  If the node is an accidental crossing which is not topologically protected, the two bands should easily be separated by a gap and the two solutions $|n_1\ket$ and $|n_2\ket$ should be equal to each other or at most differ by a relative phase. If the node is indeed a topologically protected one, the  two solutions should be different indicating that there exists a singular point in between the two states at $k_x=k_1$. In all the systems that we study here we could still use a Berry phase between the two states $e^{-i \alpha}=\frac{\bra n_1|n_2\ket}{|\bra n_1|n_2\ket |}$ to denote the topological invariant here where when the Berry phase is an undetermined one, i.e. the two states are orthogonal to each other, the system is topologically nontrivial as the two states can not be connected without passing a singularity.

In this single 4D case, $|n_1\ket=\frac{1}{\sqrt{2}}\big(0,0,-i,1\big)$ and $|n_2\ket=\frac{1}{\sqrt{\frac{1}{v_s^2}-1}}\Big(\frac{\sqrt{m^2+k_x^2}}{v_s(m+i k_x)},i,0,0\Big)$. We could easily see that $\bra n_1|n_2\ket=0$, which means that the Berry phase is undetermined and the two states cannot connect together smoothly without passing through a singular point in between. Thus the two bands cannot be separated easily by a gap without going through a topological phase transition. Similar behavior of an undetermined Berry phase has also happened for the holographic nodal line semimetals \cite{Liu:2018djq, Landsteiner:2019kxb}. 


This result confirms that the four nodes in this case are topologically nontrivial protected by the special spacetime symmetry in the $y$ and $z$ directions. At the same time, the Berry phase accumulated through the whole circle around this node would be trivial indicating that it is indeed topologically trivial without  the special spacetime symmetries in the $y$ and $z$ directions. 
\vspace{.3cm}

\noindent {\bf The 2D+2D system}

\begin{figure}[h!]
  \centering
  \includegraphics[width=0.440\textwidth]{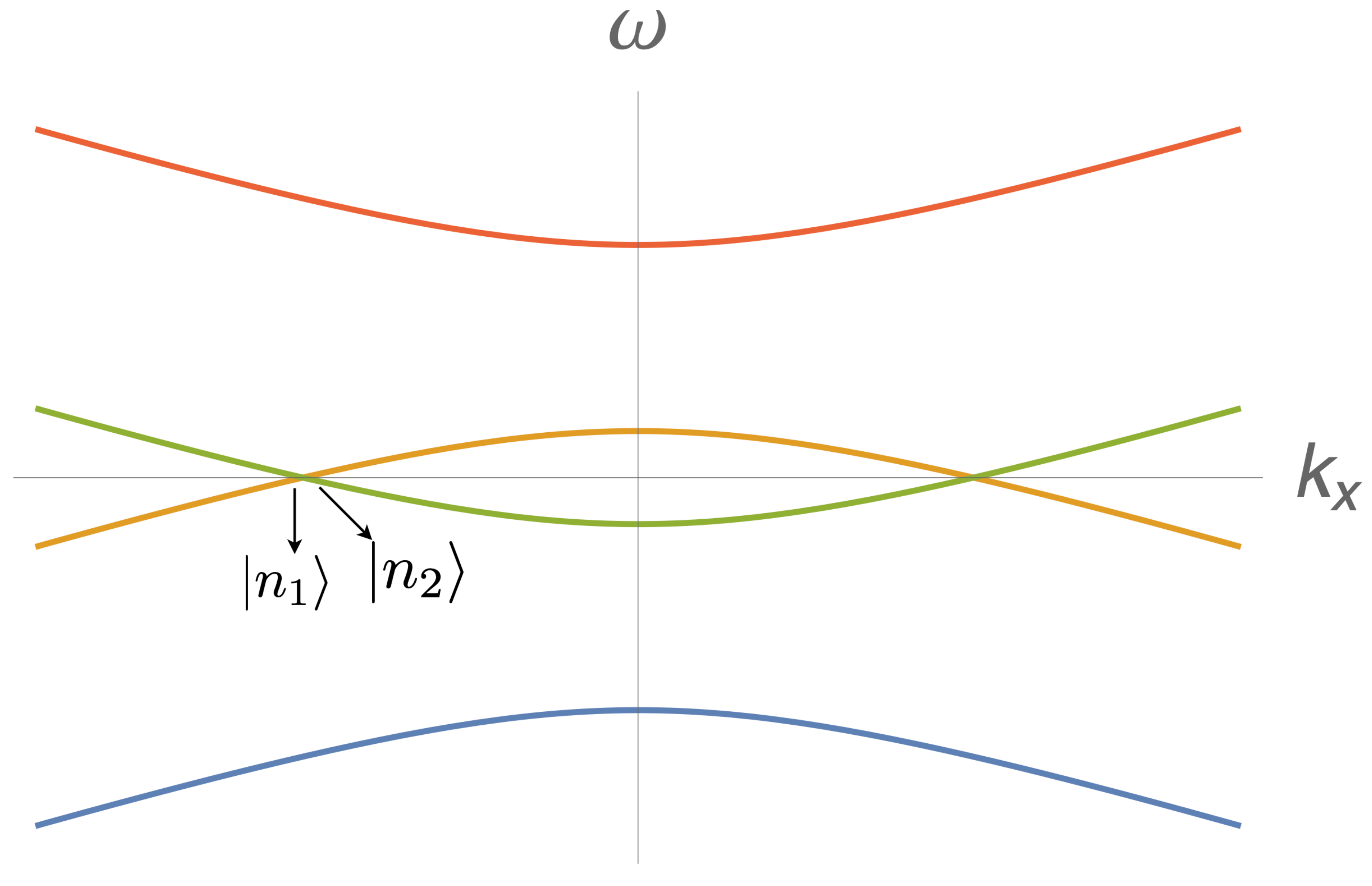}
  \caption{\small The left and right limits of eigenstates $|n_1\ket$ and $|n_2\ket$ at the crossing point of the modified hydrodynamics with dynamical equation (\ref{eq:2dhydro}). }
  \label{fig:TI2d}
\end{figure}

This is a topological band crossing state without the protection of a symmetry. For the 2D+2D case, there is only one spatial direction, i.e. the $x$ direction. As shown in figure \ref{fig:TI2d} different bands are plotted in different colors. There are two nodes when $m<b$. Again, we denote the two solutions of the same blue band on the two sides of the singular node as $|n_1\ket$ and $|n_2\ket$. Similar to the 4D case, we have $\bra n_1|n_2\ket=0$ from numerical calculations, which means that the Berry phase is undetermined. This proves that the two nodes are topologically nontrivial as the two bands cannot separate without passing through a topological phase transition, and in this case it does not require the existence of any symmetry.

\vspace{.3cm}

\noindent {\bf The 3D+3D/4D+4D case I}

\begin{figure}[h!]
  \centering
  \includegraphics[width=0.440\textwidth]{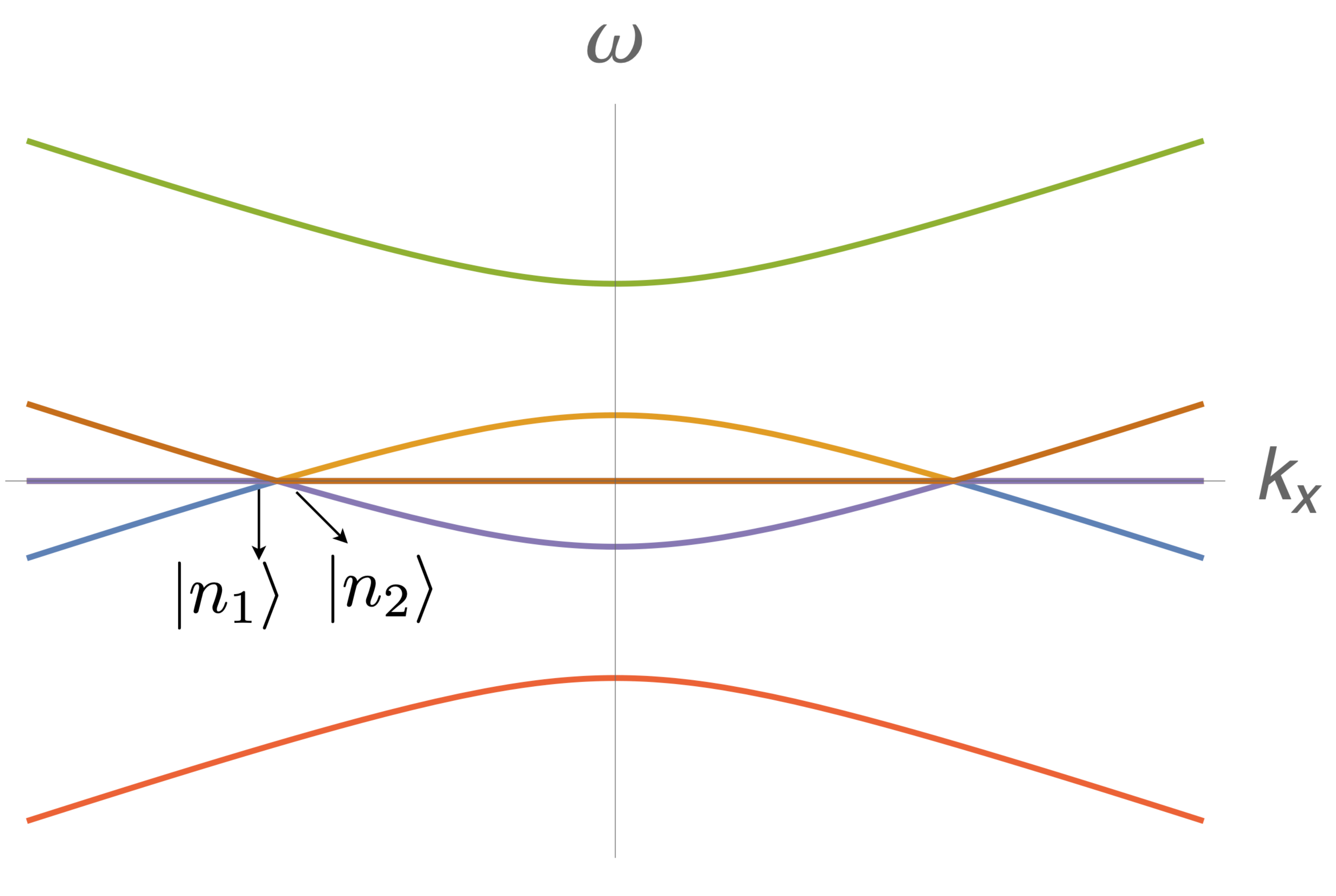}~~~
  \includegraphics[width=0.440\textwidth]{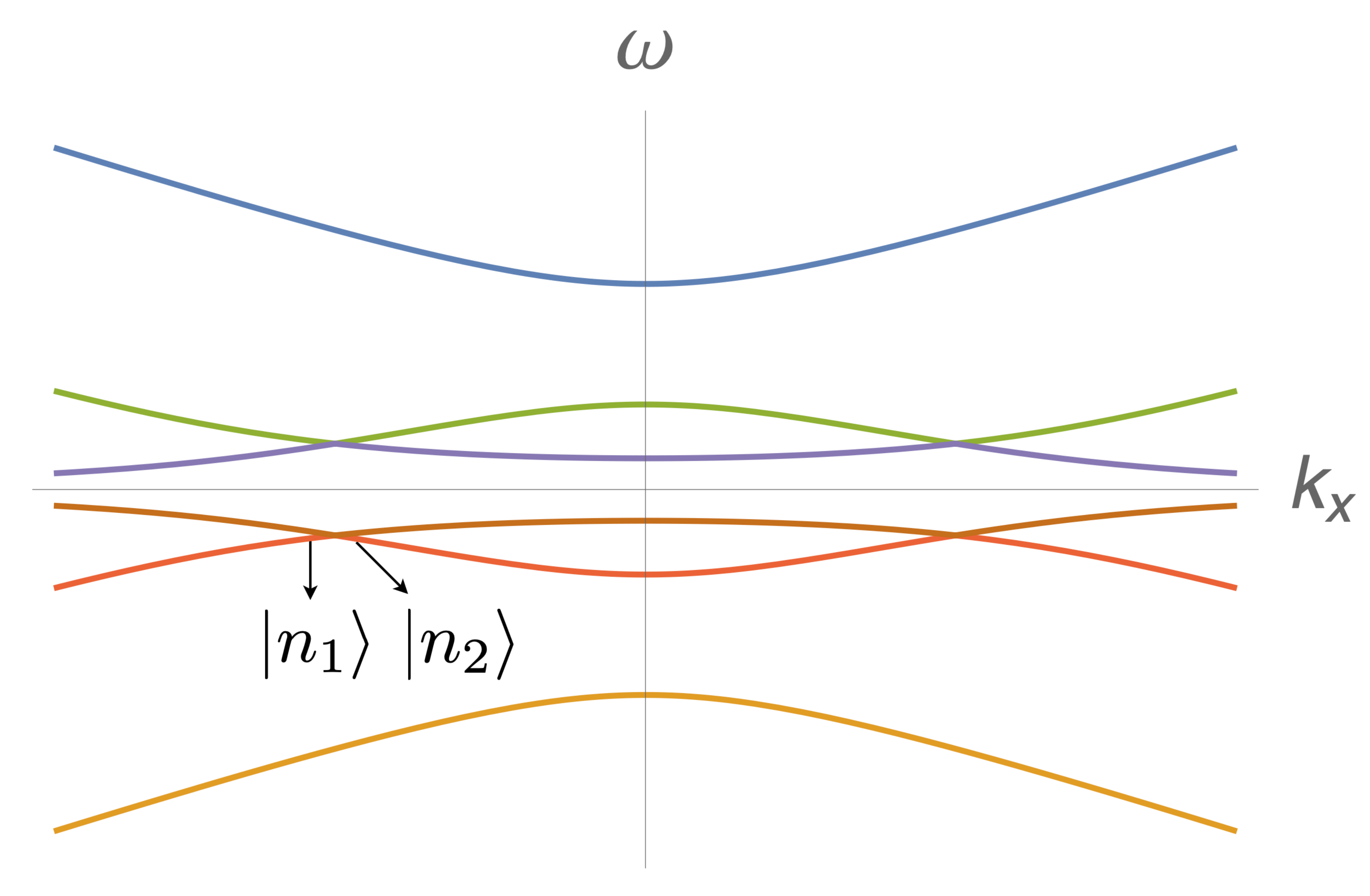}
  \caption{\small The left and right limit of the eigenstates $|n_1\ket$ and $|n_2\ket$ at  the crossing point of the modified hydrodynamics with dynamical equation (\ref{eq:3D3D}). The left plot is for the case of a single $m$ term while the right plot is for the case with $m$ terms in both $x$ and $y$ directions.}
  \label{fig:TI3d3d}
\end{figure}

The structure of the spectrum in this case has been described in section \ref{subsec:moreposs}. Here we calculate the Berry phase numerically and confirm the topological structure in that section. In the 3D+3D case, as shown in the left plot of figure \ref{fig:TI3d3d}, for $m<b$ there are two nodes in the $\omega, k_x$ plane at $k_y=0$ which look qualitatively the same as in the 2D+2D case. In this 3D+3D case, the two nodes in fact form two circles with the extra $k_y$ direction. Again, we use $|n_1\ket$ and $|n_2\ket$ to denote the two states from the same band at the left and right limits of the node $k_x=k_1$.  At the high symmetric point $k_y=0$ we have checked numerically that $\bra n_1|n_2\ket=0$, which again gives undetermined Berry phase, meaning that the two nodes in this case are topologically nontrivial protected by the special spacetime symmetry that forbids the $m$ term in the $y$ direction. This symmetry should be similar to the one of the single 4D case and the explicit form of this symmetry will be left for future work.

Besides these two nodes, we have shown that with an $m$ term in the $y$ direction, the two nodes disappear and form four nodes as shown in the right plot of figure \ref{fig:TI3d3d}, 
and now we have two separated circles with the extra spatial dimension. We have checked explicitly by numerical calculations of the Berry phase that these new four circles are still topologically nontrivial, which do not require any protection of symmetry for the 3D+3D  case. We summarize this behavior in figure \ref{fig:3d3dill}. where we can see that two circles of nodes are pinned together at two nodes. The two nodes are topologically nontrivial under the protection of a special spacetime symmetry in the $y$ direction and the two circles are topologically nontrivial without the need of any symmetry.

For the 4D+4D case, as we have two more spatial directions, depending on the existence of one or two extra mass terms, the topological structure would be more complicated. However, qualitatively the behavior is the same as the 3D+3D case above and we skip the details here.
\vspace{.3cm}




\noindent {\bf The 3D+3D/4D+4D systems with maximal $b$ terms} 

In fact the 2D+2D case above also belongs to this category, i.e. it has maximal $b$ terms in its spatial direction: the $k_x$ direction. These three cases are qualitatively the same which are topologically nontrivial without protection of any symmetry and have crossing nodes forming codimension $1$ surfaces in the momentum space. Thus in all three cases, the results for the Berry phase are also similar. As the nodes form a codimension $1$ surface, there is only one extra independent direction left: the two sides of the surface. To check whether these nodes are topologically nontrivial we only need to check if the two states on the two sides of the surface could be the same or not. 


It has been checked numerically that the at the two sides of the crossing nodes, the eigenvectors of the effective Hamiltonians are orthogonal to each other in the limit close to the nodes, meaning that the Berry phase is undetermined and the two states cannot be the same by small deformations. Thus the codimension $1$ nodes in these cases are topologically nontrivial and it does not require the protection of any symmetry.

In summary, in all cases of this section we have undetermined Berry phase, which makes the nature of all these cases qualitatively the same. The nodes are either topologically nontrivial (when the nodes form codimension $1$ surfaces) or topologically nontrivial under protection of certain spacetime symmetries of extra spatial dimensions (when the nodes form surfaces with codimension larger than $1$).



\section{Transport properties}
\label{sec:tp}

In this section we show how the extra energy momentum non-conervation terms affect the transport properties of the systems besides changing the spectrum of the hydrodynamic modes. Without other conserved currents, the most interesting transport coefficient in these systems is the heat transport. In the following we provide details for the calculation of the heat transport for the single 4D case in \cite{Liu:2020ksx}. and generalize to the 2D+2D case, while other cases would have qualitatively similar properties.

\subsection{Single 4D heat transport}
\label{sec:tran4d}
Heat transport can be computed from the linear response theory. We will follow the calculations in \cite{Kovtun:2012rj, Davison:2014lua} for the momentum dissipative hydrodynamics system to compute the heat transport for our new hydrodynamical system. 

We first consider the single 4D system studied in \cite{Liu:2020ksx} and in section \ref{subsec:single4D}. We focus on the case with $m_1=m_2=m$ and $b_1=b_2=b$. Perturbing the system of (\ref{eq:4dhydro}) by the fluctuations $(\delta T/T, \delta u^x,\delta u^y,\delta u^z)$ which couple to $(\delta\epsilon,\delta\pi_x, \delta\pi_y, \delta\pi_z)$ in the conservation equation (\ref{eq:4dhydro}), and performing Laplace and Fourier transformations in time and spatial directions respectively we obtain 
\bea
\begin{split}
(\omega \delta\epsilon-i \delta\epsilon^{(0)})-(k_x\delta\pi_x+k_y\delta\pi_y+k_z\delta\pi_z)-im\delta\pi_x&=0\,,\\
(\omega \delta \pi_x-i\delta \pi_x^{(0)})-v_s^2k_x\delta\epsilon-\frac{i\eta}{\epsilon+P}\vec{k}^2\delta\pi_x+\frac{i\left(\frac{1}{3}\eta+\zeta\right)}{\epsilon+P}k_xk_i\delta\pi_i+im v_s^2 \delta\epsilon&=0\,,\\
(\omega\delta \pi_y-i\delta \pi_y^{(0)})-v_s^2k_y\delta\epsilon-\frac{i\eta}{\epsilon+P}\vec{k}^2\delta\pi_y+\frac{i\left(\frac{1}{3}\eta+\zeta\right)}{\epsilon+P}k_yk_i\delta\pi_i-ib v_s\delta\pi_z&=0\,,\\
(\omega\delta \pi_z-i\delta \pi_z^{(0)})-v_s^2k_z\delta\epsilon-\frac{i\eta}{\epsilon+P}\vec{k}^2\delta\pi_z+\frac{i\left(\frac{1}{3}\eta+\zeta\right)}{\epsilon+P}k_zk_i\delta\pi_i+ib v_s\delta\pi_y&=0\,,
\end{split}
\eea
where $k_i\delta\pi_i=k_x\delta\pi_x+k_y\delta\pi_y+k_z\delta\pi_z$ and $\epsilon^{(0)}, \delta \pi_i^{(0)}$ are the perturbations at initial time. 

Solving these equations and following \cite{Kovtun:2012rj, Davison:2014lua} we obtain the Green function at $k_y=k_z=0$ 
\bea
\begin{split}
G_{\pi_x\pi_x}(\omega, k_x)&=-(\epsilon+P)\frac{(k_x^2+m^2)v_s^2+i\frac{\eta}{\epsilon+P}\omega k_x^2}{(k_x^2+m^2)v_s^2+i\frac{\eta}{\epsilon+P}\omega k_x^2-\omega^2}\,,\\
G_{\pi_y\pi_y}(\omega, k_x)&=G_{\pi_z\pi_z}(\omega, k_x)=
-(\epsilon+P)\frac{b^2v_s^2+k_x^2\frac{\eta}{\epsilon+P}(k_x^2\frac{\eta}{\epsilon+P}+i\omega)}{b^2 v_s^2+(i\omega+\frac{\eta}{\epsilon+P}k_x^2)^2}\,,\\
G_{\pi_y\pi_z}(\omega, k_x)&=-G_{\pi_z\pi_y}(\omega, k_x)=\frac{i\omega (\epsilon+P)dv_s}{b^2 v_s^2+(i\omega+\frac{\eta}{\epsilon+P}k_x^2)^2}\,.
\end{split}
\eea

From the Kubo formula 
\be\label{eq:kfheat}
\kappa(\omega, k_x)=\frac{i}{\omega T}\big(G^R_{\pi\pi}(\omega,k_x)-G^R_{\pi\pi}(0,k_x)\big)\ee 
we have
\bea
\begin{split}
\kappa_{xx}(\omega, k_x)&=-\frac{i\omega(\epsilon+P)}{T\left((k_x^2+m^2)v_s^2+i\frac{\eta}{\epsilon+P}\omega k_x^2-\omega^2\right)}\,,\\
\kappa_{yy}(\omega, k_x)&=\kappa_{zz}(\omega, k_x)=
-\frac{k_x^2\eta+i\omega(\epsilon+P)}{T\left(b^2 v_s^2+(i\omega+\frac{\eta}{\epsilon+P}k_x^2)^2\right)}\,,\\
\kappa_{yz}(\omega, k_x)&=-\kappa_{zy}(\omega, k_x)=\frac{ (\epsilon+P)bv_s}{T\left(b^2 v_s^2+(i\omega+\frac{\eta}{\epsilon+P}k_x^2)^2\right)}\,.
\end{split}
\eea

From the above formulas, it is easy to see that when $m=b=0$, all components of the DC heat transport diverge. For generic $m$ and $b$, we have vanishing DC heat transport $\kappa_{xx}(0,0), \kappa_{yy}(0,0)$ and 
$\kappa_{zz}(0,0)$ while $\kappa_{yz}(0,0)=-\kappa_{zy}(0,0)=\frac{\epsilon+P}{Tbv_s}.$ These $m$ and $b$ terms eliminate the unphysical divergence of DC heat transports and lead to interesting vanishing DC heat transport behavior. 

\subsection{2D+2D heat transport}

We continue to the calculations of 2D+2D heat transports using the linear response theory. Perturbing the system of (\ref{eq:2dhydro}) by the fluctuations $(\delta T/T, \delta u^x)_{L,R}$ which couple to $(\delta\epsilon,\delta\pi)_{L,R}$ in the conservation equation (\ref{eq:2dhydro}), and performing Laplace and Fourier transformations in time and spatial directions respectively we obtain 
\bea\label{eq:2d2devoluation}
\begin{split}
\omega\delta\epsilon_L-i\delta \epsilon_L^{(0)}
&=(k_x+im_1) \delta \pi_L+ib_1\delta \epsilon_R \,,\\
\omega\delta\pi_L-i\delta \pi_L^{(0)}&=(k_x-im_1) v_{sL}^2\delta \epsilon_L+ib_1\delta \pi_R\,,\\
\omega\delta\epsilon_R-i\delta \epsilon_R^{(0)}
&=(k_x+im_2) \delta \pi_R-ib_2\delta \epsilon_L \,,\\
\omega\delta\pi_R-i\delta \pi_R^{(0)}&=(k_x-im_2) v_{sR}^2 \delta \epsilon_R-ib_2 \delta \pi_L\,.
\end{split}
\eea

We consider the simplest case with the assumption that these two 2D systems $L$- and $R$- sectors have the same constitutive equations and $m_1=m_2=m, b_1=b_2=b$ and $v_{sL}=v_{sR}=v_s$. 
With equations (\ref{eq:2d2devoluation}) one can obtain the Green function following the linear response theory \cite{Kovtun:2012rj, Davison:2014lua} 
\bea
\begin{split}
G_{\pi_L\pi_L}(\omega, k_x)&=G_{\pi_R\pi_R}(\omega, k_x)=\frac{(\epsilon+P)(\omega^2(b^2+(m^2+k_x^2)v_s^2)-(b^2-(m^2+k_x^2)v_s^2)^2)}{\left((k_x^2+m^2)v_s^2-\omega^2-b^2\right)^2-4b^2\omega^2}\,,\\
G_{\pi_L\pi_R}(\omega, k_x)&=
-G_{\pi_R \pi_L}(\omega, k_x)=\frac{-ib\omega(\epsilon+P)(b^2-(m^2+k_x^2)v_s^2-\omega^2)}{\left((k^2+m^2)v_s^2-\omega^2-b^2\right)^2-4b^2\omega^2}\,.
\end{split}
\eea

From Kubo formula (\ref{eq:kfheat}) we obtain
\bea
&&\kappa_{LL}(\omega, k_x)=\kappa_{RR}(\omega, k_x)=-\frac{i\omega (\epsilon+P)\left(b^2+(k_x^2+m^2) v_s^2-\omega^2\right)}{T\left(b^4+\left((k_x^2+m^2)v_s^2-\omega^2\right)^2-2b^2\left((k_x^2+m^2)v_s^2+\omega^2\right)\right)}\,,\nn\\
&&\kappa_{LR}(\omega, k_x)=-\kappa_{RL}(\omega, k_x)=\frac{(\epsilon+P)\left(b^2-(k_x^2+m^2)v_s^2-\omega^2\right)}{T\left(b^4+\left((k_x^2+m^2)v_s^2-\omega^2\right)^2-2b^2\left((k_x^2+m^2)v_s^2+\omega^2\right)\right)}\,.\nn
\eea

When $m=b= 0$, we have divergent DC heat transports $\kappa_{LL}(0,0),\kappa_{RR}(0,0)$. For general $m$ and $b$, we have vanishing DC heat transport. The poles of heat transports are exactly the same as the spectrum of the effective  Hamiltonian (\ref{eq:2D2DH}) in section \ref{sec:2d2d}. This is due to the fact that there are no dissipation terms at the first derivative term in the constitutive equation. 
It is easy to check that when close to the poles (\ref{eq:2d2dspec}), the heat transports diverge.  


\section{$\mathcal{O}(k^2)$ effects}
\label{sec:2nd}

Up to now we have focused on the first order in $k$ effect, where energy momentum non-conservation terms change the spectrum of the modes, while there are no dissipative terms at leading order. In this section, we consider $\mathcal{O}(k^2)$ effects, which in original hydrodynamics provide dissipative effects and lead to imaginary parts in the poles.

As an example, we consider the single 4D case studied in section \ref{subsec:single4D}. We take into account all the second order in $k$ terms in the constitutive equations. The effective Hamiltonian up to $\mathcal{O}(k^2)$ becomes
\be\label{eq:Hk2}
H=\begin{pmatrix} 
0 & ~~k_x+im & ~~k_y & ~~ k_z \\
(k_x-im)v_{s}^2 & ~~\frac{-i\eta k^2-i(\frac{\eta}{3}+\zeta)k_x^2}{\epsilon+P}& ~~ -i\frac{(\frac{\eta}{3}+\zeta)k_xk_y}{\epsilon+P} & ~~ -i\frac{(\frac{\eta}{3}+\zeta)k_xk_z}{\epsilon+P}  \\
k_yv_{s}^2 & ~~ -i\frac{(\frac{\eta}{3}+\zeta)k_xk_y}{\epsilon+P}  & ~~ \frac{-i\eta k^2-i(\frac{\eta}{3}+\zeta)k_y^2}{\epsilon+P} & ~~-i\frac{(\frac{\eta}{3}+\zeta)k_yk_z}{\epsilon+P} +ib v_s \\
k_z v_{s}^2 & ~~-i\frac{(\frac{\eta}{3}+\zeta)k_xk_z}{\epsilon+P}  & ~~-i\frac{(\frac{\eta}{3}+\zeta)k_yk_z}{\epsilon+P} -ib v_s & ~~ \frac{-i\eta k^2-i(\frac{\eta}{3}+\zeta)k_z^2}{\epsilon+P} 
\end{pmatrix}\,.
\ee

The $k^2$ terms are dissipative as we can see that they make the effective Hamiltonian matrix  non-Hermitian. Here we still keep terms at $m\sim b$ order while not $m^2\sim b^2$ order assuming that $m\sim k^2$ in this part.

From the eigenvalues of (\ref{eq:Hk2}) we find that the real part has not  changed while imaginary parts appear. At the four nodes, the imaginary parts are not zero indicating that the four nodes are dissipative in comparison to nondissipative nodes at $\omega=0$ in the usual hydrodynamics. 

\begin{figure}[h!]
  \centering
  \includegraphics[width=0.40\textwidth]{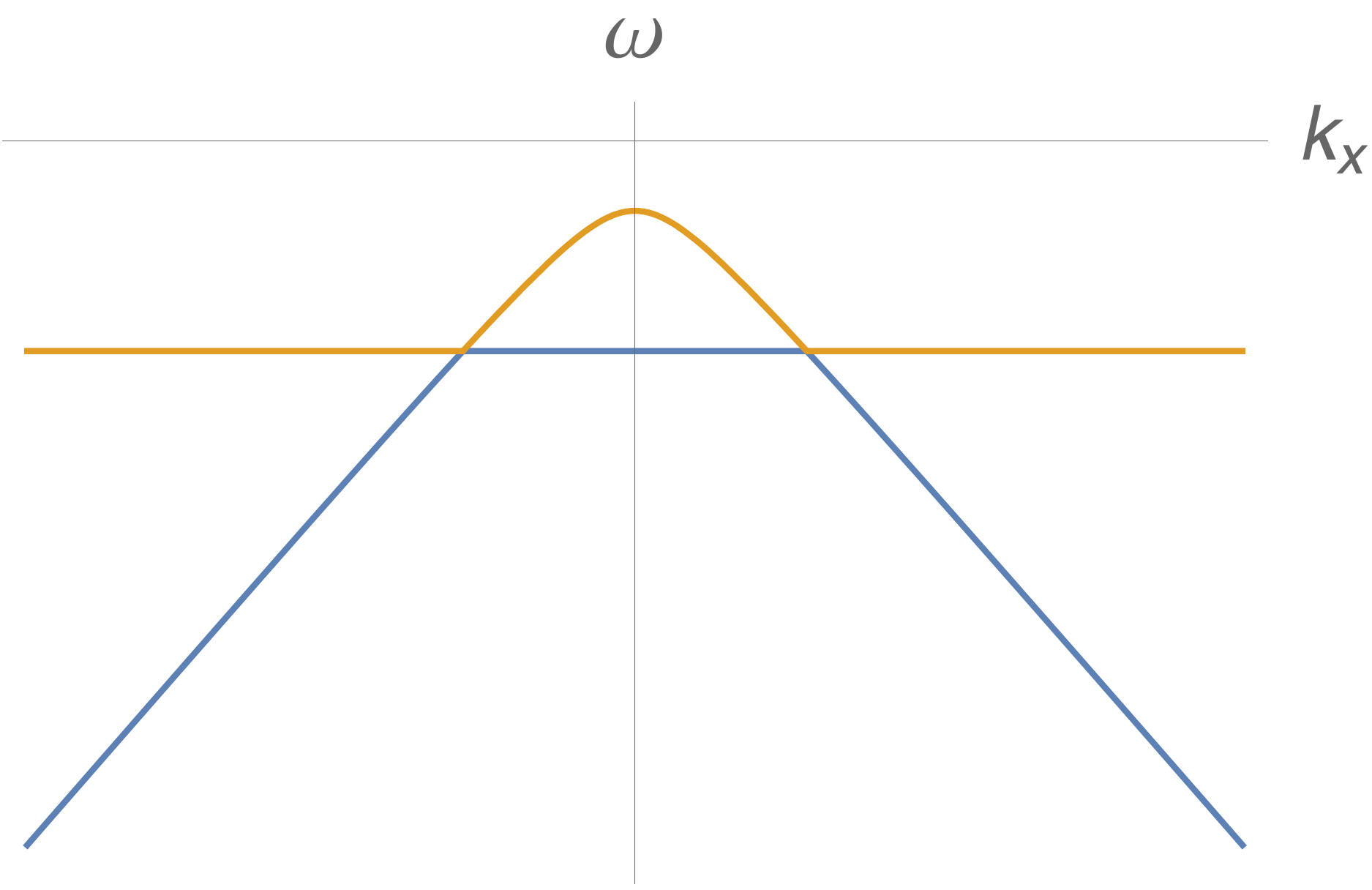}~~~~~
  \includegraphics[width=0.40\textwidth]{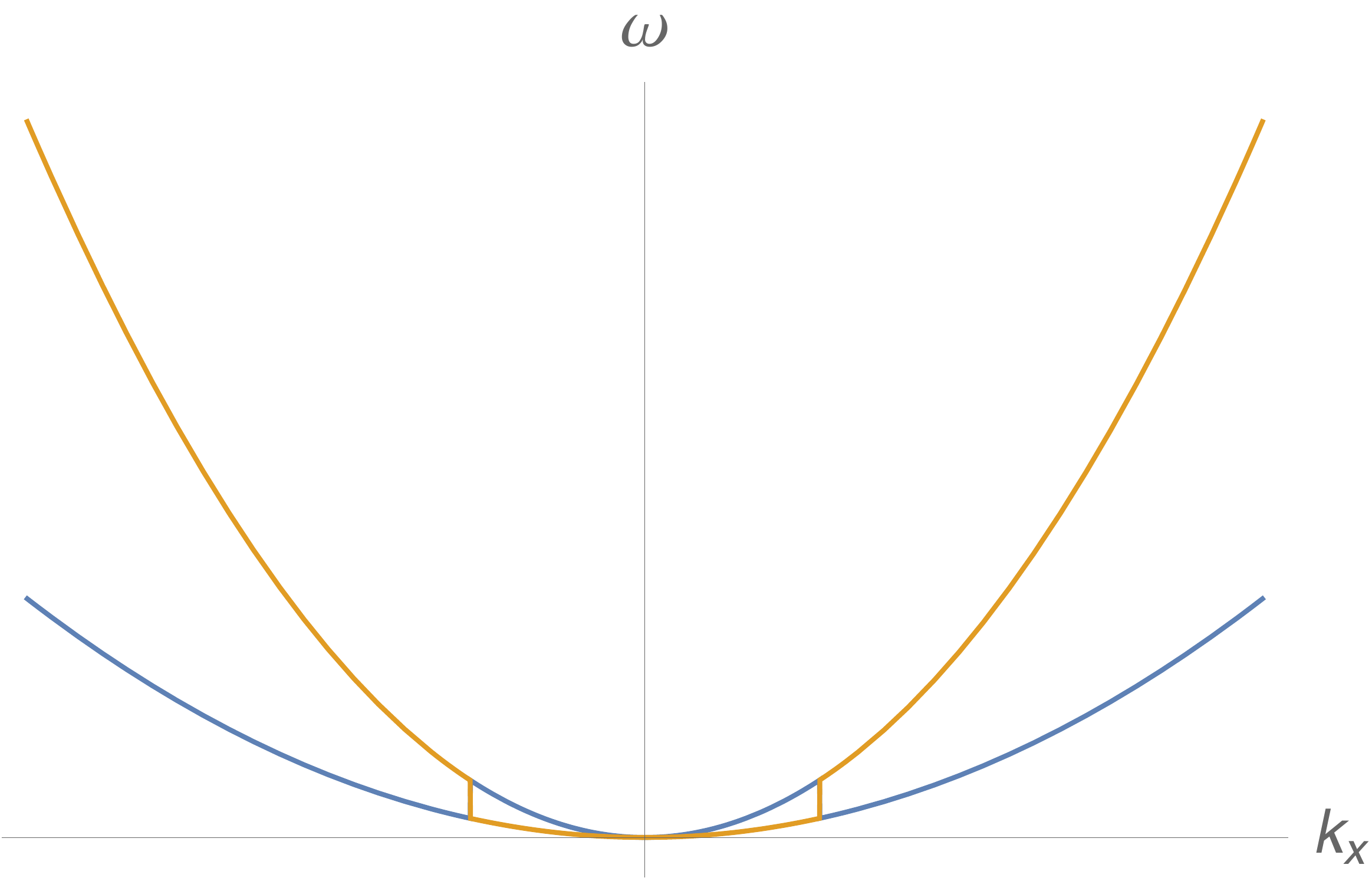}
  \caption{\small The real and imaginary parts for the lower two bands of the modified hydrodynamics with (\ref{eq:Hk2}) in the left plot of figure \ref{fig:4D1}. The same color in the two plots corresponds to the same band. }
  \label{fig:k2}
\end{figure}

As we could see from figure \ref{fig:k2}, the imaginary part for each of the band has a jump at the crossing nodes at $k_y=0$ in the $k_x$ axis, i.e. the imaginary parts of the same band are different at the left and right limits of the singular node. This behavior is similar to the behavior of the eigenstates when calculating the Berry phase and this
 provides another piece of evidence of the existence of a symmetry protected topological singular node.

For the 2D+2D case, there are no dissipative terms and for other cases the results are similar which we skip here. Note that the Berry phase results would not be affected by $k^2$ terms as these at most change the zero result of $|\bra n_1|n_2\ket|$ to a small number proportional to $k^2$ and this suggests that the two states are still far from being equivalent for which $|\bra n_1|n_2\ket |=1$.

\section{Ward identities and holographic realization}
\label{sec:wihr}

The physics of hydrodynamics has been studied extensively in holography for strongly coupled systems  \cite{Son:2007vk,Rangamani:2009xk}. Many interesting results have been obtained, e.g. the famous KSS bound for the shear viscosity over entropy density ratio \cite{Kovtun:2004de}. In this section we focus on the holographic realization of the topological hydrodynamic modes above. As a first step, we need to have a holographic system with the same non-conservation equations for the energy momentum tensor. In holography, to break translational symmetry, we could either introduce external fields into the system or start from massive gravity \cite{Vegh:2013sk} which breaks diffeomorphism itself. In this paper, we choose to view $h_{\mu\nu}$ as the gravitational field that comes from a reference frame transformation from the original inertial reference frame. We will start from an ordinary holographic system and perform a reference frame transformation to obtain a holographic system whose ward identities for the energy momentum tensor have the same form as those from the non-conservation equations in previous sections. 



Here we focus on the single 4D case with only one energy momentum tensor. We will first derive the Ward identities for the energy momentum tensor for the hydrodynamic system with non-conservation terms from section \ref{subsec:single4D}. Then we show how in holography we could get these general Ward identities without calculating out all the Green function components. 


\subsection{Ward identities}
\label{subsec:wi}

For conserved energy momentum tensor, the Ward identities are \cite{Policastro:2002tn}
\be
k_{\mu}(G^{\mu\nu\lambda\rho}-\eta^{\nu\lambda}\langle T^{\mu\rho}\rangle-\eta^{\nu\rho}\langle T^{\mu\lambda}\rangle-\eta^{\lambda\rho}\langle T^{\mu\nu}\rangle+\eta^{\mu\nu}\langle T^{\lambda\rho}\rangle)=0\,,
\ee where $G^{\mu\nu\lambda\rho}$ is the Green function for the energy momentum tensor. This Ward identity could easily be checked holographically by calculating out all components of Green functions for $T^{\mu\nu}$.

Now we calculate the Ward identities for the single 4D system with non-conservation terms of $T^{\mu\nu}$. In the case that these non-conservation terms come from a gravitational field, we could start from the covariant conservation equation $\nabla_{\mu}T^{\mu\nu}=0$ and differentiate it with respect to $ g_{\lambda\rho}$, we obtain the Ward identities in the  momentum space
\bea
\begin{split}
k_\mu\bigg[ -G^{\mu\nu,\lambda\rho}(k)+
g^{\lambda\rho} \langle T^{\mu\nu}\rangle+g^{\nu\lambda} \langle T^{\mu\rho}\rangle+g^{\nu\rho} \langle T^{\mu\lambda}\rangle-g^{\mu\nu} \langle T^{\lambda\rho}\rangle
\bigg]
&\\
-i\bigg[\big(g^{\mu\rho}\Gamma^\lambda_{~\alpha\mu}+g^{\mu\lambda}\Gamma^\rho_{~\alpha\mu}\big)\langle T^{\alpha\nu}\rangle
+\big(g^{\nu\rho}\Gamma^\lambda_{~\alpha\mu}+g^{\nu\lambda}\Gamma^\rho_{~\alpha\mu}\big)\langle T^{\alpha\mu} \rangle \bigg]
&\\
-i\Big[\Gamma^\mu_{~\mu\alpha}G^{\alpha\nu,\lambda\rho}(k)+\Gamma^\nu_{~\mu\alpha}G^{\mu\alpha,\lambda\rho}(k)\Big]=0\,,~~~&
\end{split}
 \eea where $\Gamma^\nu_{~\mu\alpha}$ is the Christoffel tensor for the metric $g_{\mu\nu}=\eta_{\mu\nu}+h_{\mu\nu}$. To the first order in $h_{\mu\nu}$ we could rewrite the Ward identities above as
\be\label{eq:ward}
k_\mu G^{\mu\nu,\lambda\rho}(k)+i\Big[\Gamma^{(1)\mu}_{~~~~\mu\alpha}G^{\alpha\nu,\lambda\rho}(k)+\Gamma^{(1)\nu}_{~~~~\mu\alpha}G^{\mu\alpha,\lambda\rho}(k)\Big]+\text{contact terms}=0\,,
\ee where the explicit form of the contact terms is omitted. With nonzero $h_{\mu\nu}$ several components of $\Gamma^{(1)\nu}_{~~~~\mu\alpha}$ would be nonzero and contribute extra terms to the Ward identities compared to the ones in hydrodynamics systems with conserved $T^{\mu\nu}$.

Note that we could as well choose to treat $h_{\mu\nu}$ as an external effective matter field as in section \ref{sec:real1}. We start from the non-conservation equation (\ref{fmunu}) and perform the same procedure as above. The final result is exactly the same as (\ref{eq:ward}).  


\subsection{Holographic realization}


In this section, we show that the Ward identities obtained in the previous subsection could be reproduced from a holographic system whose boundary theory lives in a non-inertial frame,\footnote{Previous studies on holography for field theories on curved spacetime could be found in e.g. \cite{Marolf:2013ioa}.} i.e. we start from the usual AdS Schwartzchild black hole and perform coordinate transformations so that the boundary metric becomes $g_{\mu\nu}=\eta_{\mu\nu}+h_{\mu\nu}$. In this way, we get a non-inertial reference frame version of AdS/CFT correspondence and obtain a holographic realization of the non-conservation equations for $T^{\mu\nu}$ introduced in the single 4D hydrodynamic system. More evidence, e.g. direct calculations of hydrodynamic modes and Green functions as well as holographic realizations for other systems in section \ref{sec:tophm} will be provided in future work. 


In the rest of this section we will first show how to derive Ward identities for an ordinary holographic system without calculating out all components of the Green functions. Then we use the same procedure to get the Ward identities for the holographic non-inertial reference frame system and show that they are exactly the same as (\ref{eq:ward}).

\subsubsection{Prescription for holographic Ward identities}


To calculate Green functions for the energy momentum tensor in holography, we switch on perturbations of the metric field and solve all the fields from their equations of motion. Then we substitute the solutions into the action to get the on-shell action and differentiate the on-shell action with the source terms to get corresponding Green functions. For holographic systems with simple gravity backgrounds, this calculation is simple, however, with $h_{\mu\nu}$ fields turned on and $k_x$, $k_y$, $k_z$ all nonzero, the on-shell action and the solutions would be much more complicated. Here we show that we do not need the details of the on-shell action and could still get the Ward identities for the energy momentum tensor. 

For the purpose of calculating the Green functions for $T^{\mu\nu}$ we have ten independent fields of $\delta g_{\mu\nu}$, which are the $tt,tx,ty,tz,xx,xy,xz,yy,yz,zz$ components. Here for simplicity we could denote these ten fields as $\phi_i$, $i\in\{1...,10\}$. $\delta g_{r\mu} $ could also be nonzero but will finally be eliminated from the on-shell action using its equation of motion. 

The action of $\delta g_{\mu\nu}(\vec{k})$ could be written as
\be
S\supset \int \frac{drd^4k}{(2\pi)^4}\bigg( W_1^{ij}\phi_i''(-\vec{k})\phi_j(\vec{k})+W_2^{ij}\phi_i'(-\vec{k})\phi_j'(\vec{k})+W_3^{ij}\phi_i'(-\vec{k})\phi_j(\vec{k})+W_4^{ij}\phi_i(-\vec{k})\phi_j(\vec{k})\bigg),\nn
\ee
where $\phi_i(\vec{k})$ are functions of $r$ only and here $\vec{k}=(\omega, k_x, k_y, k_z)$ and $W_{1,2,3,4}^{ij}$ are functions of $r$ whose explicit form depends on the system. Equations of motion could be derived from this action and after substituting the solutions into this action, the on-shell action that is relevant to the Green functions becomes
\be
S_{\rm on-shell}\supset \int \frac{d^4k}{(2\pi)^4} W_2^{ij}\phi_i'(-\vec{k})\phi_j(\vec{k})\Big{|}_{r_h}^{r_b}+\cdots,
\ee
where $\cdots$ are terms that are only related to the contact terms and $\phi_i(\vec{k})$ in the above formula have to be substituted by their on-shell solutions. $r_h$ and $r_b$ denote the horizon and the boundary separately. For AdS $W_{1,2,3,4}^{ij}$ could be calculated for all components, however, we show in the following that we are still able to derive the Ward identities without the explicit form of these terms, which manifest the physical origin of the Ward identities more clearly. 

Due to the diffeomorphism invariance of the gravity system, the action has to be invariant under coordinate transformations.  
Under coordinate transformations generated by $\epsilon_{\mu}(t,x,y,z)$, fields transform as
\be\label{transform}
\delta g_{\mu\nu}=\nabla_\mu \epsilon_\nu+\nabla_\nu \epsilon_\mu\,,
\ee where $\nabla$ is performed with respect to the background AdS metric.  In this way each component  $\delta g_{\mu\nu}$ cannot be invariant under general coordinate transformations, while some combinations of several components of $\delta g_{\mu\nu}$ could be a gauge invariant and the action must be written into sums of gauge invariant combinations due to its diffeomorphism invariance. 

We choose the nonzero components of $\epsilon_{\mu}$ to be $\epsilon_{t,x,y,z}(\vec{k})$ with dependence on momenta in all directions and $r$ dependence does not affect the calculations. From (\ref{transform}) six independent gauge invariants $Z_i, i=1,...,6$ under these general gauge transformations could be found in total and all possible sums of these gauge invariants are also gauge invariants. The six  independent gauge invariants are
\bea\label{Zis1}
\begin{split}
Z_1&=\frac{\delta g_{xx}}{2 k_x^2}+\frac{\delta g_{tx}}{\omega k_x}+\frac{\delta g_{tt}}{2 \omega^2}\,,~~~
Z_2=\frac{\delta g_{yy}}{2 k_y^2}+\frac{\delta g_{ty}}{\omega k_y}+\frac{\delta g_{tt}}{2 \omega^2}\,,\\
Z_3&=\frac{\delta g_{zz}}{2 k_z^2}+\frac{\delta g_{tz}}{\omega k_z}+\frac{\delta g_{tt}}{2 \omega^2}\,,~~~
Z_4=\frac{\delta g_{xx}}{2 k_x^2}-\frac{\delta g_{xy}}{k_x k_y}+\frac{\delta g_{yy}}{2k_y^2}\,,\\
Z_5&=\frac{\delta g_{xx}}{2 k_x^2}-\frac{\delta g_{xz}}{ k_x k_z}+\frac{\delta g_{zz}}{2 k_z^2}\,,~~~
Z_6=\frac{\delta g_{yy}}{2 k_y^2}-\frac{\delta g_{yz}}{k_y k_x}+\frac{\delta g_{zz}}{2 k_z^2}\,.
\end{split}
\eea
The most general form of the on-shell action could be written as
\be
S\supset \int \frac{d^4k}{(2\pi)^4} G_{ij}(r)Z_{i}'(-\vec{k})Z_{j}(\vec{k})\Big{|}_{r_h}^{r_b}\,,
\ee 
where $G_{ij}$ is a real and symmetric function matrix with $G_{ij}=G_{ji}$, $i, j=1,\dots, 6$. The Green functions could be calculated from the on-shell action above as functions of the variables $G_{ij}$, for example,  
\be G_{tt,tt}= \frac{\delta^2 S}{ \delta (\delta g_{tt}^b) \delta(\delta g_{tt}^b )} =\frac{1}{2\omega^4}(G_{11} + 2 G_{12} + 2 G_{13} + G_{22} + 2 G_{23} + G_{33})\,,
\ee 
and  
\be G_{tx,tx}= \frac{1}{4} \frac{\delta^2 S}{ \delta (\delta g_{tx}^b) \delta (\delta g_{tx}^b) }=\frac{1}{2k_x^2\omega^2}G_{11}\,,
\ee 
where $\delta g^b_{\mu\nu}$ denotes the boundary value of these fields.

Green functions have 55 independent components while $G_{ij}$ have 21 independent variables. Expressing all Green function components using $G_{ij}$ and eliminating $G_{ij}$ from these expressions, we are left with 34 identities. In this system, the Ward identity we see from the field theory derivation has 40 identities, 6 of which are independent that could be derived from other 34 identities. Thus the number of total independent Ward identities are the same from the two sides. We have checked through tedious calculations that after eliminating $G_{ij}$ the remaining 34 Ward identities are exactly the Ward identities (\ref{eq:ward}).

Note that the appearance of any $k_{\mu}$ in the expression of the action indicates that a derivative is taken in that direction. From the combinations, it seems that the total derivatives including the $r$ derivatives may exceed two which should be the highest order of derivatives in Einstein gravity. This is because the on-shell action is obtained from solving the constraint equations of $\delta g_{r\mu}$ and substituting the expressions of $\delta g_{r\mu}$ into the action, which brings higher orders into the action. 

\subsubsection{Ward identities in holographic non-inertial frame systems}

Now let us come to the system with energy momentum non-conservation as in section  \ref{subsec:single4D}. To obtain the deformed metric in the bulk after a reference frame transformation to the non-inertial frame, we first choose the reference frame transformation $\xi_{\mu} $ to be the same as in the field theory part. The AdS background now becomes $g_{\mu\nu}^{\text{bulk}}=g^{AdS}_{\mu\nu}+h_{\mu\nu}^{\text{bulk}}$, where $h_{\mu\nu}^{\text{bulk}}=\nabla_\mu \xi_\nu+\nabla_\nu \xi_\mu$ with $\nabla_\mu$ performed with respect to the bulk background metric.

At leading order in $m$ and $b$, only nonzero components of $h_{\mu\nu}^{\text{bulk}}$ at the boundary are 
\bea
\begin{split}
h_{tt}^{\text{bulk}}= m x g_{tt}^{\text{bulk}},~ h_{tx}^{\text{bulk}}=\frac{1}{2}m(1+v_s^2)g_{tt}^{\text{bulk}},~h_{xx}^{\text{bulk}}=m xg_{xx}^{\text{bulk}}, \\
h_{ty}^{\text{bulk}}=-\frac{1}{2}b v_s z g_{tt}^{\text{bulk}}, h_{tz}^{\text{bulk}}=\frac{1}{2}b v_s y g_{tt}^{\text{bulk}}.
\end{split}
\eea 
Note that in the bulk the expressions for these fields would be more complicated, but we only need the boundary expressions in deriving the Green functions.

With the new metric, the form of the on-shell action would be different from the AdS one, nevertheless, it can still be written as sums of gauge invariant terms. However, now the combinations (\ref{Zis1}) are not gauge invariant under coordinate transformations anymore. We need to derive the new gauge invariant combinations in this new background. 

We choose $\epsilon_{\mu}$ to be $\epsilon_{\mu}(t,x,y,z)$, where $\mu=t,x,y,z$. As the background metric now depends on $t, x,y,z$, we keep the $t, x,y,z$ dependence of $\epsilon_{\mu}$.  We could write out the transformations of all components of $\delta g_{\mu\nu}$ under this general coordinate transformation and look for all independent combinations of the components which remain unchanged under this coordinate transformation.  After we find out the gauge invariant combinations we can transform to the $k$ space as the background metric only has contributions at $k=0$ though it has nontrivial dependence on the coordinates.   Up to leading order in $m$ and $b$, the new gauge invariant combinations could be found to be 
\bea
\begin{split}
Z_1&=\frac{\delta g_{xx}}{2 k_x^2}+\frac{\delta g_{tx}}{\omega k_x}+\frac{\delta g_{tt}}{2 \omega^2}-\frac{i m \delta g_{tt}}{4k_x \omega^2}+\frac{i m \delta g_{tx}}{2k_x^2\omega}-\frac{i m v_s^2 \delta g_{xx}}{4 k_x \omega^2}\,, \\
Z_2&=\frac{\delta g_{yy}}{2 k_y^2}+\frac{\delta g_{ty}}{\omega k_y}+\frac{\delta g_{tt}}{2 \omega^2}-\frac{i m v_s^2 \delta g_{xx}}{4 k_x \omega^2}-\frac{i b v_s \delta g_{zz}}{4 k_y k_z \omega}\,, \\
Z_3&=\frac{\delta g_{zz}}{2 k_z^2}+\frac{\delta g_{tz}}{\omega k_z}+\frac{\delta g_{tt}}{2 \omega^2}-\frac{i m v_s^2 \delta g_{xx}}{4 k_x\omega^2}+\frac{i b v_s \delta g_{yy}}{4 k_y k_z \omega}\,,\\
Z_4&=\frac{\delta g_{xx}}{2 k_x^2}-\frac{\delta g_{xy}}{k_x k_y}+\frac{\delta g_{yy}}{2k_y^2}-\frac{i m \delta g_{xx}}{4 k_x^3}\,, \\
Z_5&=\frac{\delta g_{xx}}{2 k_x^2}-\frac{\delta g_{xz}}{ k_x k_z}+\frac{\delta g_{zz}}{2 k_z^2}\,,\\
Z_6&=\frac{\delta g_{yy}}{2 k_y^2}-\frac{\delta g_{yz}}{k_y k_x}+\frac{\delta g_{zz}}{2 k_z^2}-\frac{i m  \delta g_{xx}}{4 k_x^3}\,.
\end{split}
\eea

The on-shell action at the boundary is now
\be
S\supset \int \frac{d^4k}{(2\pi)^4} G_{ij}(r)Z_{i}'(-\vec{k})Z_{j}(\vec{k})\Big{|}_{r_b}+...\,,
\ee where $\cdots$ terms are horizon contributions and contact terms related parts. $G_{ij}$ is still a real and symmetric function matrix, while the values should be different from the AdS ones. Now the Green functions could again be obtained from the on-shell action. As an example we write out two of the components here
\bea 
 \begin{split}
G_{tt,tt}=\frac{\delta^2 S}{ \delta (\delta g_{tt}^b) \delta(\delta g_{tt}^b )}=&\frac{1}{2\omega^4}(G_{11} + 2 G_{12} + 2 G_{13} + G_{22} + 2 G_{23} + G_{33})\\
&~~ -\frac{im}{ k_x \omega^4}\bigg(G_{12} + G_{13} + G_{22} + 2 G_{23} + G_{33} \bigg)\,,
\end{split}
\eea 
and 
 \bea 
 \begin{split}
 G_{tx,tx}&= \frac{1}{4} \frac{\delta^2 S}{ \delta (\delta g_{tx}^b) \delta (\delta g_{tx}^b) }\\
& =\frac{1}{2k_x^2\omega^2}G_{11}-\frac{im}{2}\bigg( -\frac{G_{11}v_s^2}{k_x\omega^4} - \frac{G_{12}v_s^2}{k_x\omega^4} - \frac{G_{13}v_s^2}{k_x\omega^4} + 
 \frac{  G_{11}}{k_x^3\omega^2}- \frac{G_{14}}{k_x^3\omega^2} - \frac{G_{15}}{k_x^3\omega^2} \bigg)\,.
 \end{split}
 \eea 
 There are extra $m$ and 
 $b$ terms in the expressions for the Green functions now. Again we could eliminate all $G_{ij}$ and get the resulting Ward identities. Here we take a simpler method: by substituting the resulting Green functions which are functions of $G_{ij}$ into the field theory Ward identities, we could see that all the $G_{ij}$ cancel out in the Ward identities leaving only components of the Green functions and the field theory Ward identities indeed hold in holography ignoring all contact term contributions. This confirms that the holographic set-up indeed gives the holographic system of the hydrodynamic system with topological modes. 



Note that the Ward identities in the new non-inertial reference frame come from covariance of the energy momentum conservation. Here in holography, the on-shell action would still transform back to the AdS one after a backward reference frame transformation. The difference in the Ward identities reflects non-inertial effects.

More details about Green functions and the hydrodynamic modes from Green functions would be reported in a future work. We will also consider other possibilities of holographic realizations, e.g. viewing $h_{\mu\nu}$ as external fields while not gravitons from reference frame transformations or from massive gravities. Holographic systems for other dimensions and for systems with gapped hydrodynamic topological modes will be looked at, too.

\section{Conclusions and discussions}
\label{sec:cd}

We have studied topological hydrodynamic modes in relativistic hydrodynamics by introducing non-conservation terms for the energy momentum tensor. There are several systems where we could get topologically nontrivial hydrodynamic modes. Depending on different forms of the $T^{\mu\nu}$ non-conservation terms, the resulting crossing nodes in the spectrum of hydrodynamic modes could either be topologically nontrivial under the protection of special spacetime symmetries in certain dimensions or directly topologically nontrivial. We confirm their nontrivial topology from the calculation of nontrivial topological invariants for all these systems. We also studied the dissipative $O(k^2)$ effects and show that different from the original $\omega={\bf k}=0$ crossing nodes for relativistic hydrodynamics, the crossing nodes in these systems are mostly dissipative. The discontinuous imaginary parts for the hydrodynamic modes provides another piece of evidence of the nontrivial topology of the crossing nodes. 

The non-conservation terms for $T^{\mu\nu}$ could come from an effective external rank two symmetric matter field or from a gravitational field. In the latter case, the system could be viewed as to be in a non-inertial reference frame indicating that topologically nontrivial modes could arise from trivial modes by transforming to a specific non-inertial reference frame. We propose a holographic realization of one of these system, starting from ordinary AdS/CFT correspondence and perform a transformation to a non-inertial frame. We derive the Ward identities for the non-conserved equation of $T^{\mu\nu}$ and match them to the holographic realization of this system. 
 
There are many open questions and possible generalizations at this moment which we hopefully will address in the future. The following is an incomplete list of them. 
\begin{itemize}

\item We have focused on band crossing hydrodynamic modes. Is it possible to also find topologically nontrivial modes without band crossings in a classical relativistic hydrodynamic system?

\item What would be possible experimental observable effects besides the transport coefficients discussed in this paper and the sudden increase of amplitude at the crossing node? Would the graphene system be a possible arena for the test of the theoretical predictions in this paper?

\item Is it possible to have similar topological modes in non-relativistic hydrodynamics\footnote{Similar topological modes with deformed effective Hamiltonians in a non-relativistic system has been studied in \cite{np-earthphysics}  for Lamb waves/ acoustic-gravity waves in earth physics. We thank Karl Landsteiner for bringing this reference to us.} in non-inertial frame, which might be easier for laboratory tests? 

\item  We have shown that topologically trivial hydrodynamic modes could turn into nontrivial ones in a non-inertial reference frame. Would this behavior also hold for other topological states of matter, e.g. topological electronic systems? If yes, could this property be tested in a laboratory resting in a non-inertial reference frame?

\item Is it possible to find a similar $h_{\mu\nu}$ field from analog gravity systems, i.e. from certain materials?

\item We have focused on hydrodynamic systems without any other conserved charges except the energy momentum tensor. Would the structure be more complicated with more conserved charges?

\item Is it possible to find other holographic realizations of the single 4D system, e.g. from external matter field, or massive gravity \cite{Kiritsis:2008at}?

\item What are possible holographic set-ups for the 2D+2D systems? Possibly a bi-metric gravitational theory \cite{Schmidt-May:2015vnx} in the bulk could provide us a holographic example.

\item All the effective Hamiltonians studied in this paper are similar to Hermitian matrices. In our search for effective Hamiltonians that could give rise to interesting spectrum that we need, we focused on Hermitian matrices as most non-Hermitian matrices would give complex eigenvalues. This raises the interesting question if non-Hermitian matrices with PT symmetry could be found which also give similar while real spectrum and topological states here. Could we also have similar holographic non-Hermitian physics as in \cite{Arean:2019pom}? 

\item Though we have shown that the holographic set-up indeed reproduces the Ward identities for the system with $m$ and $b$ terms, what is the topological structure in the bulk? Is this related to some new global properties of spacetime?

\item In holography, hydrodynamic modes on the boundary corresponds to gravitational field in the bulk. What would the work in this paper imply for the physics of gravitational waves in asymptotically flat spacetime? Is it possible to find possible topological modes in gravitational waves?

\item It would be interesting to study the effect field theory description of hydrodynamics with nontrivial topological modes along the lines developed in \cite{Crossley:2015evo, Haehl:2015foa, Glorioso:2018wxw}. 
\end{itemize}



\subsection*{Acknowledgments}
We would like to thank Matteo Baggioli, Rong-Gen Cai, Hyun-Sik Jeong, Shu Lin, Wen-Bin Pan, Koenraad Schalm and especially Karl Landsteiner for useful discussions. This work is supported by the National Key R\&D Program of China (Grant No. 2018FYA0305800). The work of Y.L. was also supported by the National Natural Science Foundation of China grant No.11875083. The work of Y.W.S. has also been partly supported by starting grants from University of Chinese Academy of Sciences and Chinese Academy of Sciences, and by the Key Research Program of 
Chinese Academy of Sciences (Grant No. XDPB08-1), 
the Strategic Priority Research Program of Chinese Academy of Sciences, 
Grant No. XDB28000000, and by the National Natural Science Foundation of China Grant No. 12035016.. 

\appendix
\section{Hydrodynamics with momentum dissipation}
\label{sec:app}
In this appendix we show the effects of momentum dissipation terms in the dispersions of hydrodynamic modes for comparison. In contrast to the single 4D topological system in the main text, momentum dissipation terms that have been studied extensively in hydrodynamics and holography \cite{Hartnoll:2007ih, Davison:2014lua,  Grozdanov:2018fic, Hartnoll:2016apf}, contribute to imaginary parts of the spectrum, thus indicating that they are dissipative terms, different from the terms for topological modes above. { Note that we added this appendix only for readers not to mix our system with the momentum dissipation systems which have been studied a lot in holography. In the system in this appendix, momentum dissipation could be caused by external fields as well as broken diffeomorphism invariance and in all cases that have been studied $\Gamma$ is real while not imaginary. Even if the spectrum in that case could be real, that has no relation to topology, at least not any that we could directly see. Our system has a nontrivial topological structure not because their spectrum is real, i.e. we are not trying to find a real spectrum in hydrodynamics and claim that this is topologically nontrivial because it has real spectrum. Real spectrum is a very basic and necessary requirement for us to analyze the topological structure of the system but it does not mean that as long as we have real spectrum, we would have a nontrivial topological structure. Whether the system has a nontrivial topological structure depends on the structure of the effective Hamiltonian or equivalently the structure of the spectrum. }

 For relativistic hydrodynamics with weak momentum dissipation, the conservation equations change to \cite{Hartnoll:2007ih, Davison:2014lua, Hartnoll:2016apf} 
\bea
\partial_\mu \delta T^{\mu t}&=&0\,,\nn\\
\partial_\mu \delta T^{\mu i}&=&\Gamma \delta T^{t i}\,,\nn
\eea
where $\Gamma\ll 1/T.$ The momentum is almost conserved and weakly broken. 

Up to the first order in derivatives, the four dimensional conservation equation can be written as 
\be
i\partial_t\Psi=H\Psi
\ee
where
\be
\Psi=\begin{pmatrix} 
\delta \epsilon_L  \\
\delta \pi_L \\
\delta \epsilon_R  \\
\delta \pi_R
\end{pmatrix}\,,~~~~~~
H=\begin{pmatrix} 
0 & ~~k_x & ~~ k_y & ~~ k_z \\
v_{s}^2 k_x& ~~-i\Gamma & ~~ 0 & ~~ 0 \\
v_{s}^2 k_y & ~~ 0 & ~~-i\Gamma & ~~ 0 \\
v_{s}^2 k_z & ~~0 & ~~ 0 & ~~ -i\Gamma
\end{pmatrix}\,.
\ee It is easy to see that $H$ is not Hermitian and the spectrum is 
\be \omega=\bigg(-i\Gamma, ~-i\Gamma, ~\frac{1}{2}\big( -i\Gamma-\sqrt{4v_s^2k^2-\Gamma^2}\big), ~\frac{1}{2}\big( -i\Gamma+\sqrt{4v_s^2k^2-\Gamma^2}\big)\bigg)\,.\ee

These terms contribute to imaginary parts of the spectrum, indicating that they are dissipative terms, different from the terms for topological modes studied in this paper.

\end{document}